\pdfoutput=1
\documentclass[12pt,twoside,final]{huthesis}

\usepackage{aas_macros}
\usepackage{epsfig,bm,epsf,float,pdfpages}
\usepackage{amsmath, amssymb, natbib, mydefs, deluxetable,afterpage}
\begin{document}
% --- UNDERLYING SPACING FOR WHOLE DOCUMENT:
 %\ssp
\dsp

% ----- Title Page -----
\title{Magnetic Influences on the Solar Wind}
\author{Lauren N. Woolsey}
\degreemonth{May}
\degreeyear{2016}
\degree{Doctor of Philosophy}
\field{Astronomy and Astrophysics}
\department{Astronomy}
\advisor{Steven R. Cranmer}

\maketitle
\copyrightpage

% ---- Frontmatter ----
\begin{abstract}
The steady, supersonic outflow from the Sun we call the solar wind was first posited in the 1950s and initial theories rightly linked the acceleration of the wind to the existence of the million-degree solar corona. Still today, the wind acceleration mechanisms and the coronal heating processes remain unsolved challenges in solar physics. In this work, I seek to answer a portion of the mystery by focusing on a particular acceleration process: Alfv\'en waves launched by the motion of magnetic field footpoints in the photosphere. The entire corona is threaded with magnetic loops and flux tubes that open up into the heliosphere. I have sought a better understanding of the role these magnetic fields play in determining solar wind properties in open flux tubes. After an introduction of relevant material, I discuss my parameter study of magnetic field profiles and the statistical understanding we can draw from the resulting steady-state wind. In the chapter following, I describe how I extended this work to consider time dependence in the turbulent heating by Alfv\'en waves in three dimensional simulations. The bursty nature of this heating led to a natural next step that expands my work to include not only the theoretical, but also a project to analyze observations of small network jets in the chromosphere and transition region, and the underlying photospheric magnetic field that forms thresholds in jet production. In summary, this work takes a broad look at the extent to which Alfv\'en-wave-driven turbulent heating can explain measured solar wind properties and other observed phenomena.
\end{abstract}

\newpage
\tableofcontents
\listoffigures
\begingroup
\let\clearpage\relax
\listoftables
\endgroup

\begin{citations}
\vspace{0.8in}
\ssp
\noindent
Chapter 2 and portions of Section 1.3 appear in the following:
\begin{quote}
	``Turbulence-driven Coronal Heating and Improvements to Empirical Forecasting of the Solar Wind,''
	L. N. Woolsey \& S. R. Cranmer. 
	ApJ {\bf 787}, 160 (2014), {\tt 1404.5998};
\end{quote}
Chapter 3 appears in its entirety in the following:
\begin{quote}
	``Time-Dependent Turbulent Heating of Open Flux Tubes in the Chromosphere, Corona, and Solar Wind,''
	L. N. Woolsey \& S. R. Cranmer. 
	ApJ {\bf 811}, 134 (2015), {\tt 1509.00377};
\end{quote}
Section \ref{sec:tempestclassroom} appears in its entirety in the following:
\begin{quote}
	``Solar wind modeling: a computational tool for the classroom,''
	L. N. Woolsey. 
	JRAEO {\bf 3}, (2015), {\tt 1505.00727};
\end{quote}
Appendix \ref{ap:ism} appears as the following:
\begin{quote}
	``An Essay on Interactive Investigations of the Zeeman Effect in the Interstellar Medium,''
	L. N. Woolsey. 
	JAESE {\bf 2}, 1 (2015), {\tt 1506.00932};
\end{quote}
Appendix \ref{ap:profiles} appears as an appendix to:
\begin{quote}
	``Turbulence-driven Coronal Heating and Improvements to Empirical Forecasting of the Solar Wind'',
	L. N. Woolsey \& S. R. Cranmer
	ApJ {\bf 787}, 160 (2014), {\tt 1404.5998};
\end{quote}

\noindent
Electronic preprints (shown in {\tt typewriter font}) are available
on the Internet at the following website (place number after final / in URL):
\begin{quote}
	{\tt http://arXiv.org/abs/}
\end{quote}
\end{citations}

\begin{acknowledgments}

\begin{quote}
\hsp
\raggedleft
Keep your face always toward the sunshine\\and shadows will fall behind you.\\
-Walt Whitman
\end{quote}
Of everything I've written for my doctorate over the last five years, these two pages may have been both the toughest and most rewarding to fill. There's no good way to order everyone, nor will I ever be able to include every name that ought to be here.

Before I joined the Harvard Department of Astronomy, my trajectory was set by my teachers and mentors. I do not have room to list each, but for a representative sample, I want to thank Christian Johnston (mathematics; middle school), Adrian Jones (physics; high school), and Douglas Hamilton (astronomy; college). 

Additionally, my ability to dive headfirst into research when I arrived is credited to incredible research advisors as an undergraduate, especially Herbert Frey at NASA Goddard, Douglas Hamilton at UMD, and Leonard Strachan at SAO (now NRL).

With all of the lead-up out of the way, I can now take a moment to thank the astronomy department and the entire Harvard-Smithsonian Center for Astrophysics for welcoming me into the program and providing a healthy environment for this work to grow. This includes my peers throughout my time here, all of the folks that attend SSP coffee, the solar-stellar X-ray group downstairs, rockstar administrators Peg Herlihy and Robb Scholten, and the faculty I learned from and taught with (especially Dave Charbonneau, Steve Cranmer, Bob Kirshner, and Phil Sadler). 

I am grateful for funding from the National Science Foundation (NSF) Graduate Research Fellowship Program, which helped provide the flexibility to easily do my doctoral thesis under the primary advisement of a Smithsonian astrophysicist, Steven R. Cranmer, even when he moved halfway across the United States! 

To Steve: {\it thank you}. You have been a phenomenal advisor. I think we probably both learned things over these last several years, and while I may have been your first graduate student, you are a natural mentor. I look forward to continuing to learn from you and collaborate with you long after this dissertation is officially accepted.

To my committee members Alyssa Goodman, Lars Hernquist, Justin Kasper, and John Raymond: thank you for reading this dissertation and helping me improve it. 

To Aleida, Sam, Mark, and all of my solar physics graduate student peers: thank you for always making SHINE and other conferences worth attending. Keep pushing!

To thesixtyone.com: thank you for providing a writing soundtrack to keep me sane and a sense of accomplishment on days when no progress seemed to be made.

To the Physical Sciences Department at Grand Rapids Community College: thank you for being my next step, the light at the end of my dissertation tunnel vision.

To my family: thank you for always supporting all of my endeavors, from graduate school to game design. I love you forever. I am thrilled you are attending both my defense and commencement, the celebrations would not have been complete without you. A special thanks to Mom for acting as editor and catching typos in this dissertation, even when you claim not to understand any of the content.

To Sean McKillop: the entirety of these acknowledgements would not have been enough space to express how fortunate I am to have you in my life. You have been by my side through graduate school, and I would not have finished the PhD without your love and support. I am endlessly grateful that you took the leap of faith and moved up here with me, and it has been pure joy to call you my boyfriend, my fianc\'ee, and now my husband. You will always be my best friend. Even as I am completing this chapter (five chapters) of my life, I am looking forward to the next ones with you.

\end{acknowledgments}

\newpage
\startarabicpagination

% ----- The Text -----

%% ch1_introduction.tex
\chapter{Introduction}

The solar wind is a constant presence throughout the heliosphere, affecting comet tails, planetary atmospheres, and the interface with the interstellar medium. It represents a vital part of space physics, as well as one of the most important mysteries that remains in the field. The solar atmosphere is characterized by a temperature profile that has puzzled scientists for decades. The photosphere, the visible surface of the Sun, is roughly 6000 K, while the diffuse corona above it is heated to millions of degrees. Observations have not been able to determine the mechanism(s) responsible for the extreme coronal heating, though many physical processes have been suggested. The same processes that might explain the temperature of the corona can also account for the acceleration of the solar wind. Only a small fraction of the mechanical energy in the Sun's sub-photospheric convection zone needs to be converted to heat in order to power the corona. However, it has proved exceedingly difficult to distinguish between competing theoretical models using existing observations. There are several recent summaries of these problems and controversies \cite[see, e.g.,][]{2012RSPTA.370.3217P, 2012SSRv..172...69M, 2014arXiv1412.2307C, 2015RSPTA.37340256K}.

\section{Historical Foundations}
While the study of the solar wind certainly has roots deep in the history of astronomy, the initial foundation that modern discussion of the field is laid upon was put into place in the 1940s. In fact, early work by Hannes Alfv\'en (1941, \cite{1941alfven}) was so widely accepted that it was considered common knowledge within the decade \citep{2014FrASS...1....2P}. Important pieces of the puzzle were brought to the table by studies of emission lines in the solar corona done by Grotrian (1939, \citep{1939NW.....27..214G}) and Edl\'en (1943, \citep{1943ZA.....22...30E}), but Alfv\'en was the first to truly discover and argue that the solar corona must be extremely hot compared to the photosphere (i.e. a million degrees). In the same paper, he notes, ``it is necessary to introduce forces due to the action of the sun's general magnetic field upon the charged particles'' \cite{1941alfven}. Alfv\'en also noted that the granulation pattern on the solar photosphere was likely producing turbulence and consequently magnetohydrodynamic (MHD) waves (which have since been named Alfv\'en waves) that brought energy up into the corona \cite{1947MNRAS.107..211A}. 

A coronal heating mechanism, such as turbulence, was also needed to explain the excess pressure above the predicted blackbody radiation pressure that was observed by the effect on cometary tails and the existence of iron ionization states that could not be created at the photospheric temperature. These effects implied an outflow from the Sun of hundreds of kilometers per second \cite{1948ZA.....25..161B, 1951ZA.....29..274B, 1957Obs....77..109B}. In 1957, Sydney Chapman and Harold Zirin showed that a million-degree corona would produce a scale height that implies a non-zero pressure at infinity \cite{1957SCoA....2....1C}. This framework allowed Eugene Parker to develop the theoretical proof that, for a spherically-symmetric, one-fluid isothermal outflow, the non-vanishing pressure at infinity was larger than reasonable values for the interstellar medium; the outflow must therefore expand supersonically into space \cite{1958ApJ...128..664P}. The temperature profile for Parker's simplified corona held a constant temperature until a radius $r = R_{\rm P}$, where it then decreases: 
\begin{equation}T(r) = T_{0}\left(\frac{R_{\rm P}}{r}\right)^{1/(n+1)}.\end{equation} 
Defining a dimensionless parameter $\lambda = GM_{\odot}m/2kT_{0}R_{\rm P}$, the pressure at infinity is given by \begin{equation} p(\infty) = p_{0} {\rm ~exp}\left[\frac{-\lambda (n+1)}{n}\right],\end{equation} where, for $n > 0.5$, the pressure at infinity is too great to allow hydrostatic equilibrium to hold at large distances \cite{1958ApJ...128..664P}. \begin{subequations}
Parker's equations of momentum and internal energy conservation are listed below:
\begin{align}
	Nu\frac{du}{dr} = -\frac{d}{dr}(2NkT/m) - \frac{GNM_{\odot}}{r^{2}},\\
	\frac{d}{dr}\left(r^{2}Nu\right) = 0,
\end{align}
\end{subequations}
where $N$ is the number density, $k$ is the Boltzmann constant, $u$ is the outflow speed of the wind, and $m$ is the average molecular mass. There is a family of solutions to these equations, and Parker showed that the solar wind must follow a ``critical'' solution that passes through a specific critical point where the outflow changes from subsonic to supersonic. There was concern throughout the community about the likelihood that the Sun would always find the critical solution, when there were so many ``solar breeze'' solutions that never become supersonic. However, Velli (1994, \citep{1994ApJ...432L..55V}) has since shown that this critical solar wind solution, as well as Bondi accretion for the opposite case, is stable for steady-state solutions and that solar breeze solutions evolve towards the critical solution when any perturbation is introduced into the system.

An important note about Parker's milestone work is that it did not include the MHD waves Alfv\'en discovered in the previous decade. The additional wave pressure source in the momentum equation from transverse, incompressible Alfv\'en waves creates a more complex solution topology. Multiple mathematically critical points can arise when more realistic temperature profiles are introduced, though only one remains physically meaningful. The revised momentum equation can be written as
\begin{equation}\label{parkermomentum}\left(u - \frac{u_{c}^{2}}{u}\right)\frac{du}{dr} = -\frac{GM_{\odot}}{r^{2}} - u_{c}^{2}\frac{d{\rm ln}B}{dr} - a^{2}\frac{d{\rm ln}T}{dr}.\end{equation}
Here, $B$ is the radial magnetic field as a function of height, which goes as $B(r) \propto r^{-2}$ at large heights; $a$ is the isothermal sound speed. For a given form of the critical speed $u_{c}$, critical radii are found by determining the location of roots of the right-hand side (RHS) of Eq. \eqref{parkermomentum}. At each critical point, the slope of the outflow is found using L'H\^opital's Rule, which gives \begin{equation} \frac{du}{dr} = \frac{d{\rm RHS}/dr}{2\frac{du}{dr} - 2\frac{du_{c}}{dr}},\end{equation} since $u = u_{c}$ at the critical point but $\frac{du}{dr} \ne \frac{du_{c}}{dr}$ at this point. Solving for the slope $\frac{du}{dr}$ at the critical point, one finds 
\begin{equation}
\label{critpoint}
\left. \frac{du}{dr}\right|_{r=r_{c}} = \frac{1}{2}  \left[ \frac{du_{c}}{dr} \pm \sqrt{\left(\frac{du_{c}}{dr}\right)^{2} + 2\frac{d{\rm RHS}}{dr}}\right].
\end{equation} 
The true critical point using Eq. \eqref{critpoint} must satisfy a few criteria: 1) the physical case for solar wind outflow is the positive case, as accretion would take the negative case, thus forming an ``X'' through the critical point (see Figure \ref{fig:xpoint}); 2) the contents of the square root must be positive to create such an X-point in the first place; and 3) the correct X-point of several possibilities lies at the global minimum of the integrated RHS \cite{1976SoPh...49...43K}. 

\afterpage{%
\begin{figure}
\includegraphics[width=\textwidth]{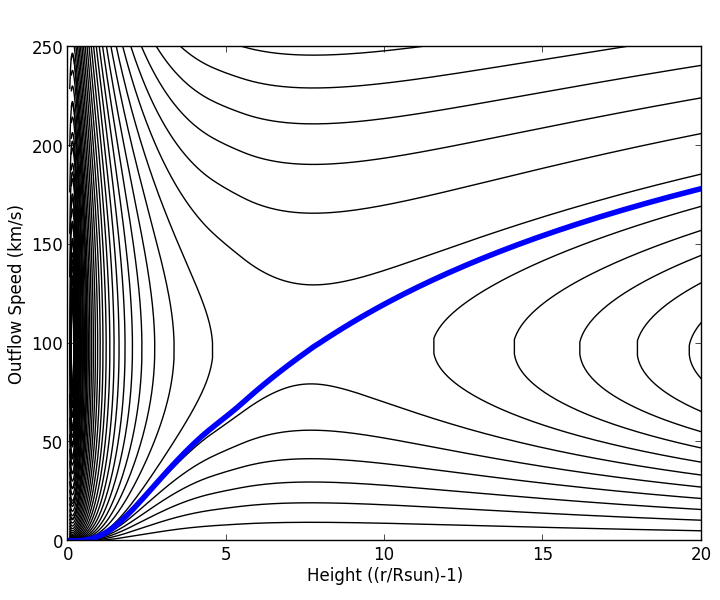}
\caption[Example solution set of Parker's momentum equation]{Example solution set of Parker's momentum equation, Equation \eqref{parkermomentum}. The highlighted blue line shows the critical solution found by the wind. A line from upper left to bottom right, through the ``X-point'' critical point, would represent Bondi accretion.}
\label{fig:xpoint}
\end{figure}
\clearpage
}

The first criterion is self-explanatory. The second brings up the discussion of the range of possible topologies of critical points beyond the X-point shown in Figure \ref{fig:xpoint} \cite{1977JGR....82...23H}. Finally, the third criterion is more complex: if there are multiple critical points but only one of which is an X-point, the outermost root of the RHS is the location of the X-point. However, if there are multiple X-points, the global minimum of the function $F(r) = \int_0^r \! {\rm RHS} \, dr'$ is the location of the X-point for the stable wind solution \cite{1976SoPh...49...43K, 2003ApJ...598.1361V}. As I will describe in later chapters, much of the work Parker published decades ago has stood the test of time, though our understanding of the physics has deepened. This includes his initial hydrodynamic modeling of the solar wind as well as the general idea of nanoflare heating \cite{1988ApJ...330..474P}, though I will show in Chapter 4 that magnetic reconnection may not be the only way to explain such phenomena.

%%%%%%%%%%%%%%%%%%%%%%%%%%%%%%%%%%%%%%%%%%%%%%%%%%%%%%%%%%%%%%%%%%%%%%%%%%%%%%%%
\section{Current Understanding}

Much of the historical foundation I have just described remains a solid fixture on which our current understanding of the solar wind builds. There are still many open questions, and while this dissertation does not seek to answer all of them, I will provide a short overview of these areas of study. I will focus on the current state of our knowledge about methods of wind acceleration, observational constraints and correlations, and the different types of solar wind.

The scientific community has not reached any consensus on which physical processes power the solar wind. The myriad proposed mechanisms can be categorized by their primary use of either magnetic reconnection and the opening of loops of magnetic field that have footpoints at the Sun's photosphere (reconnection/loop-opening models, RLO) or the generation of magnetoacoustic and Alfv\'en waves and the turbulence that is created by counter-propagating waves (wave-driven/turbulence models, WTD). Several reviews have been written to compare and contrast the variety of suggested models \cite{1993SoPh..148...43Z, 1996SSRv...75..453N, 2006SoPh..234...41K, 2009LRSP....6....3C}. 

RLO models often require closed field lines. This means that both footpoints of bundles of magnetic field, called flux tubes, are anchored to the photosphere. Interactions between neighboring closed loops or between closed and open field lines lead to magnetic reconnection, which releases stored magnetic energy when the magnetic topology is reconfigured. Reconnection in closed-field regions has been modelled in streamers \cite{1999JGR...104..521E, 2011ApJ...731..112A} and in the quiet Sun on supergranular scales \cite{1992sws..coll....1A, 1999JGR...10419765F, 2003JGRA..108.1157F, 2006ApJ...642.1173S, 2011ApJ...731L..18M, 2013ApJ...770....6Y}. However, the complex and continuous evolution of the so-called ``magnetic carpet'' \citep{1998ASPC..154..345T} of open and closed field lines may not provide enough energy to accelerate the outflow to match {\it in situ} observations of wind speed \cite{2010ApJ...720..824C}.

Alternatively, WTD models are also used to explain heating and wind acceleration, especially in regions of the Sun where the flux tubes are primarily open, that is, they are rooted to the photosphere by only one footpoint. In this case, Alfv\'en waves and magnetoacoustic oscillations can be launched at the footpoints when the flux tube is jostled by convection at the photosphere. The speed of these waves changes with height and the waves are partially reflected. Counter-propagating waves then interact and generate MHD turbulence. This turbulence generates energy at large scales, and the break-up of eddies causes energy to cascade down to smaller scales where, at a certain scale, the energy can be dissipated as heat\footnote{Determining the specific dissipation mechanism(s) constitutes an entire field of study in itself.}. Such models have naturally produced winds with properties that match observed outflows in the corona and further out in the heliosphere \cite{1986JGR....91.4111H, 1991ApJ...372L..45W, 1999ApJ...523L..93M, 2006JGRA..111.6101S, 2007ApJS..171..520C,2010ApJ...708L.116V}. This paradigm for solar wind acceleration has, however, also been challenged \cite{2010ApJ...711.1044R}, so perhaps the answer to the entire question of coronal heating is more complex than previously thought.

I should also mention that this grouping of proposed heating mechanisms (RLO vs WTD) is sometimes reframed in the context of timescales of magnetic field driving. The granulation pattern created by energy being brought to the surface through the convection envelope in the solar interior, the pattern that Alfv\'en recognized as a potential source of turbulence, is what drives the random walk of magnetic field footpoints. If the characteristic timescale for this driving is fast compared to the Alfv\'en crossing time (i.e. the time it takes for transverse Alfv\'en waves to travel along the magnetic field), waves can be launched to carry energy higher into the solar atmosphere. This is considered alternating current (AC) heating. If instead the photospheric driving is slow compared to this crossing time, the magnetic field in loops or flux tubes can undergo braiding. The braided field can then exhibit direct current (DC) heating \citep{1972ApJ...174..499P}. In each of these cases, however, I have glossed over the fine details of how the energy is actually dissipated as heat, as this is outside the scope of this dissertation. I will discuss some of the observational constraints we have on the details of dissipation and heating.

The corona, while extremely hot, is also incredibly diffuse. In the open-field regions, where the solar wind can escape, the plasma quickly becomes collisionless. This allows us to study how electrons, protons, and heavier ions each undergo heating. One of the features that is most prominent in {\it in situ} measurements is that these particles are not in thermal equilibrium \citep[see, e.g.][]{1999SSRv...87....1M, 2006LRSP....3....1M, 2008PhRvL.101z1103K}. Spectroscopic observations from instruments on the {\it Solar and Heliospheric Observatory} (SOHO) \citep{1995SoPh..162..313K, 1995SoPh..162..189W} also reflect the preferential heating experienced by heavy ions over electrons and protons, as well as anisotropic heating where temperatures in the directions perpendicular to the magnetic field direction ($T_{\perp}$) can be as much as ten times the parallel temperatures ($T_{\parallel}$) in coronal holes \citep[see, e.g.,][]{2002SSRv..101..229C, 2008ApJ...678.1480C, 2009ApJ...691..794L}.

The sense of the anisotropy that ions experience is opposite what is expected from adiabatic expansion, where $T_{\parallel} > T_{\perp}$ to conserve the first adiabatic moment \citep{1956RSPSA.236..112C}. The temperature anisotropy, therefore, gives some indication of the heating mechanisms that the ions experience in addition to the adiabatic expansion, such as the dissipation of low-frequency turbulence \citep[see, e.g.,][and references therein]{2004ApJ...617..667D, 2010ApJ...720..503C} or resonant cyclotron interactions with high-frequency waves \citep[see, e.g.,][and references therein]{2002JGRA..107.1147H, 2006JGRA..111.6105G}. It should be mentioned, however, that the deviations from the expected anisotropy from adiabatic expansion can be partially ``fixed'' when a wave pressure gradient is considered \citep{1999JGR...104.9963O}.

Several recent studies of ion properties have focused on protons and alpha particles measured by the Faraday cups on the {\it Wind} spacecraft \cite{1995SSRv...71...55O}. Maruca (2012, \citep{2012PhDT.......127M}) processed millions of spectra from nearly 16 years of observations to produce a cleaner data set for which greater analysis could be done on the plasma instabilities that form thresholds in the observed temperature anisotropies, alpha to proton anisotropy ($T_{\alpha}/T_{p}$), and the alpha particle differential flow \citep{2013ApJ...777L...3B}. These instabilities, \citep[see, e.g.,][and references therein]{1976JGR....81.1241G, 2006GeoRL..33.9101H} can tell us more about the solar wind plasma properties, heating, and collision histories \citep{2008PhRvL.101z1103K, 2011PhRvL.107t1101M, 2013ApJ...776...45C, 2013PhRvL.110i1102K}. 

As I mentioned previously, the details of the dissipation of the turbulent cascade as heat is a deep and complex field of study. In a weakly collisional plasma, going from the injection of energy at large scales to the production of heat is a two-stage process: there must first be collisionless damping of the turbulent fluctuations and then the energy can be converted to heat \citep{2015RSPTA.37340145H}. The remainder of this dissertation focuses on the fluid nature of turbulence, i.e. the nonlinear interactions that begin the cascade of energies to scales where damping can occur. I therefore return now to the larger-scale properties of the solar wind, where mysteries still loom.

It has been somewhat challenging thus far to discuss modeling paradigms and observational constraints of the solar wind without mentioning one of the most striking aspects: the appearance of a bimodal distribution of speeds at 1 AU. The existence of separate components of the outflow has been observed since {\it Mariner 2} began collecting data in interplanetary space \cite{1962Sci...138.1095N, 1966JGR....71.4469N}. The fast wind has asymptotic wind speeds above roughly 600 km s$^{-1}$ at 1 AU and is characterized by low densities, low variability, and photospheric abundances. The slow wind, however, has speeds at or below 450 km s$^{-1}$ and is chaotic, with high densities and enhanced abundances of elements with low first ionization potentials \cite{1995SSRv...72...49G}. 

Fast wind streams are widely (if not universally) accepted to originate from coronal holes, which are characterized by unipolarity, open magnetic field, and lower densities \cite[and references therein]{1977chhs.conf.....Z, 2009LRSP....6....3C}. The origin of the slow wind is more of a mystery, though it has often been attributed to sources in the streamer belt.\footnote{For a cartoon that compares coronal holes and streamers, see Figure \ref{fig:fourlines}a in the final chapter.} A robust theory for the slow wind identifies a separatrix web (``S-web''), which traces the streamer belt and pseudostreamers \citep{2011ApJ...731..112A, 2011ApJ...731..110L}. Observations are consistent with the existence of this global network \citep{2012JGRA..117.4104C}, and current simulations prepare for the additional constraints that Solar Probe Plus will provide \citep{2015TESS....110804H}. 

Over the past six years, the fact that both RLO and WTD modeling paradigms have continued to match observations at ever-increasing spatial and temporal resolution seems to point to a compromise of sorts. It must still be determined if the different populations of solar wind can be explained by simply a difference in their region of origin (i.e. coronal holes versus S-web) or if multiple acceleration mechanisms exist simultaenously. It is possible that the two populations exist because they are heating and/or accelerated by different physical processes \citep{2011JGRA..116.3101M}. It is certainly the case that RLO models are used more often for the slow wind, and that WTD models often concentrate on the fast wind, as is the case for this dissertation. 

To complicate the matter further, even the two-state paradigm is in question. One of the cases where the fast/slow dichotomy breaks down is in pseudostreamers. I will discuss pseudostreamers further in \S \ref{sec:pseudo}, but it is worth noting here that pseudostreamers are a structure seen in the corona that separate two coronal holes of the same polarity and do not have a central current sheet \citep{2007ApJ...658.1340W}. The outflows that seem to come from the open field regions along the pseudostreamers have properties intermediate to the fast and slow wind types, acting as a``hybrid'' outflow and can serve as a test case for proposed acceleration mechanisms to address \citep{2012ApJ...749..182W}. Additionally, there is building evidence for a third quasi-stationary state for the solar wind that arises from coronal hole boundaries \citep{1997ApJ...489..992B,2003AIPC..679...33M, 2005JGRA..110.4104S}.

To discuss this boundary wind, we have to move beyond the basic categorization of wind population using measured speeds at 1 AU. Other differences between wind populations are the relative abundances of different elements which is often referred to as the first ionization potential (FIP) effect. An overabundance of iron was noted in early UV images of the corona \citep{1968Sci...162...95G}, and the full FIP effect has been studied for decades \citep[and references therein]{1993AdSpR..13..377M,1995SSRv...72...49G, 2007ARA&A..45..297Z}. Low-FIP elements are enhanced in the slow solar wind, where the temperature where elements no longer switch charge states (i.e. they ``freeze in'' ratios of one charge state to another) is often higher than in the fast wind: 1.7 MK versus 1.2 MK. Along with abundance ratios, charge state ratios provide useful information. One of the most commonly used charge state ratios is for oxygen, where ${\rm O^{7+}/O^{6+}}$ is well-correlated with source region, as it freezes in quickly in the low corona \citep{1986SoPh..103..347B,2000JGR...10518327Z}. High-speed streams and the coronal holes they originate from tend to have ${\rm O^{7+}/O^{6+}} < 0.1$ and the slow wind has a higher, but more variable, ratio. The oxygen charge state ratio can be used as a probe of coronal temperatures \citep{1983ApJ...275..354O}, and is used to estimate the electron temperature, which is observed to be anti-correlated with bulk wind speed (first by {\it ISEE 3} \citep{1989SoPh..124..167O} and definitively with {\it Ulysses/SWICS} \citep{1995SSRv...72...49G,2003JGRA..108.1158G}). With this understanding of the known inverse correlation between charge state ratios and wind speed, let us return to the results for the boundary wind.

Stakhiv et al. (2015 \citep{2015ApJ...801..100S}) analyzed the fast-latitude scans made by {\it Ulysses} when the spacecraft passed over the poles of the Sun, providing measurement of outflows that would never reach the ecliptic and Earth. This fact is paramount, as all current solar wind spacecraft are either Earth-orbiting or are at the L1 Lagrange point, which in both cases allows us only to probe wind that originates within a certain latitude from the equator. With the polar passes of {\it Ulysses}, there is a continuum of observed speeds and charge states. The boundary wind is defined as having speeds between 500 km s$^{-1}$ and 675 km s$^{-1}$, with a composition that is closer to the fast wind population than the slow wind \citep{2015ApJ...801..100S}. Due to the similar composition, it is likely that the boundary wind is heated and accelerated by the same mechanism as the fast wind, but undergoes a different rate of expansion from the typical fast wind. The role that flux tube expansion plays on wind acceleration is consistent with Alfv\'en waves \citep{1991ApJ...372L..45W}, and it is the primary motivator for the work described in Chapter 2. 

As we learn more about the heating and acceleration of these wind populations, there may emerge a better way to characterize the different types of wind. There is room to improve the way measurements are categorized. For example, a recent scheme looks at four parameters--proton number density, proton temperature, the solar wind speed, and the magnetic field strength--to separate measured plasma into four types based on source region \citep{2015JGRA..120...70X}. The last of these properties is one that I focus on better understanding in this work. Even with the current generation of ground- and space-based instruments, it is the most difficult to measure accurately.

%%%%%%%%%%%%%%%%%%%%%%%%%%%%%%%%%%%%%%%%%%%%%%%%%%%%%%%%%%%%%%%%%%%%%%%%%%%%%%%%
\section{Measuring Magnetic Fields}

Magnetic fields are notoriously difficult to directly detect, yet we know they play a crucial role in a wide variety of astrophysical contexts. There are only a handful of reliable ways to measure magnetic field strength and direction of a distant source. A common method is the polarization of starlight or scattered light. Linear polarization can provide a partial direction in the plane of the sky, limiting the potential direction of the vector to two possible opposite directions. This helps, but the ambiguity is a significant issue. Many physical processes can cause random directions to look aligned over an area, but the effects of a magnetic field are significant when the field is coherent, i.e. the vectors all point in a single direction across a region.

The other primary ways to measure the true direction of the magnetic field that can provide constraints on coherence are Faraday rotation and the Zeeman effect (Zeeman splitting). If there is a strong linearly-polarized background source, Faraday rotation through a medium can provide information about the magnetic field in that medium \cite{1976ARA&A..14....1H,2012A&A...542A..93O}. The Zeeman effect can measure the strength and direction of magnetic fields along the line of sight. The strong Zeeman effect is when spectral lines are measurably split, as is the case in the solar photosphere (but not in the corona). The weak Zeeman effect also has a variety of applications, and I cover one such case in Appendix \ref{ap:ism} \citep{2015arXiv150600932W}. 

A discussion of magnetic fields at the surface of the Sun is a natural starting point to an investigation of the identified but unknown specific influence of these fields on the solar wind. ``If the Sun had no magnetic field, it would be as uninteresting as most astronomers think it is.''\footnote{Dr. Robert B. Leighton (c. 1965)} As I will describe throughout this work, it is the fact that the Sun's atmosphere is threaded with magnetic field in a wide variety of structures that makes it such an interesting and complex object of study. The map of the magnetic field strength on the Sun's surface due to the splitting of spectral lines is called a magnetogram. Magnetograms are used as a boundary condition for extrapolations of full magnetic field profiles, since we can only make reliable direct measurements in the photosphere, and must extrapolate to determine how the magnetic field changes in the chromosphere and corona.

One of the simplest methods of extrapolation, and the one used throughout this dissertation, is called Potential Field Source Surface (PFSS) extrapolation \cite{1969SoPh....6..442S,1969SoPh....9..131A}. PFSS modeling assumes that ${\bf \nabla} \times {\bf B} = 0$ and that the source surface is a surface of zero potential. Field lines that reach this height, typically set at $r = 2.5~ R_{\odot}$ \citep{1977ApJ...215..636L}, are forced to be radial and defined as ``open'' to the heliosphere. The PFSS model is by far not the only way to extrapolate from magnetograms, but it works well for getting a general sense of the field strength and structure throughout the solar cycle. Nonpotential models range in computational complexity from magnetohydrostatic solutions like the Current Sheet Source Surface model \citep{1995JGR...100...19Z}, to quasi-static magneto-frictional modeling \citep{2000ApJ...539..983V, 2014SoPh..289..631Y}, to full MHD simulations \citep[e.g.][]{1999PhPl....6.2217M,2003AdSpR..32..497O,2011SoPh..274..361R}. Reviews and comparisons of these various frameworks show that while the PFSS model is not the most accurate, it is still a viable tool that remains in use in solar wind forecasting \citep{2006ApJ...653.1510R, 2012LRSP....9....6M, 2014ApJ...782L..22P, 2015SoPh..290.2791E}.

In the following chapters, we will use the most common current version of the PFSS model, which takes global synoptic magnetograms from a full Carrington Rotation (CR) as input \cite{1984PhDT.........5H,1986SoPh..105..205H,1992ApJ...392..310W}. From the assumption of the potential field, we can write
\begin{subequations}
\begin{align}
{\bf B} = - {\bf \nabla}\Psi , \\
 {\bf \nabla}^{2} \Psi = 0 .
\end{align}
\end{subequations}
By setting a spherical source surface, the total potential, $\Psi$, can be broken up into terms for the contributions on either side of the bounded region (i.e.the region between the photosphere and the source surface), such that $\Psi = \Psi_{\rm in} + \Psi_{\rm out}$, where we can write these terms as a sum over spherical harmonics:
\begin{subequations}
\begin{align}
\Psi_{\rm in} = \sum^{\infty}_{l=0} r^{-(l+1)} \sum^{l}_{m=-l} f_{{\rm in},lm} Y_{lm}(\theta,\phi) , \\
\Psi_{\rm out} = \sum^{\infty}_{l=0} r^{l} \sum^{l}_{m=-l} f_{{\rm out},lm} Y_{lm}(\theta,\phi) ,
\end{align}
\end{subequations}
where $\Psi_{\rm in}$ is the potential inside the inner photospheric boundary, $\Psi_{\rm out}$ is the potential outside the source surface outer boundary, and $Y_{lm}$ are spherical harmonics. One can solve the equations for the real part of the potential in terms of a set of coefficients; for a synoptic map in heliospheric coordinates, the magnetic field vector can be re-written in terms of coefficients $g$ and $h$.\footnote{For detailed derivation, see http://wso.stanford.edu/words/pfss.pdf} Wilcox Solar Observatory (WSO) uses the Stanford PFSS model, and reports the values of $g$ and $h$ for each CR synoptic map with $X \times Y$ grid points, where
\begin{equation}
\left(  {g_{lm} \atop h_{lm}}   \right) = \frac{2l + 1}{XY} \sum^{X}_{i=1} \sum^{Y}_{j=1} B_{r}(R_{0},\theta_{i},\phi_{j})P^{m}_{l}(\cos{\theta_{i}}) {\cos \atop \sin}m \phi_{j}.
\end{equation}

The raw data can vary from one observatory to another. When comparing computed PFSS models using data from Mount Wilson Observatory and WSO, considering a saturation correction factor produces a better empirical fit \citep{1995ApJ...447L.143W}. Along with instumental effects, changes may be due also to the use of different normalization methods, spectral lines for Zeeman analysis, or spatial resolution \citep[see, e.g.,][]{2014SoPh..289..769R}.

\afterpage{
\begin{figure}[p!]
\includegraphics[width=\textwidth]{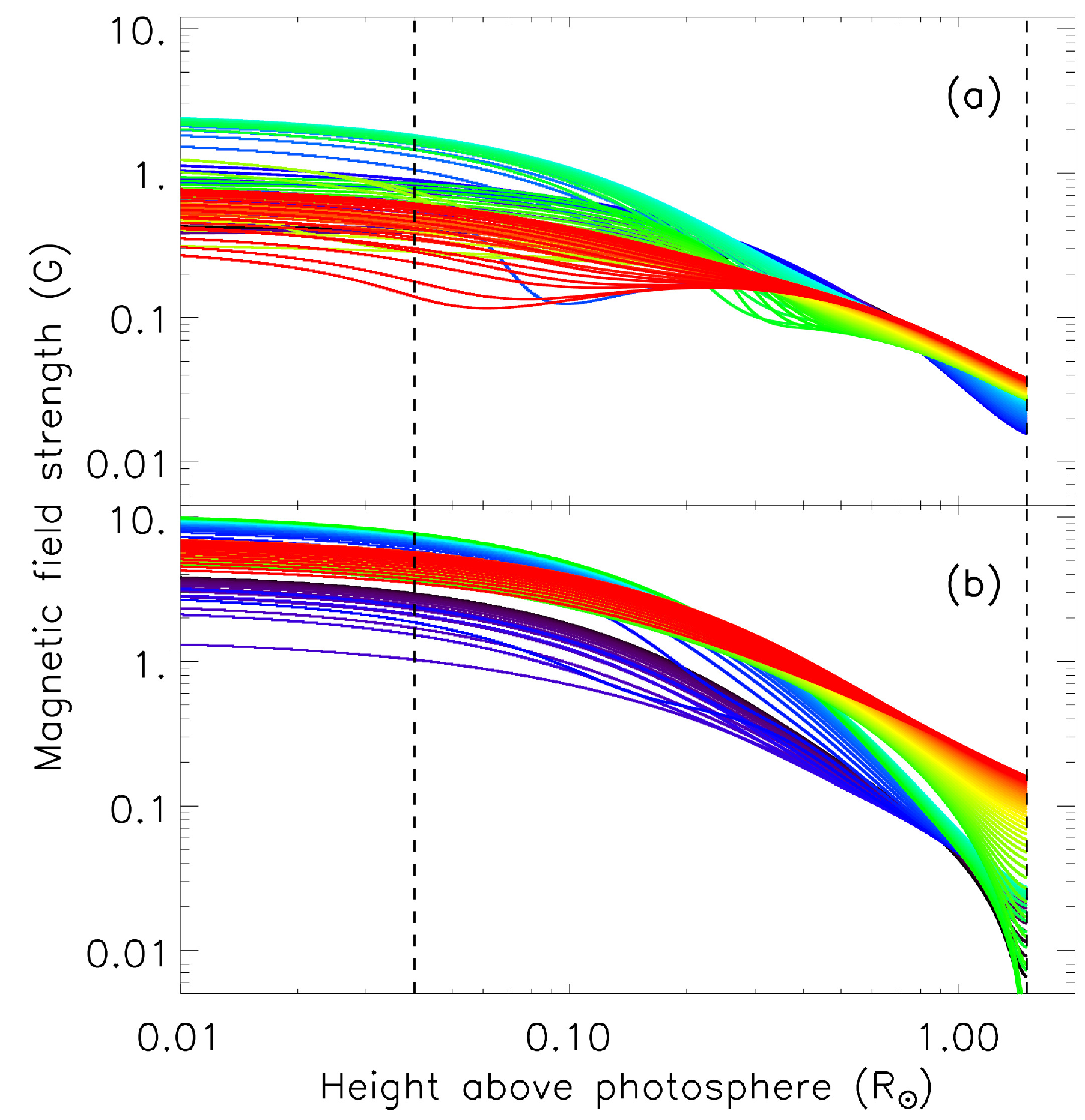}
\caption[Extrapolations of magnetic field structure across the solar cycle]{PFSS extrapolations from WSO magnetograms for cycle 23 at (a) solar minimum in May 1996 (CR 1909) and at (b) solar maximum, in March 2000 (CR 1960). Magnetic field lines are traced from the source surface along the equator, where longitude is denoted by color. Dashed lines show heights used in calculating the expansion factor, $z_{\rm base} = 0.04~ R_{\odot}$ and $z_{\rm ss} = 1.5~ R_{\odot}$.}
\label{fig:wso_pfss}
\end{figure}
\clearpage
}

In order to investigate the full range of magnetic field structures that exist throughout the solar cycle, we examine PFSS extrapolations from synoptic magnetograms taken by WSO, which uses the Zeeman effect on an Fe I line at 5250\AA\ \citep{1986SoPh..105..205H}. Although WSO has one of the lowest resolutions, we use it only to obtain an overall sense of how the observed magnetic field strength changes from solar minimum to solar maximum. Figure \ref{fig:wso_pfss} shows two representative data sets from solar cycle 23. What is most important to note is that the flux tubes extrapolated from observations do not always decrease monotonically. Two flux tubes with identical strengths at the source surface may look significantly different at heights in the chromosphere and corona. These differences at intermediate heights could have a significant impact on the properties of the resulting solar wind. It is for this reason that we consider a wide array of magnetic field models in the work described in the following chapters. 

The two months of synoptic data, defined by their CR value, that are presented in Figure \ref{fig:wso_pfss} do not reflect all possible magnetic field geometries, but they provide an idea of how the magnetic field changes throughout a solar cycle. In order to investigate the entire parameter space of open magnetic geometries, we looked at the absolute maximum and minimum field strengths at several heights between the source surface ($z_{\rm ss} = 1.5~ R_{\odot}$) and a base height of $z_{\rm base} = 0.04 R_{\odot}$ (the scale of supergranules) for the previous three solar cycles to define our parameter space. We then created a grid of models spanning strengths slightly beyond those observed from solar minimum to solar maximum to use the in the parameter study described in Chapter 2.

Those heights, $z_{\rm base}$ and $z_{\rm ss}$ are shown with vertical dashed lines in Figure \ref{fig:wso_pfss}. They are used to compute the {\it expansion factor}, a measure of the geometry of the flux tube \citep{1990ApJ...355..726W}. Figure \ref{fig:expfac} shows an example open flux tube that would have an expansion factor $f_{s}=1$ (see Equation 2.1), which represents radial expansion. The flux tube shown, however, has a very different structure than simple monotonic expansion. The shortcomings of using expansion factor as a predictor of solar wind properties are one of the driving forces behind the first of the projects described in this dissertation.

\afterpage{%
\begin{figure}[p!]
\includegraphics[width=\textwidth]{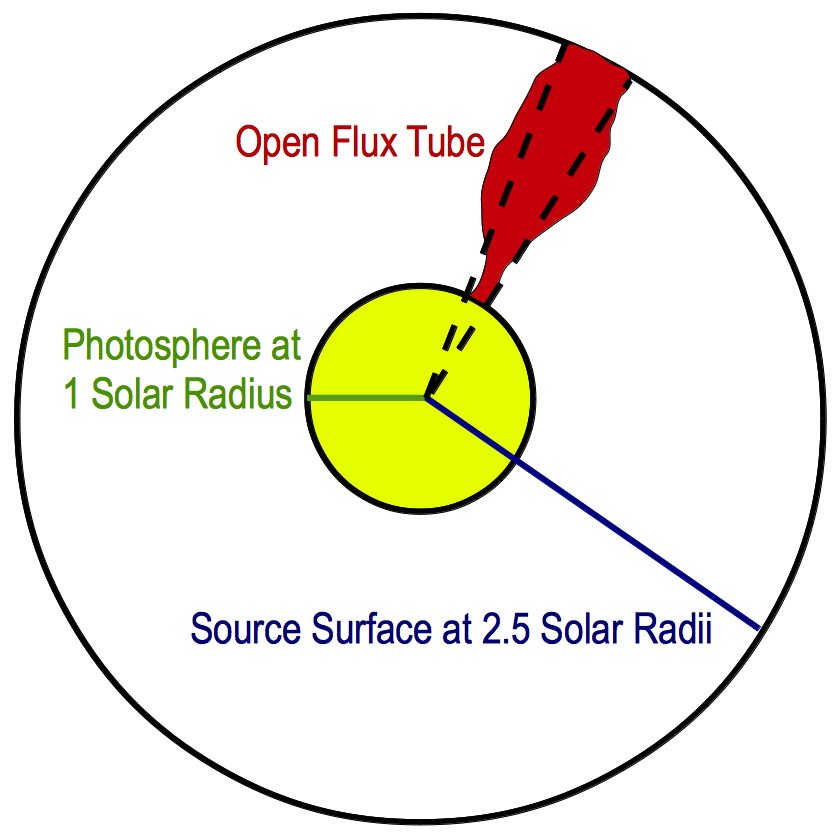}
\caption[Diagram of expansion factor and source surface]{An example flux tube is drawn to have an expansion factor of 1, but a non-monotonically-increasing cross-sectional area.}
\label{fig:expfac}
\end{figure}
\clearpage
}

%%%%%%%%%%%%%%%%%%%%%%%%%%%%%%%%%%%%%%%%%%%%%%%%
\section{Structure of this Work}
Current observations cannot distinguish between competing theoretical models, as many models have a variety of free parameters that can be adjusted to fit observations without specifying all of the physics. To compare the validity of these models at different points in the solar cycle and for different magnetic field structures on the Sun, the community needs flexible tools that predict wind properties using a limited number of input parameters that are all based on observations and fundamental physics. 

A set of such tools are presented throughout this dissertation, which falls naturally into three parts. These parts are relatively independent yet coherently build upon one another and are laid out in separate chapters as follows: in Chapter 2, I discuss the different types of magnetic field structures found in the solar atmosphere throughout the solar cycle and how they affect general properties of the wind through a parameter study of one-dimensional modeling. In Chapter 3, I delve into the greater detail needed to forecast solar wind properties, by considering the time dependence of these phenomena and modeling a set of epitomic open flux tubes in three dimensions. In Chapter 4, I look to observations of transient, small-scale network jets identified in the chromosphere to connect the model results to observable counterparts. 

Finally, I discuss the results contained in this thesis and the conclusions we can draw from them in Chapter 5. Since science can always progress, I also discuss future work I plan to pursue and avenues open to other researchers based on the results of this project.
%% ch3_paramstudy.tex
\chapter{Parameter Study of Magnetic Field Profiles}
%%%%%%%%%%%%%%%%%%%%%%%%%%%%%%%%%%%%%%%%%%%%%%%%
For several decades, the solar physics community has relied heavily on a single measure of the magnetic field geometry to forecast the solar wind properties at 1 AU, a ratio of field strengths called the expansion factor. Wang \& Sheeley \citep{1990ApJ...355..726W} defined the expansion factor relative to the source surface radius \citep{1969SoPh....6..442S,1969SoPh....9..131A} as: 
\begin{equation}
\label{eq:test} 
f_{s} = \left(\frac{R_{\rm base}}{R_{\rm ss}}\right)^{2} \left[\frac{B(R_{\rm base})}{B(R_{\rm ss})}\right].
\end{equation} 
In Equation \eqref{eq:test}, the subscript base signifies the radius of the photospheric footpoint of a given flux tube and the subscript ss refers to the source surface at $r$ = 2.5~ R$_{\odot}$ (see Figure 1.3 for a diagram of the geometry). Potential Field Source Surface (PFSS) modeling assumes that ${\bf \nabla} \times {\bf B} = 0$, the source surface is a surface of zero potential.  Using the expansion factor of Equation \eqref{eq:test}, Wang \& Sheeley determined an empirical relationship between $f_{s}$ and the radial outflow at 1 AU \citep{1990ApJ...355..726W}. They binned observed expansion factors and gave typical outflow speeds for each bin. For $f_{s} < 3.5$, measured speeds at 1 AU ($u_{\rm AU}$) were roughly 700 km s$^{-1}$ and for $f_{s} > 54$, the measured speeds were closer to 330 km s$^{-1}$. The key point from this simple model is that the fastest wind comes from flux tubes with the lowest expansion factors and vice versa, an observation hinted at in previous studies \citep[see, e.g.,][]{1977JGR....82.1061L}.

The Wang-Sheeley empirical relation was used throughout the field for a decade before it was modified \citep{2000JGR...10510465A}. The adjusted method used a two-step process to make four-day advanced predictions, first defining the relation between expansion factor and wind speed {\it at the source surface} and then propagating that boundary condition of the solar wind to the radius of the Earth's orbit, including the effects of stream interactions. The initial step relies on a similar empirical fit to assign a velocity at the source surface based on the expansion factor in Equation \eqref{eq:test}, and was originally set by the following expression: 
\begin{equation}
\label{eq:argepizzo} 
u(f_{s}) = 267.5 + \left(\frac{410}{f_{s}^{2/5}}\right).
\end{equation}
Because this combined Wang-Sheeley-Arge (WSA) model is often the exclusive method used for forecasting the solar wind, it is important for us to consider the efficacy of this method correctly matching observations. 

Early comparisons of the WSA model and observations gave correlation coefficients often at or below 0.5 for a given subset of the observations, and over the full three-year period they considered, the best method used had an overall correlation coefficient of 0.39 \citep{2000JGR...10510465A}. Comparisons of the measured and predicted wind speed using expansion factor led to a correlation coefficient of 0.56 \citep{2005AdSpR..35.2185F}. Expansion on WSA with semi-empirical modeling predicted solar maximum properties well, but produced up to 100 km s$^{-1}$ differences in comparison to observations during solar minimum \citep{2007ApJ...654L.163C}. This suggests that the community could benefit from a better prediction scheme than this simple reliance on the expansion factor. More recently, the updated WSA model has been used in conjunction with an ideal MHD simulation code called ENLIL \citep{2004JGRA..109.2116O, 2011JGRA..116.3101M}. Even with the more sophisticated MHD code, it is still very difficult to make accurate predictions of the wind speed based on only a single measure of the magnetic field geometry at the Sun, $f_{s}$. Prediction errors are often attributed to the fact that these models do not account for time evolution of the synoptic magnetic field, but we also believe that the limitations of the simple WSA correlation may be to blame as well.

\section{The Curious Case of Pseudostreamers}
\label{sec:pseudo}
There is a specific structure type observed on the Sun for which the WSA model is consistently inaccurate.
At solar minimum, the Sun's magnetic field is close to a dipole, with large polar coronal holes (PCH) where the field is open to the greater heliosphere and a belt of helmet streamers where the northern and southern hemispheres have opposite polarity radial magnetic field strength. However, if an equatorial coronal hole (ECH) is present with the same polarity of the PCH of that hemisphere, there is an additional structure that has a shape similar to a helmet streamer but has the same polarity on either side of it that fills the corona between the ECH and PCH. Early work \citep[][and references therein]{1998JGR...103.2021E, 1999SoPh..188..277E} refers to these structures as ``streamer belts without a neutral line,'' and this is an important distinction. These structures contain no large current sheets, whereas helmet streamers are nearly always part of the heliospheric current sheet (HCS). The term ``pseudostreamers'' was coined to distinguish these structures for discussions of observations of the solar wind emanating from them \citep{2007ApJ...658.1340W}. The $v-f_{s}$ relationship vastly overestimates the wind speed from pseudostreamers because these structures are characterized by squashed expansion but produce slow wind \citep[see also][]{2012ApJ...749..182W}. Further work found that comparisons using the parameter $B_{\odot}/f_{s}$, where $B_{\odot}$ is the mean photospheric magnetic field strength of the flux tube, yielded a more accurate prediction of the wind speed \citep{2005AdSpR..35.2185F}. Comparing this parameter to our definition of $f_{s}$ in Equation \eqref{eq:test} suggests that only the magnetic field at the source surface is needed to describe the relationship between magnetic field geometry and solar wind properties. Theoretical interpretation can explain why this parameter ($B_{\odot}/f_{s}$) can describe solar wind accelerated by Alfv\'en waves \citep{2006ApJ...640L..75S}.

\section{Analysis of the Model Grid}

We use a grid of models to provide the dynamic range of field strengths over the solar cycle. We include geometries consistent with the open field lines in and around structures observed in the corona such as helmet streamers and pseudostreamers. Using a standard coronal hole model as a baseline \citep{2007ApJS..171..520C}, we specified the magnetic field strength at four set heights (z = 0.002, 0.027, 0.37, and 5.0 $R_{\odot}$) and connected these strengths using a cubic spline interpolation in the quantity $\log{B}$. Thus, we include all combinations of sets of magnetic field strengths at ``nodes'' between the chromosphere and a height beyond which flux tubes expand into the heliosphere radially such that $B \propto r^{-2}$. To account for the way in which magnetic fields are thought to trace down to the intergranular network, we add two hydrostatic terms in quadrature to the potential field at heights below $z \approx 10^{-3}$ R$_{\odot}$ to account for photospheric G-band bright points and non-potential enhancements in the low corona \citep{2013ApJ...767..125C}.

The resulting 672 models are shown in Figure \ref{fig:zt25}. They span the full range of field strengths measured at 1 AU as found in the OMNI solar wind data sets. The central 90\% of the OMNI data lie between $3 \times 10^{-6}$ and $7 \times 10^{-5}$ Gauss, and our models have magnetic field strengths at 1 AU between $10^{-6}$ and $10^{-4}$ Gauss.

\afterpage{%
\begin{figure}[p!]
\includegraphics[width=\textwidth]{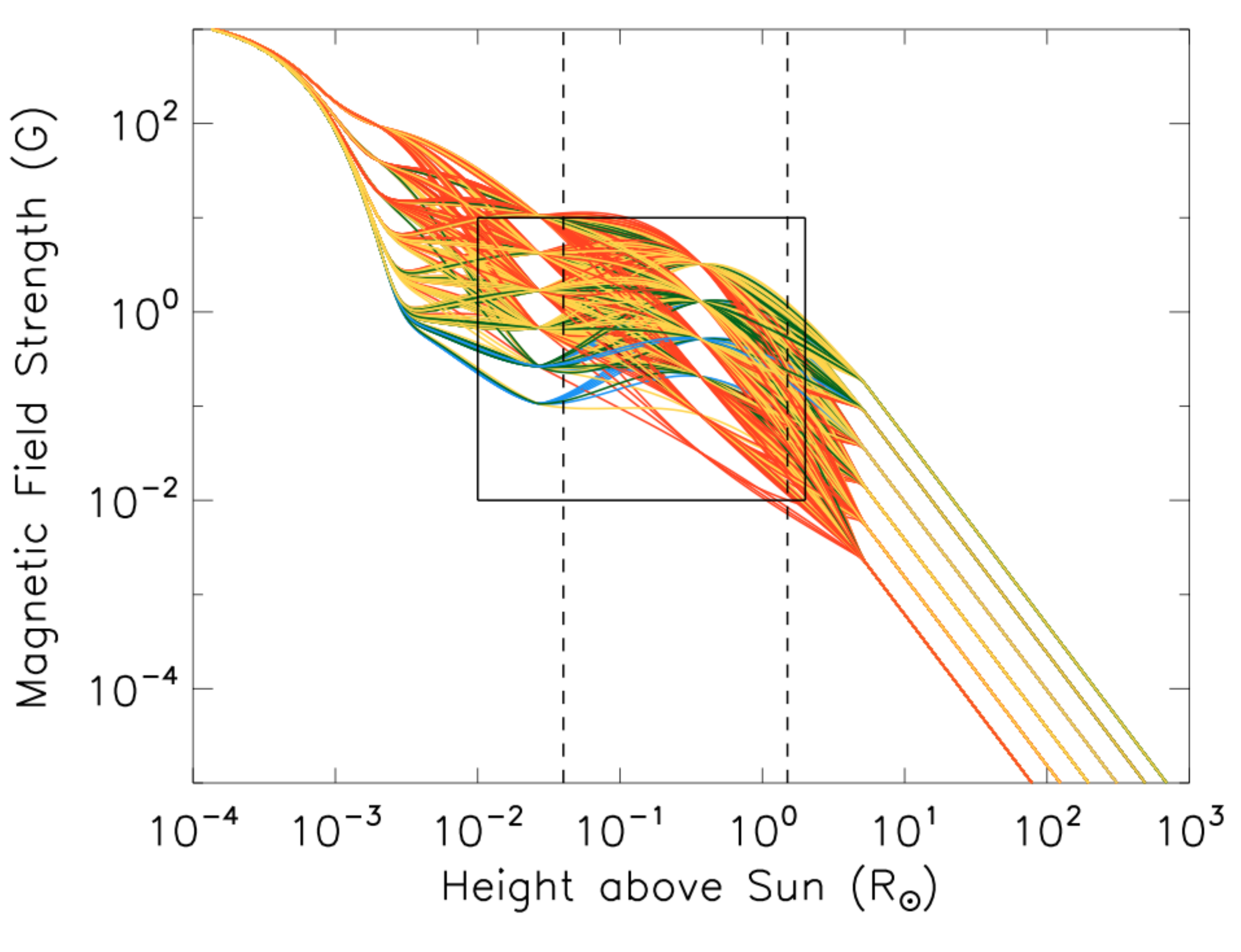}
\caption[Grid of models used in ZEPHYR parameter study]{Input flux tube models for ZEPHYR, where color indicates 200 km s$^{-1}$ wide bins of wind speed at 1 AU. Red signifies the slowest speed bin while blue shows the highest speed models (see Figure \ref{fig:tempest_temps} for key). Dashed vertical lines are plotted at the heights used for the expansion factor calculations ($z_{base} = 0.04 R_{\odot}$ and $z_{ss} = 1.5 R_{\odot}$). The black box provides the relative size of the plot ranges in Figure \ref{fig:wso_pfss} for reference.}
\label{fig:zt25}
\end{figure}
\clearpage
}

We use the MHD one-fluid code ZEPHYR, which has been shown to accurately match observations of the solar wind \citep{2007ApJS..171..520C}. The authors based their magnetic field geometry on a standard configuration with modifications, which has a similar shape and boundary conditions to our grid of models \citep{1998A&A...337..940B, 2005ApJS..156..265C}. 
\begin{subequations}
The equations of mass, momentum, and internal energy conservation solved by ZEPHYR are listed below:
\begin{align}
	\frac{\partial \rho}{\partial t} + \frac{1}{A}\frac{\partial}{\partial r}(\rho u A) = 0\\
	\frac{\partial u}{\partial t} + u\frac{\partial u}{\partial r} + \frac{1}{\rho}\frac{\partial P}{\partial r} = -\frac{G M_{\odot}}{r^{2}} + D\\
	\frac{\partial E}{\partial t} + u\frac{\partial E}{\partial r} + \left(\frac{E + P}{A}\right)\frac{\partial}{\partial r}(uA) = Q_{rad} + Q_{cond} + Q_{A} + Q_{S}
\end{align}
\label{eq:zephyrset}
\end{subequations}
In these equations, the cross-sectional area $A$ is a stand-in for $1/B$ since magnetic flux conservation requires that, along a given flux tube, $BA$ is constant. $D$ is the bulk acceleration from wave pressure and $Q_{A}$ and $Q_{S}$ are heating rates due to Alfv\'en and sound waves. ZEPHYR assumes the number densities of protons and electrons are equal. This one-fluid approximation means that we are unable to include effects such as preferential ion heating, but the base properties of the solar wind produced are accurate. As ZEPHYR solves for a steady-state solution, we can neglect the time-derivatives. The bulk acceleration from wave pressure can be given by \citep{1977ApJ...215..942J}: 
\begin{equation}
D = -\frac{1}{2\rho}\frac{\partial U_{A}}{\partial r} - \left(\frac{\gamma + 1}{2\rho}\right)\frac{\partial U_{S}}{\partial r} - \frac{U_{S}}{A\rho}\frac{\partial A}{\partial r}, 
\end{equation}
where $U_{A}$ and $U_{S}$ are the Alfv\'enic and acoustic energy densities, respectively. With the above equations and wave action conservation, ZEPHYR uses two levels of iteration to converge on a steady-state solution for the solar wind. The only free parameters ZEPHYR requires as input are 1) the radial magnetic field profile and 2) the wave properties at the footpoint of the open flux tube. In this chapter, we do not change the wave properties at the photospheric base from the previous standard \cite{2007ApJS..171..520C}. The process of analyzing grids of models for this project led to some code fixes and we use this updated version of ZEPHYR for all work presented here \citep[see also][]{2013ApJ...767..125C}.

\subsection{Finding the physically signficant critical point}
\label{sec:critpt}
For this project, we have revised the method by which ZEPHYR determines the true critical point from its original version. The revised method is described below. We solve for heights where the right-hand side (RHS) of the equation of motion,
\begin{equation}
\left(u - \frac{u_{c}^{2}}{u}\right)\frac{du}{dr} = \frac{-GM_{*}}{r^{2}} - u_{c}^{2}\frac{dlnB}{dr} - a^{2}\frac{dlnT}{dr} +\frac{Q_{A}}{2\rho (u+V_{A})}
\label{eq:parkereq}
\end{equation}
(rewritten from Equation \eqref{eq:zephyrset} neglecting contributions from sound waves) is equal to zero, i.e. heights where the outflow speed is equal to the critical speed, whose radial dependence is defined by
\begin{equation}
u_{c}^{2} = a^{2} + \frac{U_{A}}{4\rho}\left(\frac{1+3M_{A}}{1+M_{A}}\right).
\label{eq:ucrit}
\end{equation}
Here, $a$ is the isothermal sound speed, $a^{2} = kT/m$, where ``isothermal'' means that in the definition, $\gamma = 1$, and $M_{A}$ is the Alfv\'enic Mach number, $M_{A} = u/V_{A}$. At heights where $u(r) = u_{c}(r)$, the critical slope must have two non-imaginary values at the true critical point in order to create a saddle point, and for the solar wind we take the positive slope. The physically meaningful critical point, if there is more than one, lies at the global minimum of the integrated RHS of Equation \eqref{eq:parkereq}, a criterion discussed in \S 1.1.

\section{ZEPHYR Results for the 672-Model Grid}
We now present some of the most important relations between the solar wind properties output by ZEPHYR and the input grid of magnetic field profiles. For the number of iterations we allowed in ZEPHYR, a subset of the models converged properly to a steady-state solution (i.e. a model is considered converged if the internal energy convergence parameter $\langle \delta E \rangle$ as defined by \cite{2007ApJS..171..520C} is $\le 0.07$). We analyze only the results of these converged 428 models (out of the total 672) in the figures presented in the following subsections. Recall that the solar wind forecasting community relies heavily on a single relation between one property of the solar wind, speed at 1 AU, and one ratio of magnetic field strengths, the expansion factor, Equation \eqref{eq:test}. We compare the WSA relation to our results in \S \ref{sec:zwsa}, present a correlation between the Alfv\'en wave heating rate and the magnetic field strength in \S \ref{sec:zheating} and discuss important correlations between the magnetic field and temperatures in \S \ref{sec:ztempest}.

\subsection{Revisiting the WSA model}
\label{sec:zwsa}
Our models follow the general anti-correlation of wind speed and expansion factor seen in observations \citep{1990ApJ...355..726W}, as shown in Figure \ref{fig:zephyr_wsa}. There is nevertheless a large scatter around any given one-to-one relation between $u_{\infty}$ and $f_{s}$, which is highly reminiscent of the observed solar wind. That our models reproduce an observation-based relation is an important and successful test of the validity of ZEPHYR. The concordance relation previously found, $u_{\infty} = 2300/(\ln{f_{s}} + 1.97)$ \citep{2013ApJ...767..125C}, runs through the middle of the scatter for expansion factors greater than 2.0.
%\placefigure{fig:zephyr_wsa}
\afterpage{%
\begin{figure}[p!]
\includegraphics[width=\textwidth]{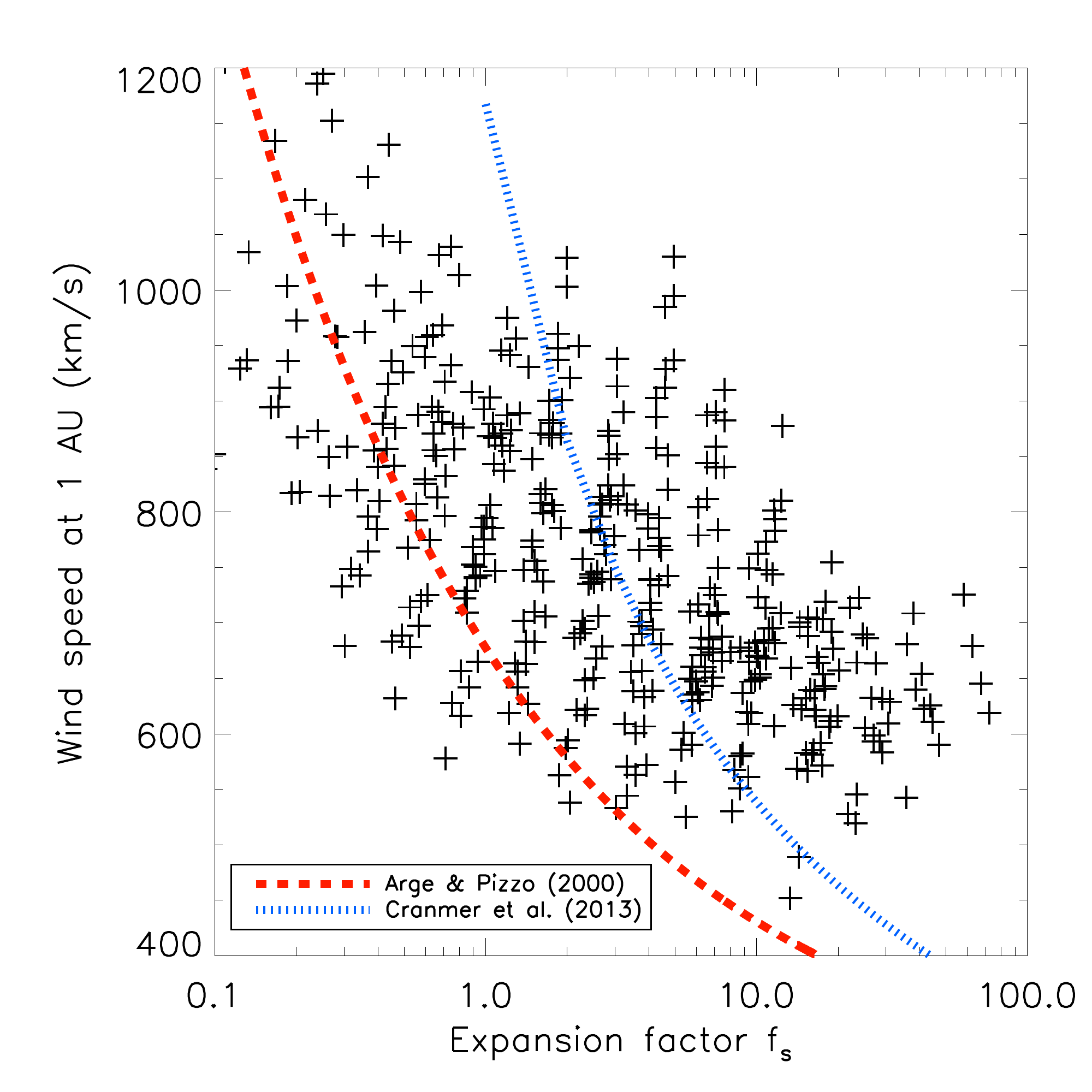}
\caption[Anti-correlation between expansion factor and wind speed]{Expansion factor anti-correlation is reproduced by ZEPHYR. The red dashed line shows the source surface velocity relation to expansion factor given in Equation \eqref{eq:argepizzo}, which is expected to be lower than the wind speed at 1 AU. The blue dotted line shows a previously-published concordance relation \cite{2013ApJ...767..125C}.}
\label{fig:zephyr_wsa}
\end{figure}
\clearpage
}

\subsection{Alfv\'en wave heating rate}
\label{sec:zheating}
The turbulent heating rate by Alfv\'en waves can be written in terms of the Els\"asser variables $Z_{-}$ and $Z_{+}$ as the following:
\begin{equation}
Q_{A} = \rho \epsilon_{\rm turb}\frac{Z_{-}^{2}Z_{+} + Z_{+}^{2}Z_{-}}{4L_{\perp}}.
\end{equation}
The effective turbulence correlation length $L_{\perp}$ is proportional to $B^{-1/2}$ and $\epsilon_{turb}$ is the turbulence efficiency \citep{2007ApJS..171..520C}. The ZEPHYR code iterates to find a value of $Q_{A}$ that is consistent with the time-steady conservation equations (Equation \eqref{eq:zephyrset}). For Alfv\'en waves at low heights where the solar wind speed is much smaller than the Alfv\'en speed, the Els\"asser variables are roughly proportional to $\rho^{-1/4}$. Putting this together with the thin flux-tube limit where the Alfv\'en speed is roughly constant, we show that Alfv\'en wave heating should produce, at low heights, the proportionality $Q_{A} \propto B$ \citep[see also][]{2009LRSP....6....3C}. Figure \ref{fig:QvB} shows this relation at a height of 0.25 solar radii. This also suggests that the magnetic field and temperature profiles should be reasonably well-correlated.
%\placefigure{fig:QvB}
\afterpage{%
\begin{figure}[p!]
\includegraphics[width=\textwidth]{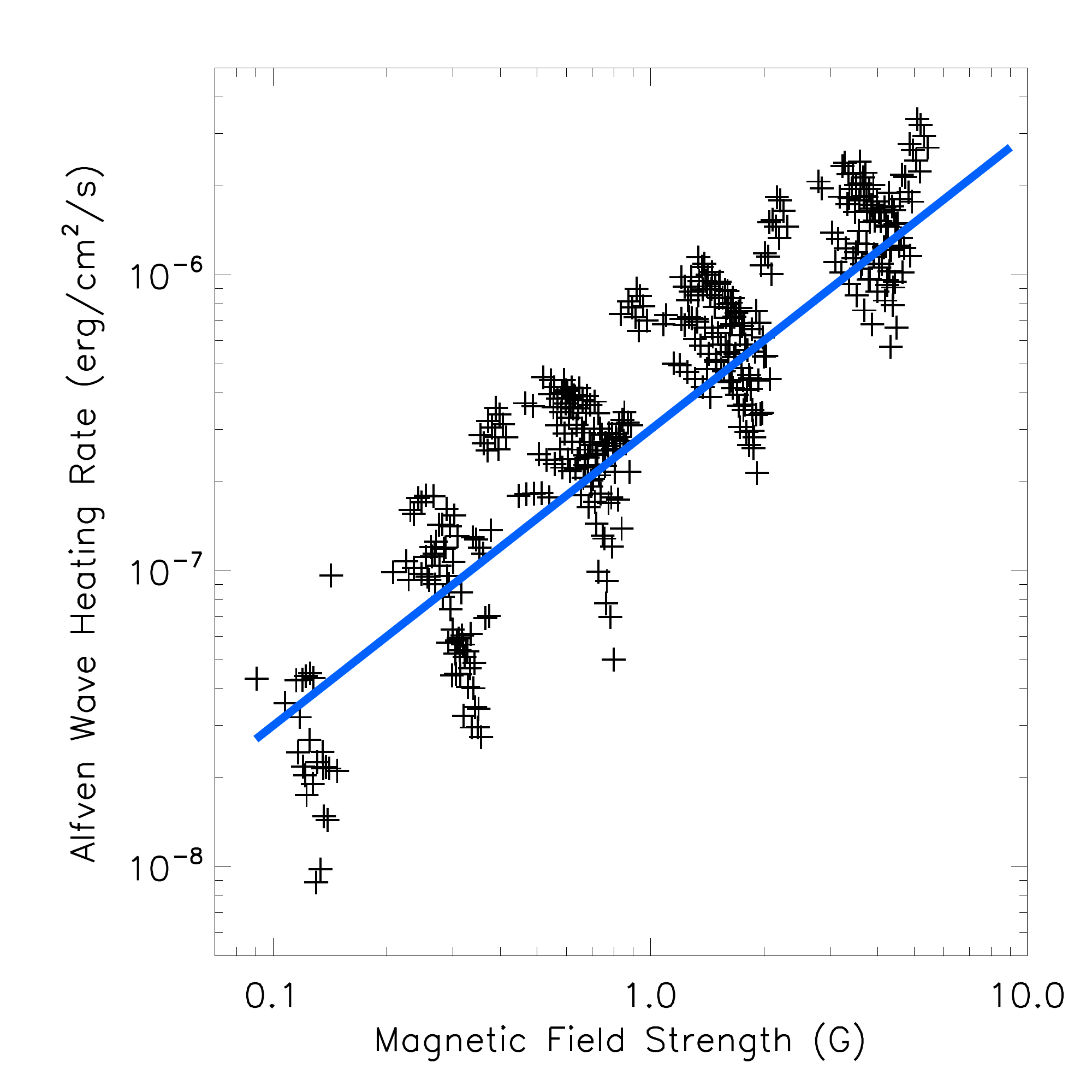}
\caption[Heating rate as a function of magnetic field strength]{Heating rate versus magnetic field strength at a height of 0.25 solar radii. At these low heights, this relation, shown with a solid line of slope = 1, is expected from turbulent damping.}
\label{fig:QvB}
\end{figure}
\clearpage
}

\subsection{Predicted temperature profiles}
\label{sec:ztempest}
ZEPHYR uses a simplified treatment of radiative transfer to compute the heating and cooling rates throughout the solar atmosphere. The model includes terms for radiation, conduction, heating by Alfv\'en waves, and heating by acoustic waves \citep{2007ApJS..171..520C}. The photospheric base and lower chromosphere are considered optically thick and are dominated by continuum photons in local thermodynamic equilibrium that provide the majority of the heating and cooling. However, in the corona, the atmosphere is optically thin, where many spectral lines contribute to the overall radiative cooling. Further description of the internal energy conservation terms listed in Equation \eqref{eq:zephyrset}c can be found in Section 3 of \cite{2007ApJS..171..520C}.\\
\indent The temperature profiles found for each flux tube model are presented in Figure \ref{fig:tempest_temps}. The blue models have speeds greater than 1100 km s$^{-1}$, and their temperature profiles peak higher than the mean height of maximum temperature. These models probably do not correspond to situations realized in the actual solar wind, but they are instructive as examples of the implications of extreme values of $B(r)$.

Figure \ref{fig:tempcorr} shows illustrative correlations between the maximum temperature and the temperature at 1 AU with the magnetic field at $r = 2.5$ R$_{\odot}$. These are both very strong correlations (with values of the Pearson coefficient $R > 0.8$), and they can be used as an independent measure of the magnetic field near the source surface besides PFSS extrapolations from magnetogram data to test the overall validity of turbulence-driven models. If the measured solar wind exhibits a similar correlation between, e.g., temperature at 1 AU and the field strengths at a field line's extrapolated location at 2.5 R$_{\odot}$, this would provide additional evidence in favor of waves and turbulence-driven heating models.

%\placefigure{fig:tempest_temps}
\afterpage{%
\begin{figure}[p!]
\includegraphics[width=0.9\textwidth]{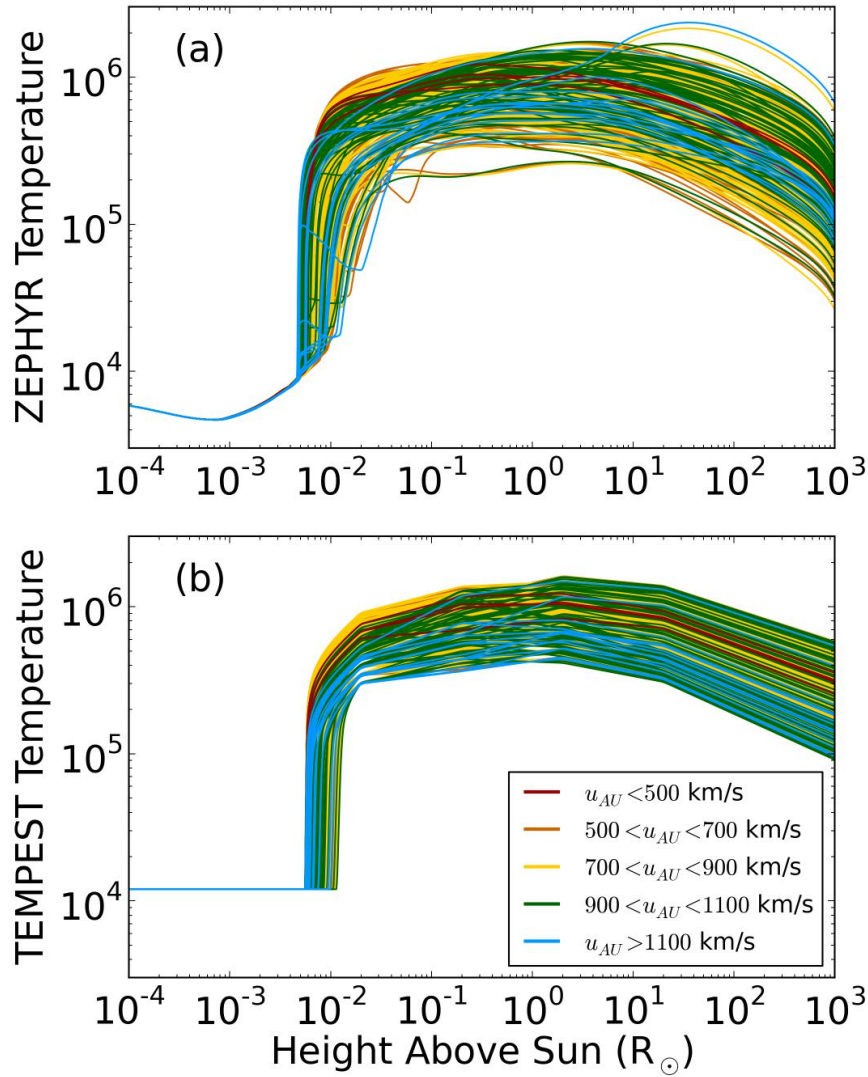}
\caption[Temperature profiles from ZEPHYR and TEMPEST]{Comparison between (a) calculated temperature profiles from internal energy conservation in ZEPHYR and (b) the temperature profiles we set up for TEMPEST based on correlations with magnetic field strengths (see Appendix \ref{ap:profiles} for more).}
\label{fig:tempest_temps}
\end{figure}
\clearpage
}

%\placefigure{fig:tempcorr}
\afterpage{%
\begin{figure}[p!]
\includegraphics[width=\textwidth]{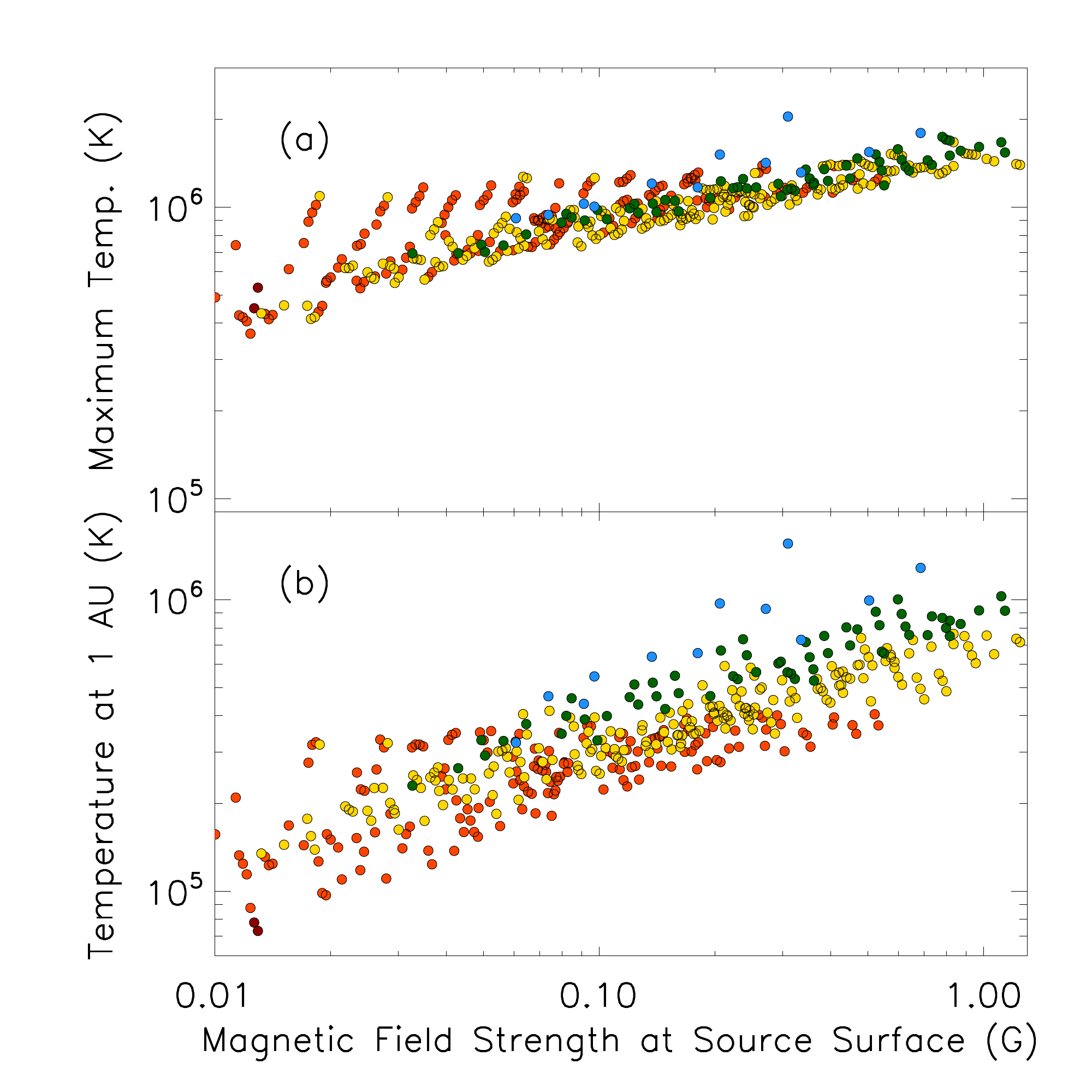}
\caption[Correlations between temperature and magnetic field]{Strong correlation of (a) the maximum temperature and (b) T at 1 AU with the magnetic field strength at the source surface. Color indicates outflow speed as in Figures \ref{fig:zt25} and \ref{fig:tempest_temps}.}
\label{fig:tempcorr}
\end{figure}
\clearpage
}

%\placefigure{fig:rpr_temp}
\afterpage{%
\begin{figure}[p!]
\includegraphics[width=\textwidth]{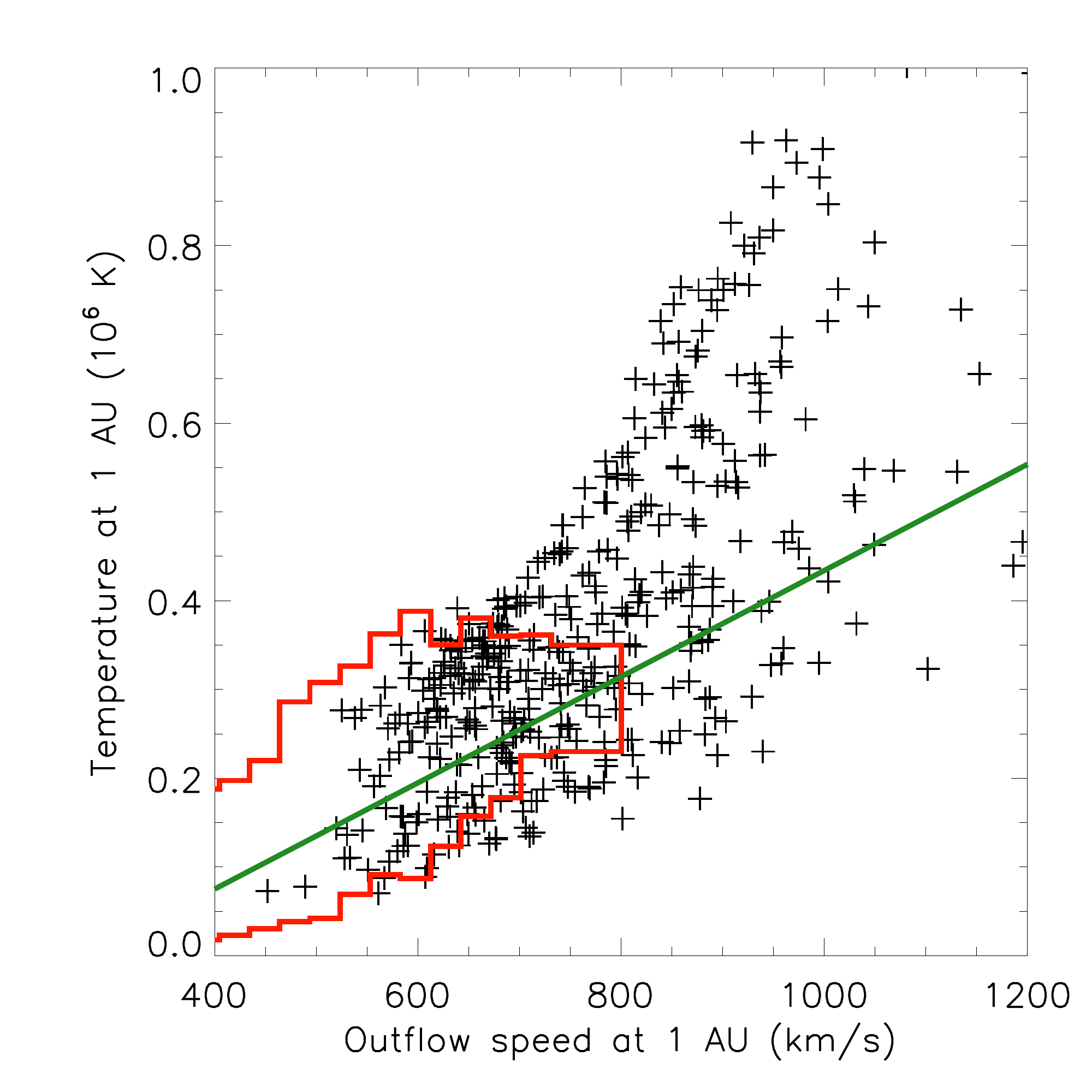}
\caption[Comparison of ZEPHYR results with observation data]{Temperature and Wind Speed at 1 AU: red outline represents several decades of OMNI data and green line is an observationally derived linear fit of relation between proton temperatures and wind speed \citep{2012JGRA..117.9102E}.}
\label{fig:rpr_temp}
\end{figure}
\clearpage
}

Additionally, we show the relation between the temperature at 1 AU and the wind speed at 1 AU in Figure \ref{fig:rpr_temp}. We plot the linear fit between observed proton temperatures and wind speeds \cite{2012JGRA..117.9102E}, which is a good fit to models with wind speeds at or below 800 km s$^{-1}$. Models with higher wind speeds may be generated from slightly unphysical magnetic field profiles in our grid. We also plot the outline of the OMNI data set, which includes several decades of ACE/Wind data for proton temperatures and outflow speed. Our models populate the same region and spread for wind speeds between 550 and 700 km s$^{-1}$. We discuss the relative lack of slow solar wind results, i.e. $u < 400$ km s$^{-1}$, in \S \ref{sec:ch3disc}.

\section{TEMPEST Development}
I developed The Efficient Modified-Parker-Equation-Solving Tool (TEMPEST) in Python, in order to provide the community with a fast and flexible tool that can be used as a whole or in parts due to its library-like structure. TEMPEST can predict the outflow speeds of the solar wind based only on the magnetic field profile of an open flux tube, which could be measured using PFSS extrapolations from magnetogram data. TEMPEST uses the modified Parker equation given in Equation \eqref{eq:parkereq}, but we neglect the small term proportional to $Q_{A}$ for simplicity. For a given form of the critical speed $u_{c}$, the critical radius, $r_{c}$, is found as described in \S \ref{sec:critpt}. At each critical point, the slope of the outflow must be found using L'H\^opital's Rule. Doing so, one finds \begin{equation}\label{eq:critslope}\left.\frac{du}{dr}\right|_{r=r_{c}} = \frac{1}{2}  \left[ \frac{du_{c}}{dr} \pm \sqrt{\left(\frac{du_{c}}{dr}\right)^{2} + 2\frac{d{\rm (RHS)}}{dr}}\right]\end{equation} 
where the postive sign gives the accelerating solution appropriate for the solar wind, and RHS is the right-hand side of Equation \eqref{eq:parkereq}. We emphasize that the actual wind-speed gradient $du/dr$ at the critical point is not the same as the gradient of the critical speed $du_{c}/dr$. Consider the simple case of an isothermal corona without wave pressure, in which $u_c$ is a constant sound speed and $du_{c}/dr = 0$.  Even in this case, Equation \eqref{eq:critslope} gives nonzero solutions for $du/dr$ at the critical point: a positive value for the transonic wind and a negative value for the Bondi accretion solution.

The magnetic field profile is the only user input to TEMPEST, and the temperature profile is set up using temperature-magnetic field correlations from ZEPHYR, as we do not include the energy conservation equation in TEMPEST. We considered the correlations between temperatures and magnetic field strengths at different heights, similar to the results shown in Figure \ref{fig:tempcorr}. We set the temperatures at evenly spaced heights $z_{T}$ in log-space, and at each height we sought to find the heights $z_{B}$ at which the variation of the magnetic field strength best correlates with the temperature at $z_{T}$. At $z_{T} = 0.02 $ R$_{\odot}$, the results from ZEPHYR give the best correlation with the magnetic field in the low chromosphere. At $z_{T} = 0.2 $ R$_{\odot}$, the temperature best correlates with the magnetic field at $z_{B} = 0.4 $ R$_{\odot}$. Since the temperature peaks around this middle height, the correlations reflect the fact that heat conducts away from the temperature maximum. At $z_{T} = $ 2, 20, and 200  R$_{\odot}$, the magnetic field near the source surface ($z_{B} \approx 2 - 3 $ R$_{\odot}$) provides the best correlation. We show the comparison between the temperature profiles of ZEPHYR and TEMPEST in Figure \ref{fig:tempest_temps} and provide the full equations in Appendix \ref{ap:profiles}.

TEMPEST has two main methods of use. The first is a vastly less time-intensive mode I refer to as ``Miranda'' that solves for the outflow solution without including the wave pressure term. I outline this in \S \ref{sec:miranda}. Miranda can run 200 models in under 60 seconds, making it a useful educational tool for showing how the magnetic field can affect the solar wind in a relative sense. It is important to note that the way that the temperature profiles are set up in TEMPEST means that Miranda already includes the effects of turbulent heating even though it does not have the wave pressure term, effectively separating the two main ways that Alfv\'en waves contribute to the acceleration of the solar wind. The second mode of TEMPEST use is the full outflow solver based on including both the gas and wave pressure terms. I call this function ``Prospero'' for ease in reference, and in \S \ref{sec:prospero} I outline the additional steps that Prospero takes after the inital solution is found using Miranda.

\subsection{Miranda: Without waves}
\label{sec:miranda}
The first step towards the full outflow solution requires calculating an initial estimate the outflow without waves in order to calculate the density profile. This first step in TEMPEST, Miranda, solves Equation \eqref{eq:parkereq} with the terms for gravity, the magnetic field gradient, and the temperature gradient, where the critical speed $u_{c}$ is set to the isothermal sound speed, $a = (kT/m)^{1/2}$. The temperature profiles that TEMPEST uses already include the effects of turbulent heating, so we can effectively separate the two primary mechanisms by which Alfv\'en waves accelerate the solar wind: turbulent heating and wave pressure. The solution found using Miranda has only the first of these mechanisms included, and therefore will produce outflows at consistently lower speeds.

With all terms in the Parker equation defined, we find the critical point as discussed in Section 3.1. Using the speed and radius of the correct critical point, we find the slope at the critical point using Equation \eqref{eq:critslope}. Once we have determined the critical point and slope, we use a 4th-order Runge-Kutta integrator to move away in both directions from this point.

The results from TEMPEST without the wave pressure term are shown in red in Figure \ref{fig:prospero} and result in much lower speeds than ZEPHYR produced, which is to be expected, as the additional pressure from the waves appears to provide an important acceleration for the solar wind. The mean wind speed at 1 AU for the results from ZEPHYR is 776 km s$^{-1}$ (standard deviation is 197 km s$^{-1}$); the mean wind speed at 1 AU after running Miranda is only 357 km s$^{-1}$ with a standard deviation of 105 km s$^{-1}$. Therefore, we now look at the solutions to the full modified Parker equation used by Prospero.
%\placefigure{fig:prospero}
\afterpage{%
\begin{figure}[p!]
\includegraphics[width=\textwidth]{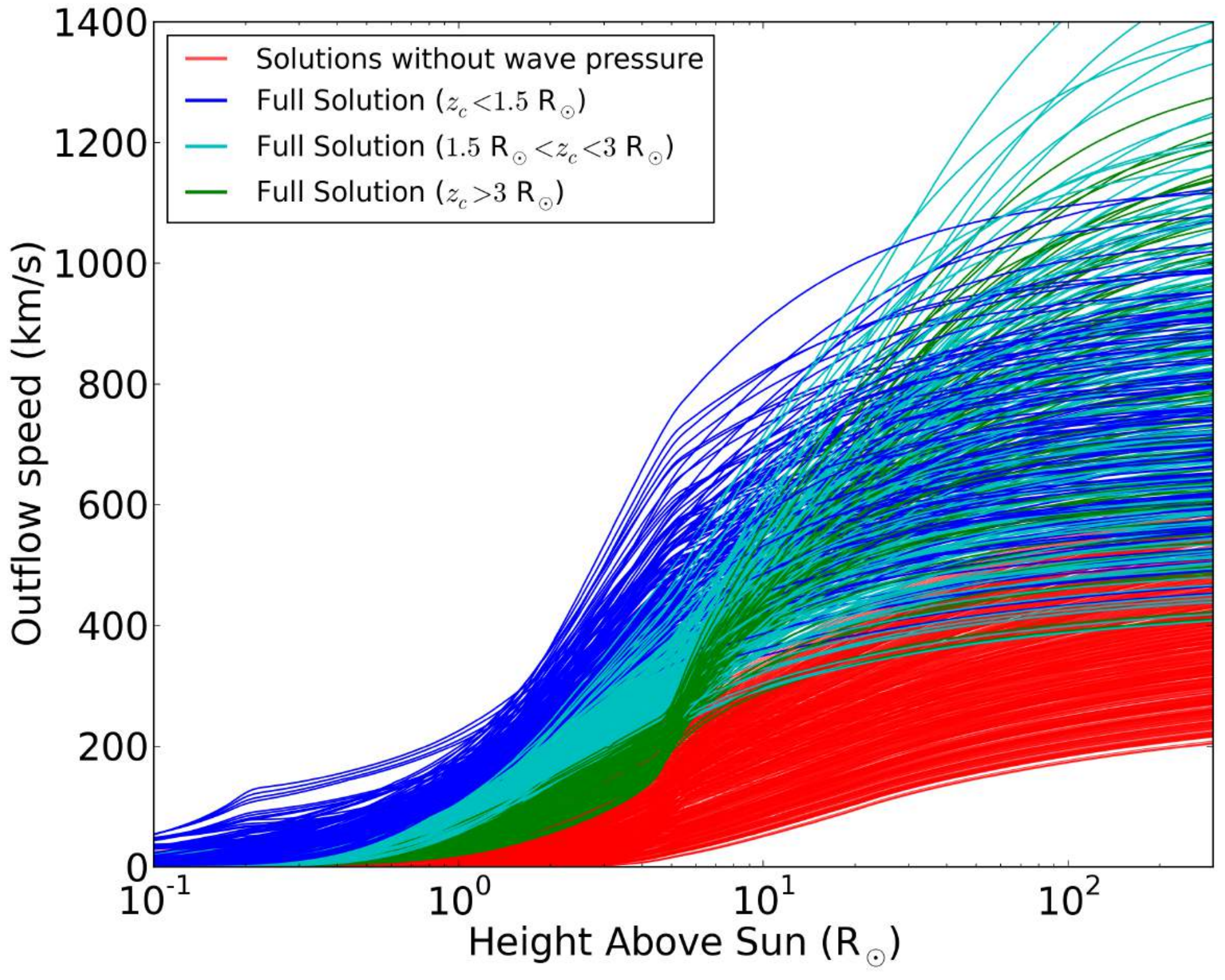}
\caption[Outflow speed profiles from TEMPEST modeling]{Results from Prospero represent the full solution. The solutions from Miranda (i.e. without the wave pressure term) are shown in red. Color indicates bands of critical location height as shown in the legend. We plot only the 428 models from the grid that were well-converged in ZEPHYR, although TEMPEST iterates until all the models find a stable solution.}
\label{fig:prospero}
\end{figure}
\clearpage
}

\subsection{Prospero: Adding waves and damping}
\label{sec:prospero}
Everything presented in the previous section remains the same for the full solution except for the form of the critical speed. With waves, we must use the full form, given by Equation \eqref{eq:ucrit}. The mass density, $\rho$, is determined by using the outflow solution and the enforcement of mass flux conservation:
\begin{equation}
\frac{u(r)\rho(r)}{B(r)} = \frac{u_{\rm TR}\rho_{\rm TR}}{B_{\rm TR}}.
\end{equation}
We set the transition region density based on a correlation with the transition region height that we found in the collection of ZEPHYR models, 
\begin{equation}
\log(\rho_{\rm TR}) = -21.904-3.349\log(z_{\rm TR}),
\end{equation}
where $\rho_{\rm TR}$ is specified in g cm$^{-3}$ and $z_{\rm TR}$ is given in solar radii. Although the pressure scale height differs at the transition region between ZEPHYR and TEMPEST (see Figure \ref{fig:tempest_temps}), we found the uncertainties produced by this assumption were small. Our initial version of TEMPEST used a constant value of this density, taken from the average of the ZEPHYR model results, and did not create significant additional disagreement between the ZEPHYR and TEMPEST results. 

We use damped wave action conservation to determine the Alfv\'en energy density, $U_{A}$. We start with a simplified wave action conservation equation to find the evolution of $U_{A}$ \citep[see e.g.][]{1977ApJ...215..942J, 2007ApJS..171..520C}:
\begin{equation}
\label{eq:waveact1}
\frac{\partial}{\partial t}\left(\frac{U_{A}}{\omega'}\right) + \frac{1}{A}\frac{\partial}{\partial r}\left(\frac{[u+V_{A}]AU_{A}}{\omega'}\right) = -\frac{Q_{A}}{\omega'}.
\end{equation}
Because we are working with the steady-state solution to the Parker equation, we are able to neglect the time derivative. The Doppler-shifted frequency in the solar wind frame, $\omega'$, can be written as $\omega' = \omega V_{A}/(u+V_{A})$ where $\omega$ is a constant and may be factored out. The exact expression for the heating rate $Q_{A}$ depends on the Els\"asser variables $Z_{+}$ and $Z_{-}$, but we approximate it following \cite{2011ApJ...741...54C} as 
\begin{equation}
\label{eq:qa}
Q_{A} = \frac{\widetilde{\alpha}}{L_{\perp}}\rho v_{\perp}^{3},
\end{equation} 
where $\widetilde{\alpha} = 2\epsilon_{\rm turb}R(1+R)/[(1+R^{2})^{3/2}]$. For TEMPEST, we define a simplified radial profile for the reflection coefficient $R$ based on correlations with the magnetic field strength in ZEPHYR (see Figure \ref{fig:refl} and Appendix \ref{ap:profiles}). We set the correlation length at the base of the photosphere, $L_{\perp\odot}$, to 75 km \citep{2005ApJS..156..265C,2007ApJS..171..520C} and use the relation $L_{\perp} \propto B^{-1/2}$ for other heights. 

\afterpage{%
\begin{figure}[p!]
\includegraphics[width=0.9\textwidth]{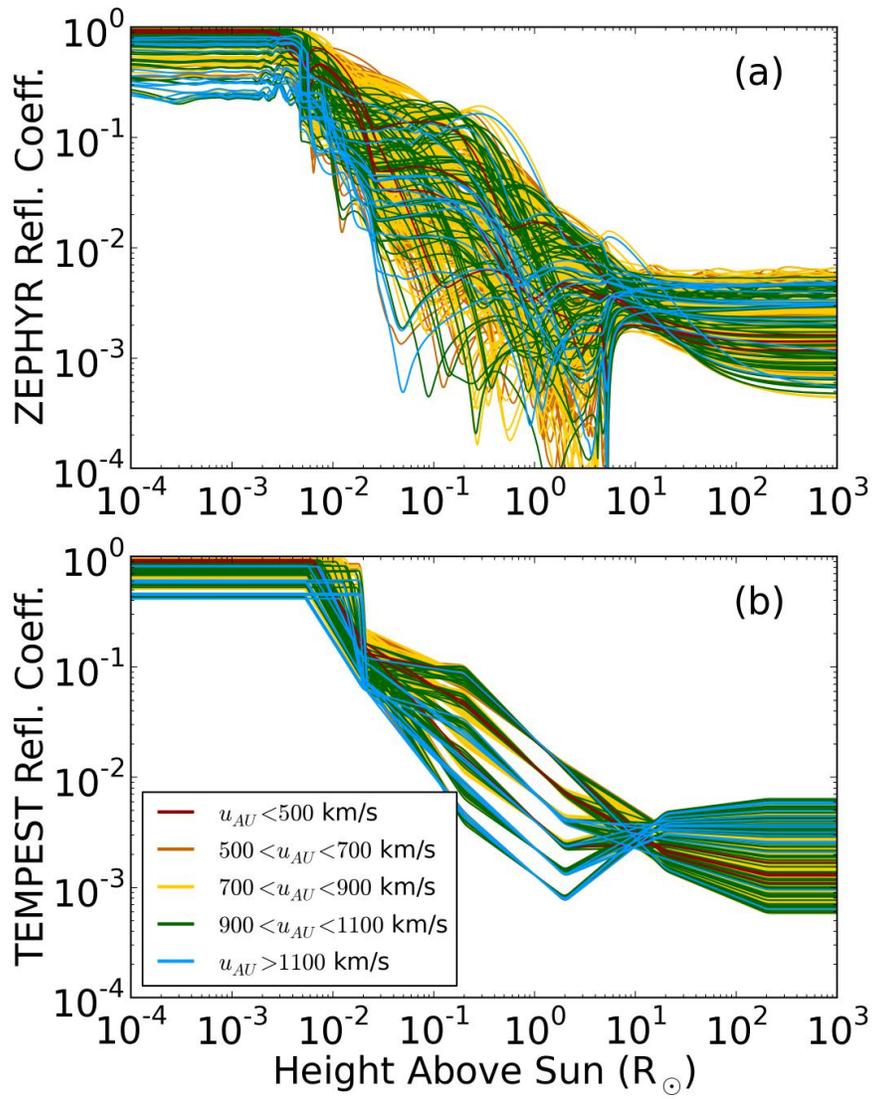}
\caption[Reflection coefficients from ZEPHYR and TEMPEST]{Comparison between (a) calculated reflection coefficients from ZEPHYR and (b) the defined reflection coefficients from TEMPEST (see Appendix \ref{ap:profiles} for more).}
\label{fig:refl}
\end{figure}
\clearpage
}

Combining Equations \eqref{eq:waveact1} and \eqref{eq:qa} using the approximations mentioned and including the conservation of magnetic flux, we define the wave action as 
\begin{equation}
S \equiv \frac{(u+V_{A})^{2}\rho v_{\perp}^{2}}{BV_{A}},
\end{equation}
such that the wave action conservation equation can now be written as the following:
\begin{equation}
\frac{dS}{dr} = -S^{3/2}\left(\frac{\widetilde{\alpha}}{L_{\perp}(u+V_{A})^{2}}\right)\sqrt{\frac{BV_{A}}{\rho}}.
\end{equation}
TEMPEST then integrates using a Runge-Kutta method to solve for $S(r)$ and uses a value of this constant at the photospheric base, $S_{\rm base} = 5 \times 10^{4} $ erg/cm$^{2}$/s/G (which was assumed for each of the ZEPHYR models), to obtain the Alfv\'en energy density needed by the full form of the critical speed, such that
\begin{equation}
U_{A}(r) \equiv \rho v_{\perp}^{2} = \frac{S(r)B(r)V_{A}(r)}{(u(r)+V_{A}(r))^{2}}
\end{equation}

To converge to a stable solution, Prospero must iterate several times. We use undercorrection to make steps towards the correct outflow solution, such that $u_{next} = u_{previous}^{0.9}*u_{current}^{0.1}$. The first iteration uses the results from Miranda as $u_{previous}$ and an initial run of Prospero using this outflow solution to provide $u_{current}$, and subsequent runs use neighboring iterations of Prospero to give the outflow solution to provide to the next iteration. The converged results from Prospero are presented in Figure \ref{fig:prospero}. The outflow speeds are between 400 and 1400 km s$^{-1}$ at 1 AU, consistent with observations (mean: 794 km s$^{-1}$, standard deviation: 199 km s$^{-1}$). For this figure, we have binned the models by color according to the height of the critical point in order to highlight the relation between critical point height and asymptotic wind speed.

A key result from the TEMPEST results is the recovery of a WSA-like correlation between expansion factor and wind speed. In Figure \ref{fig:tempest_wsa}, we plot the relation along with the results from the 428 well-converged models in our grid. The Arge \& Pizzo relation predicts the wind speed at the source surface based on the expansion factor (Equation \eqref{eq:test}). The relation should act as a lower bound for the wind speed predicted at 1 AU, since there is further acceleration above 2.5 $R_{\odot}$. This is what we see for the slower wind speeds, which is to be expected for a relation calibrated at the equator, which rarely sees the highest speed wind streams. Both the ZEPHYR and TEMPEST models naturally produce a substantial spread around the mean WSA-type relation, highlighting the need the take the full magnetic field profile into account.
%\placefigure{fig:tempest_wsa}
\afterpage{%
\begin{figure}[p!]
\includegraphics[width=\textwidth]{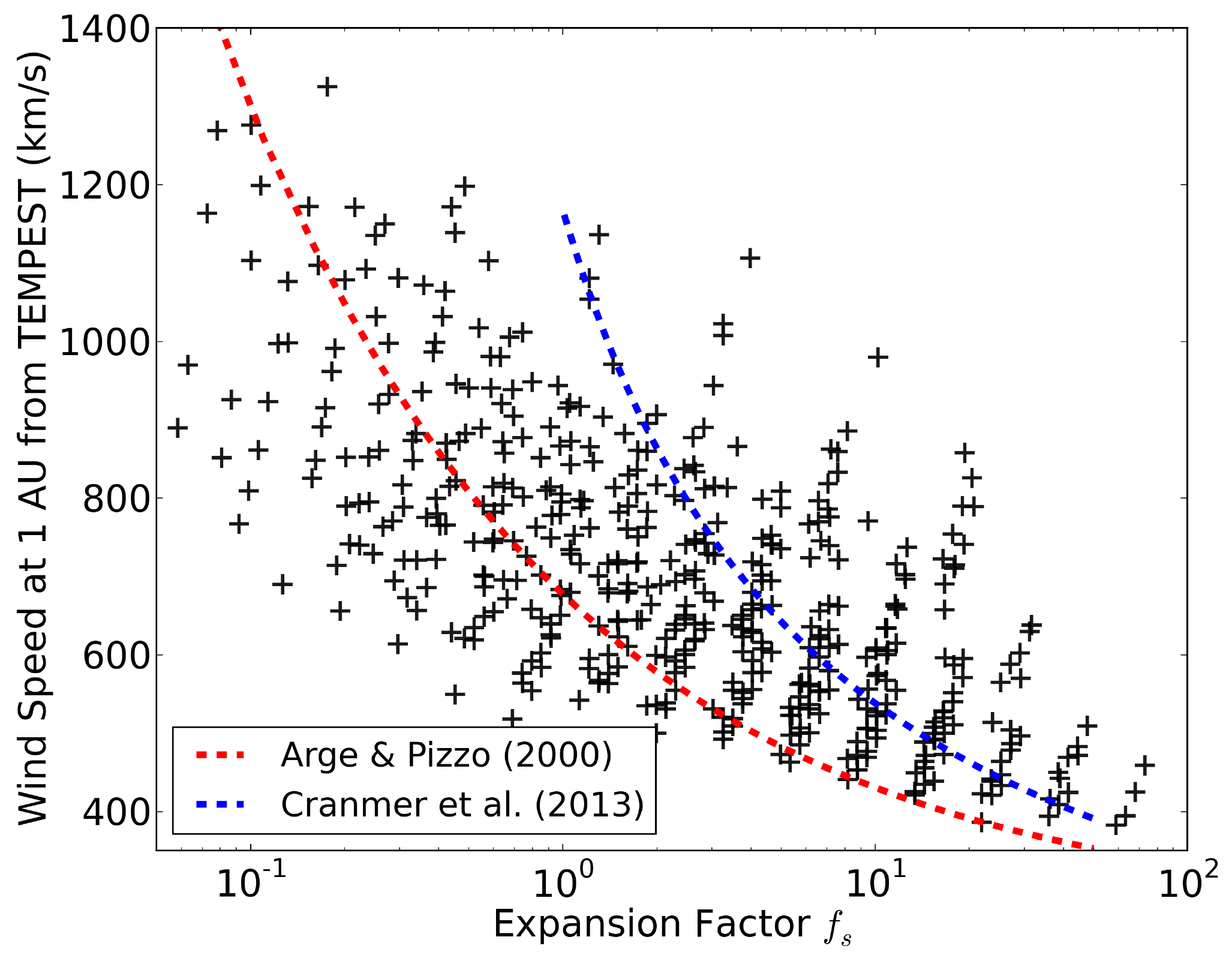}
\caption[Expansion factor versus wind speed with TEMPEST]{Here we show a plot similar to Figure \ref{fig:zephyr_wsa}, now for results from TEMPEST. Again, the red dashed line is wind speed {\it at the source surface}, given by Equation \eqref{eq:argepizzo}, which should be lower than the wind speed at 1 AU for most of the models. The blue dashed line, as in Figure \ref{fig:zephyr_wsa}, is the concordance relation given by \cite{2013ApJ...767..125C}.}
\label{fig:tempest_wsa}
\end{figure}
\clearpage
}

\section{Code Comparisons}
In Figure \ref{fig:zt}, we show directly the wind speeds determined by both modes of TEMPEST and by ZEPHYR. The solutions found by Miranda have a minimum speed at 1 AU around 200 km s$^{-1}$, similar to the observed lower limit of in situ measurements. Figure \ref{fig:zt}a highlights the two discrete ways in which Alfv\'en waves contribute to the acceleration of the solar wind. It is important to note that the scatter in comparing ZEPHYR and a WSA prediction (Figure \ref{fig:zt}b) is much greater than the scatter in the ZEPHYR-TEMPEST comparison, due to the magnetic variability ignored by using only the expansion factor to describe the geometry. 

%\placefigure{fig:zt}
\afterpage{%
\begin{figure}[p!]
\includegraphics[width=0.8\textwidth]{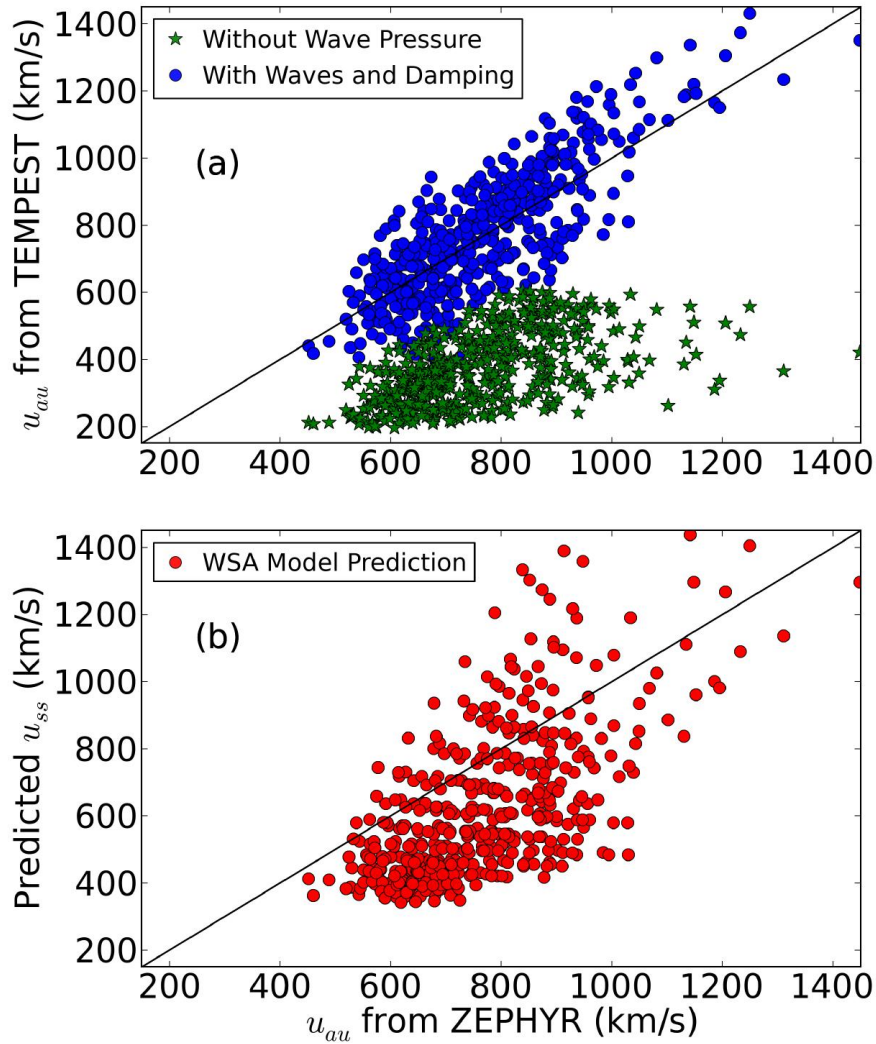}
\caption[Predicted wind speeds from TEMPEST and ZEPHYR]{(a) Wind speeds determined by TEMPEST compared to those from ZEPHYR. The black line represents agreement. Models that reached a steady state solution in ZEPHYR are highlighted in blue; green shows the initial Miranda solutions. (b) Predictions using Equation \eqref{eq:argepizzo} compared to ZEPHYR results.}
\label{fig:zt}
\end{figure}
\clearpage
}

The root-mean-square (RMS) difference in the ZEPHYR-TEMPEST comparison (the blue points in Figure  \ref{fig:zt}a) is 115 km s$^{-1}$, while the RMS difference in the ZEPHYR-WSA comparison (the red points in Figure \ref{fig:zt}b) is 228 km s$^{-1}$. We expect the overall lower speeds of the WSA model, but the scatter is much greater when using WSA to make predictions instead of TEMPEST. The average percent difference between the computed speeds for ZEPHYR and those of the full mode of TEMPEST for each model is just under 14\%. We also ran the same 628 models through a version of TEMPEST that directly reads in the temperature and reflection coefficient profiles, and the percent difference was just below 12\%. We discuss other possible improvements to TEMPEST to lower this scatter in \S \ref{sec:ch3disc}. These numbers indicate that TEMPEST, while it makes many simplifying assumptions, is a more consistent predictor of wind speeds than the traditional observationally-derived WSA approach, which does not specify any particular choice of the underlying physics that accelerates the wind. There does exist the possibility that WSA predictions better match observations than either TEMPEST or ZEPHYR, and future work will use the completed TEMPEST code, in combination with magnetic extrapolations of the coronal field, to predict solar wind properties for specific time periods and compare them with in situ measurements.

Another important distinction between these two codes is CPU run-time. TEMPEST runs over forty times faster than ZEPHYR because it makes many simplifying assumptions. We intend to take advantage of the ease of parallel processing in Python to improve this speed increase further in future versions of the code, which has been made publicly available on GitHub.\footnote{http://github.com/lnwoolsey/tempest}

\section{Discussion}
\label{sec:ch3disc}
We have used WTD models to heat the corona through dissipation of heat by turbulent cascade and accelerate the wind through increased gas pressure and additional wave pressure effects. Our primary goal for this chapter was to improve empirical forecasting techniques for the steady-state solar wind. As we have shown, the community often relies on WSA modeling, based on a single parameter of the magnetic field expansion in open flux tubes. Even with the advances of combining MHD simulations as the WSA-ENLIL model, comparisons between predictions and observations make it clear that further improvements are still necessary. An important point to make is that extrapolations from magnetograms show that many flux tube magnetic field strengths do not all monotonically decrease, so two models with identical expansion factors could result in rather different structures. We anticipate that TEMPEST could easily be incorporated within an existing framework to couple it with a full MHD simulation above the source surface.

The first code we discuss, ZEPHYR, has been shown to correspond well with observations of coronal holes and other magnetic structures in the corona \citep{2007ApJS..171..520C}. We investigate here the results of a grid of models that spans the entire range of observed flux tube strengths throughout several solar cycles to test the full parameter space of all possible open magnetic field profiles.

ZEPHYR also provides us with temperature-magnetic field correlations that help to take out much of the computation time for a stand-alone code, TEMPEST, that solves the momentum conservation equation for the outflow solution of the solar wind based on a magnetic field profile. The solar physics community has come a long way since Parker's spherically symmetric, isothermal corona, but the groundwork laid by this early theory is still fully applicable.

The special case presented by pseudostreamers is an ongoing area of our analysis. Pseudostreamers do not contribute to the heliospheric current sheet and they seem to be a source of the slow solar wind. The community does not fully understand the differences in the physical properties of the solar wind that may emanate from pseudostreamers and helmet streamers, although observational evidence suggests the slow wind is generated from these areas or the edges of coronal holes. Our results from both codes do not currently recreate the bimodal distribution of wind speeds observed at 1 AU. It is unclear whether this is due to the inclusion of many unphysical flux tube models or because the fast wind and the slow wind are generated by {\it different} physical mechanisms.

One slightly troubling feature of the ZEPHYR and TEMPEST model results is a relative paucity of truly ``slow'' wind streams ($u < 350$ km s$^{-1}$) in comparison to the observed solar wind.  However, many of the slowest wind streams at 1 AU may be the result of gradual deceleration due to stream interactions between 0.1 and 1 AU \citep{2011JGRA..116.3106M}.  Similarly, previous ZEPHYR models of near-equatorial quiet-Sun stream lines exhibit a realistic distribution of slow speeds at 0.1 AU, but they exhibited roughly 150 km s$^{-1}$ of extra acceleration out to 1 AU when modeled in ZEPHYR without stream interactions \citep{2013ApJ...767..125C}.  Clearly, taking account of the development of corotating interaction regions and other stream-stream effects is key to producing more realistic predictions at 1 AU.

Another important avenue of future work will be to compare predictions of wind speeds from TEMPEST with in situ measurements. We are already able to reproduce well-known correlations and linear fits from observations, but accurate forecasting is our goal. Other ways in which forecasting efforts can be improved that TEMPEST does not address include better lower boundary conditions on and coronal extrapolation of $\vec{B}$, moving from 1D to a higher dimensional code, and including kinetic effects of a multi-fluid model ($T_{p} \ne T_{e}$, $T_{\parallel} \ne T_{\perp}$).

Space weather is dominated by both coronal mass ejections (CMEs) and high-speed wind streams. The latter is well-modeled by the physics presented here, and these high-speed streams produce a greatly increased electron flux in the Earth's magnetosphere, which can lead to satellite disruptions and power-grid failure \citep{2011A&A...526A..20V}. Understanding the Sun's effect on the heliosphere is also important for the study of other stars, especially in the ongoing search for an Earth analog. The Sun is an indespensible laboratory for understanding stellar physics due to the plethora of observations available. The modeling we have done in this chapter marks an important step toward full understanding of the coronal heating problem and identifying sources of solar wind acceleration.
%% timedependence_phd4.tex
\chapter{Time-Dependent Turbulent Heating}
%%%%%%%%%%%%%%%%%%%%%%%%%%%%%%%%%%%%%%%%%%%%%%%%

\section{Time Dependence and Higher Dimensions}
From the many suggested physical processes to heat the corona and accelerate the solar wind, nothing has been ruled out because (1) we have not yet made the observations needed to distinguish between the competing paradigms, and (2) most models still employ free parameters that can be adjusted to improve the agreement with existing observational constraints. A complete solution must account for all important sources of mass and energy into the three-dimensional and time-varying corona. Regardless of whether the dominant coronal fluctuations are wave-like or reconnection-driven, they are generated at small spatial scales in the lower atmosphere and are magnified and stretched as they propagate up in height. Their impact on the solar wind's energy budget depends crucially on the multi-scale topological structure of the Sun's magnetic field. 

The current generation of time-steady 1D turbulence-driven solar wind models \citep[e.g.][]{2007ApJS..171..520C, 2010ApJ...708L.116V, 2011ApJ...743..197C, 2013ApJ...767..125C,2014ApJ...784..120L} contain detailed descriptions of many physical processes relevant to coronal heating and solar wind acceleration. However, none of them self-consistently simulate the actual process of MHD turbulent cascade as a consequence of the partial reflections and nonlinear interactions of Alfv\'enic wave packets. Thus, it is important to improve our understanding of the flux-tube geometry of the open-field corona and how various types of fluctuations interact with those structures by using higher-dimensional and time-dependent models to understand the underlying physical processes.

In this chapter, I compare two models of Alfv\'en-wave-driven turbulent heating: ZEPHYR \citep{2007ApJS..171..520C} and BRAID \citep{2011ApJ...736....3V}. These and other such models have been shown to naturally produce realistic fast and slow winds with wave amplitudes of the same order of magnitude as those observed in the corona and heliosphere \citep[see also][]{1986JGR....91.4111H, 1999ApJ...523L..93M, 2006JGRA..111.6101S, 2010LRSP....7....4O, 2010ApJ...708L.116V, 2011ApJ...743..197C, 2014ApJ...784..120L, 2014MNRAS.440..971M, 2014ApJ...787..160W}. The chapter will focus on three models that are typical of common structures in the solar corona, and go into the new insights one can draw from the time-dependent heating results.

\section{MHD Turbulence Simulation Methods}
\subsection{Previous time-steady modeling}
The magnetic field profiles considered are listed in Table \ref{tab:threemodels} and are representative of a polar coronal hole, an equatorial streamer, and a flux tube neighboring an active region. The radial profiles of magnetic field strength $B(r)$ for these models are shown in Figure \ref{fig:zephyr3}a \citep[see also][]{1998A&A...337..940B,2007ApJS..171..520C}. The Alfv\'en travel time from the base of the model to a height of 2 solar radii is calculated using Equation \eqref{eq:traveltime}. The expansion factor, $f_{s}$, is a ratio of the magnetic field strength at a height of 1.5 solar radii to the field strength at the photospheric base. As in Equation \eqref{eq:test}, the base is at a height of 0.04 solar radii, the size of a supergranule. The expansion factor is normalized such that $f_{s} = 1$ is simple radial expansion, and $f_{s} > 1$ is considered superradial expansion.

\begin{table}[h!]
    \caption[Three modeled flux tubes studied with BRAID]{Models in BRAID}
    \begin{tabular}{|c|c|c|c|c|}
    \hline
    {\bf Modeled structure}  & {\bf Time to 2R$_{\odot}$} & {\bf Exp. $f_{s}$} & {\bf Speed at 1AU} & {\bf Identifier} \\ \hline \hline
    Polar Coronal Hole & 770 s & 4.5 & 720 km s$^{-1}$ & PCH \\ \hline
    Equatorial Streamer & 3550 s & 9.1 & 480 km s$^{-1}$ & EQS \\ \hline
    Near Active Region & 4400 s & 41 & 450 km s$^{-1}$ & NAR  \\ \hline
    \end{tabular}
    \label{tab:threemodels}
\end{table}

Key results from these three models using ZEPHYR, the one-dimensional time-steady model described in the previous chapter, are presented in Figure \ref{fig:zephyr3}. The polar coronal hole model has the smallest expansion from the photosphere to the low corona, and the highest wind speed at 1 AU. This relationship was first seen in empirical fits to observations \citep[see, e.g.,][]{1990ApJ...355..726W,2000JGR...10510465A} and has also been seen in models \citep[see, e.g.,][]{2014ApJ...787..160W}. The equatorial streamer model represents an open flux tube directly neighboring the helmet streamers seen around the equator at solar minimum, and produces slower wind at 1 AU. The flux tube neighboring an active region has a stronger magnetic field above the transition region and an even slower wind speed at 1 AU than the equatorial streamer model. In these models, the transition region for PCH and EQS is at a height of 0.01 solar radii, and the NAR model has a transition region slightly higher, at a height of 0.015 solar radii. The density profiles were determined self-consistently with the wind speed using mass flux conservation.

\afterpage{%
\begin{figure}
\centerline{\includegraphics[width=\textwidth]{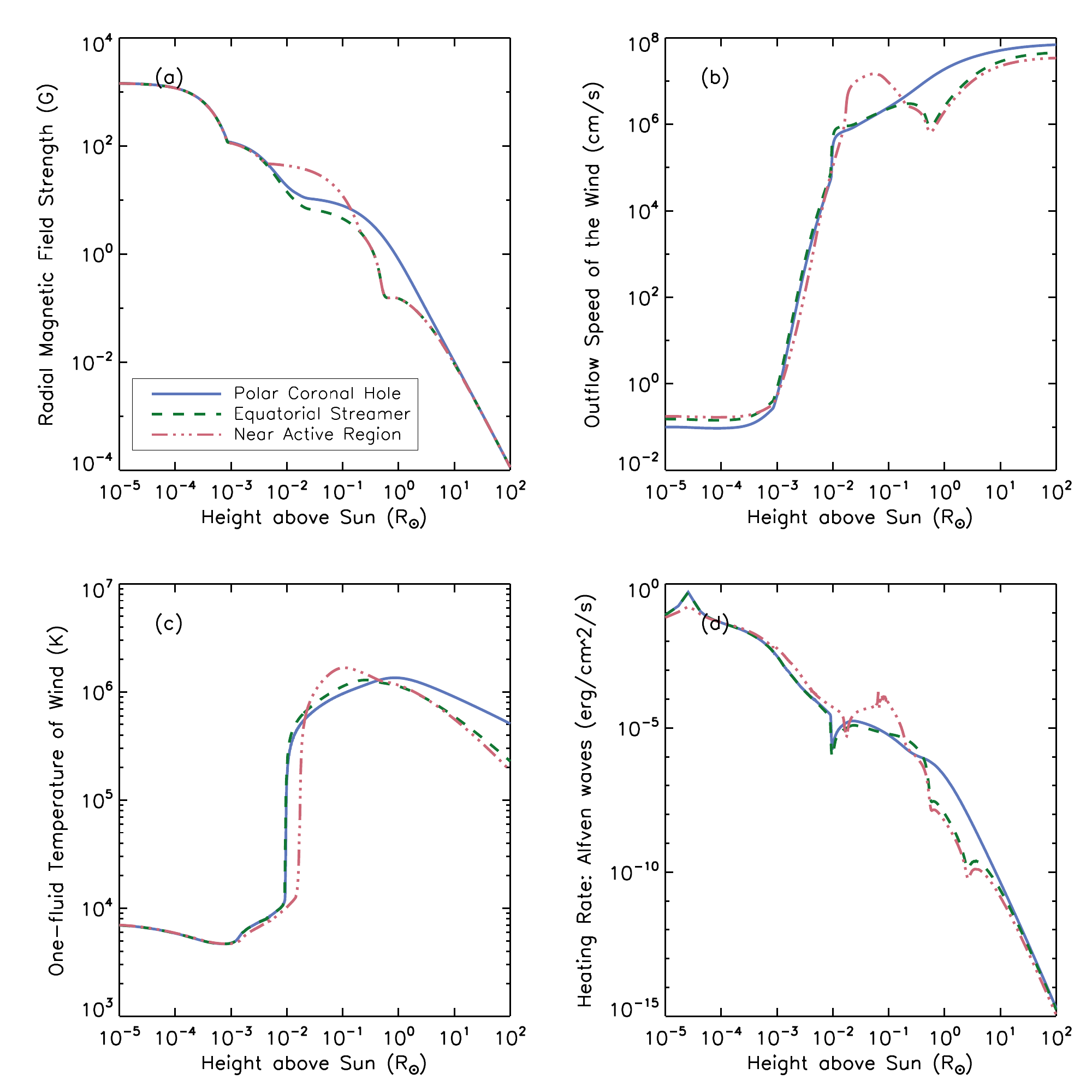}}
\caption[ZEPHYR results for three modeled flux tubes]{The primary input to ZEPHYR are (a) the magnetic profiles of three representative coronal structures. Results for the bulk solar wind properties from each flux tube include: (b) outflow speed, (c) one-fluid temperature, and (d) Alfv\'enic heating rate.}
\label{fig:zephyr3}
\end{figure}
\clearpage
}

ZEPHYR solves for a steady state solution to the solar wind properties generated by a one-dimensional open flux tube. By solving the equations of mass, momentum, and energy conservation and iterating to a stable solution, the code produces solar wind with mean properties that match observations and {\it in situ} measurements \citep{2007ApJS..171..520C,2014ApJ...787..160W}. The code's expression for the turbulent heating is a phenomenological cascade rate whose form has been guided and validated by several generations of numerical simulations and other models of imbalanced, reflection-driven turbulence \citep[see, e.g.,][]{1995PhFl....7.2886H,2007ApJ...655..269L,2009ApJ...701..652C,oughton15}. For additional details, see \S \ref{sec:braiddetails}.

ZEPHYR can only take us so far, however. It is only by modeling the fully 3D spatial and time dependence of the cascade process (together with the intermittent development of magnetic islands and current sheets on small scales) that we can better understand the way in which the plasma is heated by the dissipation of turbulence. We therefore make use of the reduced magnetohydrodynamics (RMHD) code called BRAID \citep{2011ApJ...736....3V}.

\subsection{Including time-dependence and higher dimensions}
There have been previous numerical simulations of reflection-driven RMHD turbulence with the full nonlinear terms  \citep{2003ApJ...597.1097D, 2013ApJ...776..124P} and prior studies using BRAID on closed loops \citep{2011ApJ...736....3V,2012ApJ...746...81A,2013ApJ...773..111A,2014ApJ...786...28A}. Our version of BRAID uses the three-dimensional equations of RMHD \citep{1976PhFl...19..134S,1982PhST....2...83M,1992JPlPh..48...85Z,1998ApJ...494..409B} to solve for the nonlinear reactions between Alfv\'en waves generated at the single footpoint of an open flux tube. RMHD relies on the assumption that the incompressible magnetic fluctuations $\delta{\bf B}$ in the system are small compared to an overall background field ${\bf B}_{0}$. At scales within the turbulence inertial range, observations show that $\delta{\bf B}$ is much smaller in amplitude than the strength of the surrounding magnetic field ${\bf B}$ and $\delta{\bf B}$ is perpendicular to ${\bf B}$ \citep{1990JGR....9520673M,1995SSRv...73....1T,2012ApJ...758..120C}. 

Some implementations of RMHD combine together the implicit assumptions of incompressible fluctuations, high magnetic pressure (i.e., plasma $\beta \ll 1$), and a uniform background field ${\bf B}_0$. However, the flux tubes we model in the upper chromosphere and low corona have some regions with $\beta \approx 1$ and a vertical field ${\bf B}_0$ that declines rapidly with height. It is not necessarily the case that RMHD applies in this situation. There is a self-consistent small-parameter expansion that gives rise to a set of RMHD equations appropriate for the chromosphere and corona \citep[see derivation in][]{2011ApJ...736....3V}. In these equations, the dominant, first-order fluctuations are transverse and incompressible---even when $\beta \approx 1$---and gravitational stratification is included to account for the height variation of ${\bf B}_{0}$.

BRAID uses two cross-sectional dimensions of the flux tube and a third dimension along the length of the flux tube, which aligns with the background field ${\bf B}_{0}$. Alfv\'en waves are generated at the lower boundary by random footpoint motions with an rms velocity of 1.5 km s$^{-1}$ and correlation time of 60 s. These fluctuations are generated by taking a randomized white-noise time stream and passing it through a low-pass (Gaussian) frequency filter that removes fluctuations shorter than the specified correlation time. The time stream is normalized to the desired rms velocity amplitude and split up between two orthogonal low-$k_{\perp}$ Bessel-function modes of the cylindrically symmetric system. The driver modes are shown in Figure \ref{fig:sincos} and are discussed further in Appendix B of \cite{2011ApJ...736....3V}.

\afterpage{%
\begin{figure}
\centerline{\includegraphics[height=0.85\textheight]{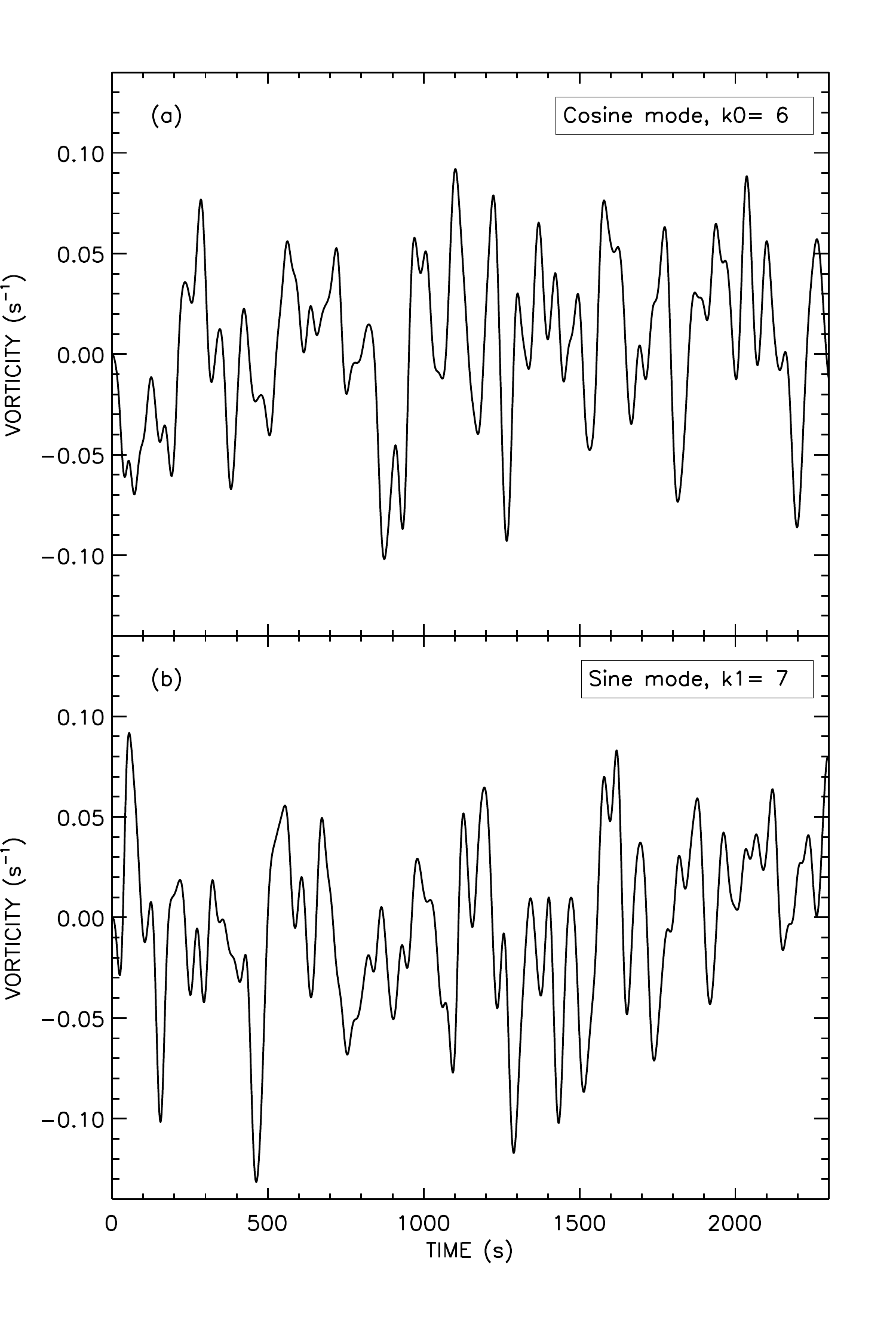}}
\caption[Driving modes for BRAID footpoint motion]{We show here vorticity $\omega$ from (a) cosine and (b) sine driver modes for the polar coronal hole model.}
\label{fig:sincos}
\end{figure}
\clearpage
}

The magnetic and velocity fluctuations can be approximated by 
\begin{subequations}
\begin{align}
{\bf \delta B} = \nabla_{\perp}h \times {\bf B},\\
{\bf \delta v} = \nabla_{\perp}f \times {\bf \hat{B}},
\end{align}
\end{subequations}
\noindent where ${\bf B}$ is the full magnetic field vector, whose magnitude varies with height, and $\hat{\bf B}$ is the unit vector along the magnetic field. Also, $h({\bf r},t)$ is a height- and time-dependent function that is analogous to the standard RMHD magnetic flux function, and $f({\bf r},t)$ is a velocity stream function (sometimes called $\psi$ in other derivations of RMHD). We can also define the magnetic torsion parameter $\alpha \equiv -\nabla_{\perp}^{2}h$ and the parallel component of vorticity $\omega \equiv -\nabla_{\perp}^{2}f$ \citep[see, e.g.,][]{1982PhST....2...83M}. The functions $h({\bf r},t)$ and $f({\bf r},t)$ satisfy the coupled equations
\begin{subequations}\begin{small}
\begin{align}
\frac{\partial{\omega}}{\partial{t}} + {\bf \hat{B}} \cdot \left(\nabla_{\perp}\omega \times \nabla_{\perp}f\right) = \nonumber\\v_{\rm A}^{2}\left[{\bf \hat{B}} \cdot \nabla{\alpha} + {\bf \hat{B}} \cdot \left(\nabla_{\perp}\alpha \times \nabla_{\perp}h\right)\right] + D_{v},\\
\frac{\partial{h}}{\partial{t}} = {\bf \hat{B}} \cdot \nabla{f} + \frac{f}{H_{B}} + {\bf \hat{B}} \cdot \left(\nabla_{\perp}f \times \nabla_{\perp}h\right) + D_{m},
\end{align}\end{small}
\end{subequations}
where $H_{B}$ is the magnetic scale length and $D_{v}$ and $D_{m}$ describe the effects of viscosity and resistivity \citep[see detailed derivation in][]{2011ApJ...736....3V}. Both $D_v$ and $D_m$ have a hyperdiffusive $k_{\perp}^{4}$ dependence so that the smallest eddies are damped preferentially and the cascade is allowed to proceed without significant damping over most of the $k_{\perp}$ inertial range.

We have extended previous work using BRAID by using an open upper boundary condition instead of a closed coronal loop. The model extends to a height of $z_{\rm top} = 2 R_{\odot}$. This height was chosen to model as much of the solar wind acceleration region as possible, without extending into regions where the wind speed becomes an appreciable fraction of the Alfv\'en speed, since the RMHD equations of BRAID don't include the outflow speed. To implement the open upper boundary condition, we set $\omega_{+}$ (which corresponds to the downward Els\"asser variable $Z_{+}$) to zero, whereas in the previous coronal loop models that used BRAID, it was set to the opposite footpoint's boundary time stream as described above \citep[][and references therein]{2014ApJ...786...28A}.

\section{Time-Averaged Results from Three Modeled Flux Tubes}
We present three models that are representative of common coronal structures (see Table \ref{tab:threemodels}). Figure \ref{fig:models} gives some of the time-steady background variables for the three models, based on the input magnetic field profiles shown in Figure \ref{fig:zephyr3}a. 

\afterpage{%
\begin{figure}
\centerline{\includegraphics[height=0.9\textheight]{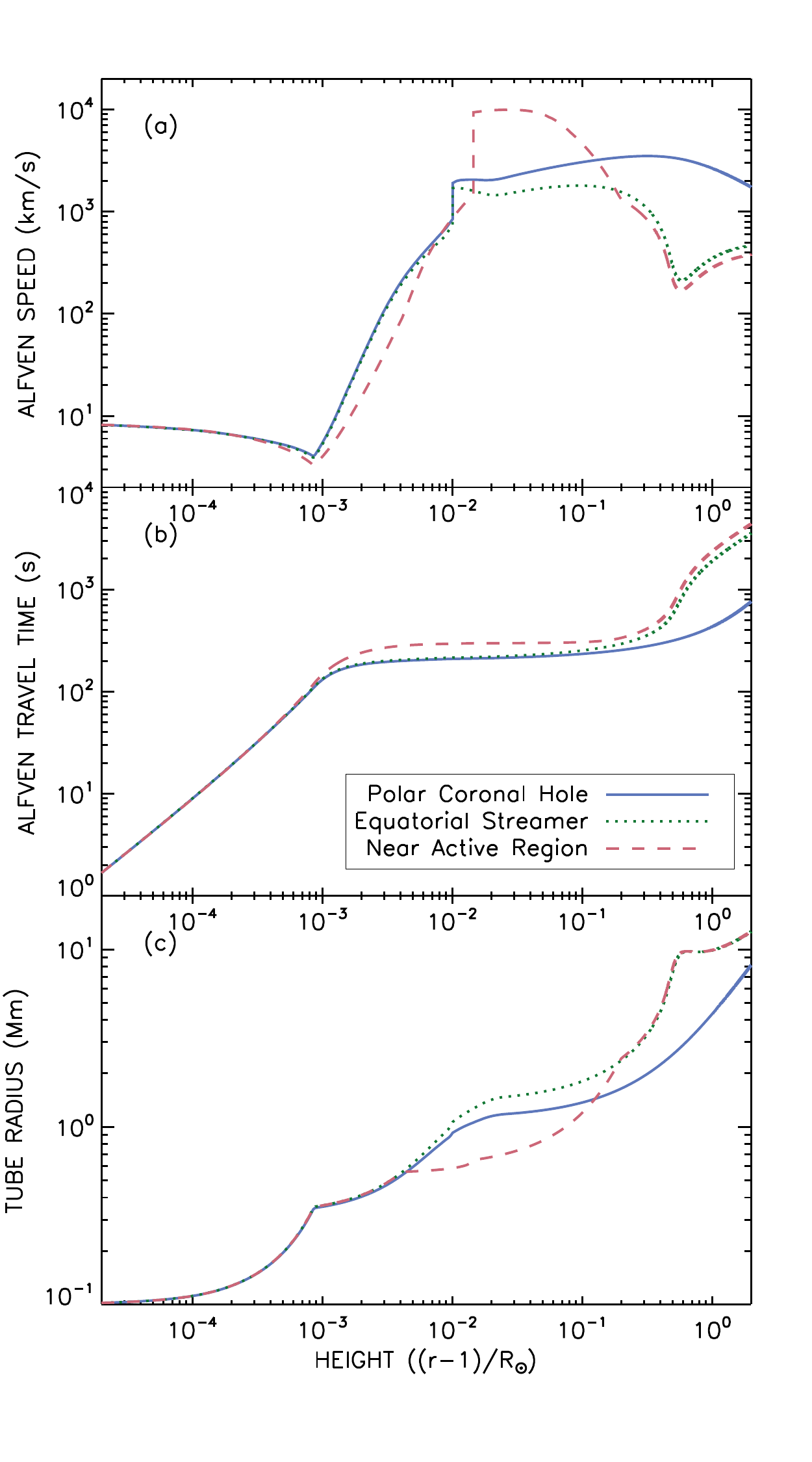}}
\vspace{-4mm}
\caption[Time-averaged radial profiles for BRAID flux tubes]{Radial profiles for (a) Alfv\'en speed, (b) Alfv\'en travel time, and (c) tube radius.}
\label{fig:models}
\end{figure}
\clearpage
}

The Alfv\'{e}n speed, $V_{\rm A} = B / \sqrt{4\pi\rho}$ shows a rapid rise in the upper chromosphere, followed by varied behavior in the corona depending on the model. The Alfv\'{e}n travel time is defined as a monotonically increasing function of height,
\begin{equation}\tau_{\rm A} (z) \, = \, \int_{0}^{z} \frac{dz'}{V_{\rm A}(z')}\label{eq:traveltime}\end{equation} where $z=0$ is the photospheric lower boundary of each model. The BRAID code uses $\tau_{\rm A}$ as the primary height coordinate. Figure \ref{fig:models} shows the modeled transverse radius of the flux tube, which is normalized to 100~km at $z=0$ (i.e., a typical length scale for an intergranular bright point) and is assumed to remain proportional to $B^{-1/2}$ in accordance with magnetic flux conservation.

Figure \ref{fig:pch} provides the time-averaged results from BRAID for the PCH. The transition region is at a travel time of roughly 210 s and is shown with a dotted line. In Figure \ref{fig:pch}a, we show the magnetic, kinetic, and total energy densities of the
RMHD fluctuations. We plot these quantities separately since no equipartition is assumed. Figure \ref{fig:pch}b shows the increase in the rms transverse velocity amplitude, $\Delta v_{\rm rms}$ with increasing height, which roughly follows the expected sub-Alfv\'enic WKB relation $\Delta v_{\rm rms} \propto \rho_{0}^{-1/4}$ below the transition region. The heating rate is also broken up into magnetic (from the $D_m$ term), kinetic (from the $D_v$ term), and total for Figure \ref{fig:pch}c. Finally, we show the magnetic field fluctuation as a function of travel time in Figure \ref{fig:pch}d. Like the rms velocity, the sub-Alfv\'enic WKB relation $\Delta b_{\rm rms} \propto \rho_{0}^{1/4}$ is followed below the transition region but diverges above it.

\afterpage{%
\begin{figure}
\centerline{\includegraphics[width=\textwidth]{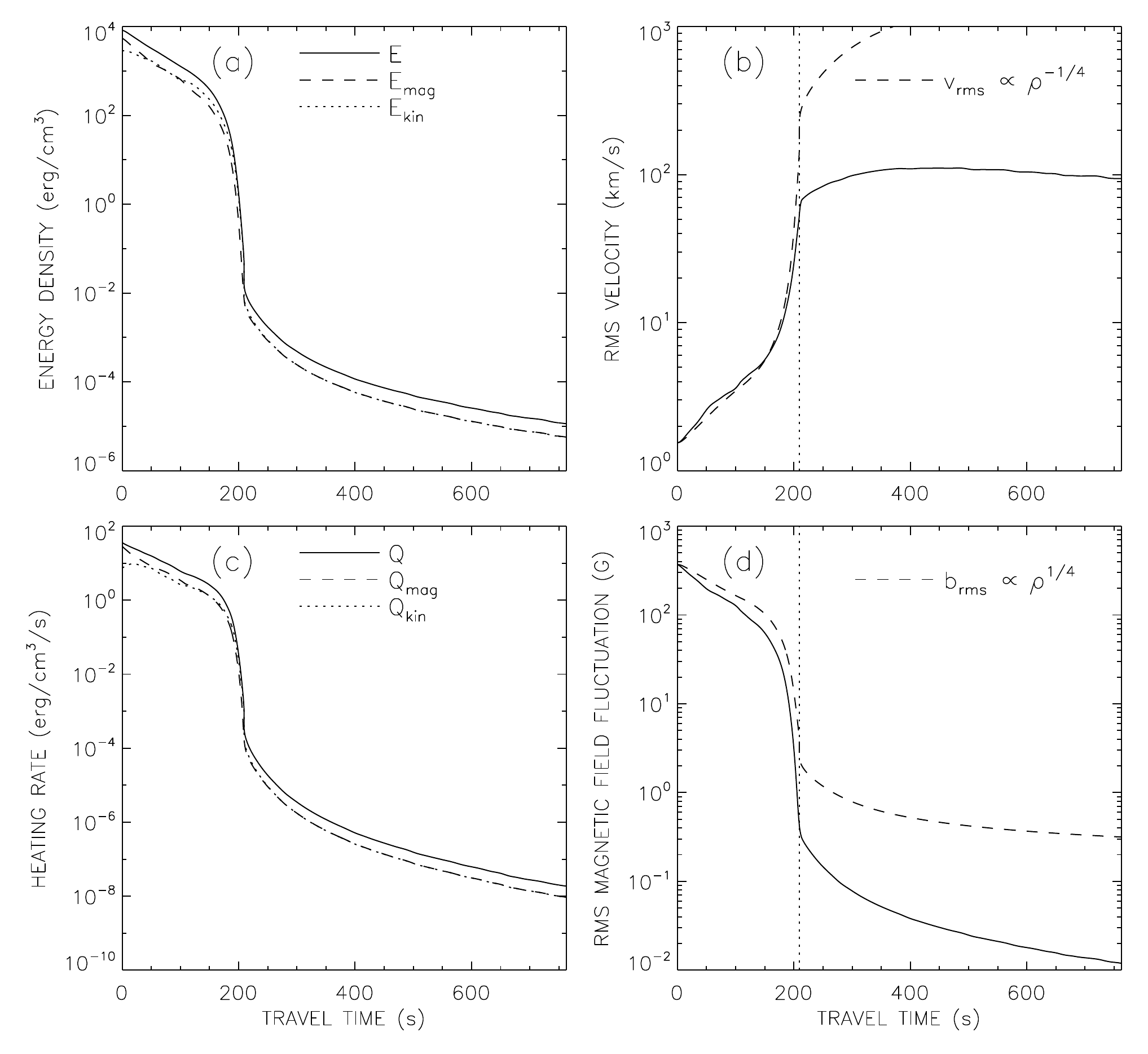}}
\caption[Time-averaged BRAID results for a polar coronal hole]{The time-averaged results for the PCH model are shown as a function of Alfv\'en travel time as a proxy for height above photosphere: (a) Energy density, (b) rms velocity, (c) heating rate, and (d) rms magnetic field fluctuation.}
\label{fig:pch}
\end{figure}
\clearpage
}

The quantities plotted in Figures \ref{fig:pch}b, \ref{fig:pch}d, and \ref{fig:elsa}b have gone through two ``levels'' of root mean square averaging. First, we take the variance over all $k_{\perp}$ modes at a given height and time, then take the square root.  Then, at each height, we take an average over the simulation-time dimension using the squares of the first level of rms values. It is those quantities that we then take the square root of and plot.

In Figure \ref{fig:elsa}a, we show the magnitude of the Els\"asser variables, ${\bf Z_{\pm}} = {\bf \delta v} \pm {\bf \delta B} / \sqrt{4\pi \rho_0}$, for the PCH model, and show the fluctuations $\Delta v_{\rm rms}$ and $\Delta b_{\rm rms}$ in Figure \ref{fig:elsa}b. Note that here $\Delta b_{\rm rms}$ is in velocity units, as the actual fluctuations are divided by $\sqrt{4 \pi \rho_{0}}$ (where $\rho_{0}$ is the background density) for comparison with the velocity fluctuations. Our boundary condition enforces that the incoming waves ($Z_{+}$) have exactly zero amplitude at the upper boundary of our model. Figure \ref{fig:elsa}a also shows the radial dependence of the Els\"asser variables from the ZEPHYR model for the coronal hole \citep{2007ApJS..171..520C}. The ZEPHYR code computes the Alfv\'enic wave energy using a damped wave action conservation equation that contains the assumption that $Z_{-} \gg Z_{+}$. Thus, when reporting the magnitudes of the Els\"asser variables here for direct comparison with the BRAID results, we make use of Equation (56) of Cranmer et al. (2007; \citep{2007ApJS..171..520C}) and correct for non-WKB effects by multiplying these quantities by a factor of $(1 + {\mathcal R}^{2})/(1 - {\mathcal R}^{2})$, where ${\mathcal R}$ is the reflection coefficient. With this correction, there is good agreement between the modeling of the PCH for both codes. We use the same correcting factor when plotting the $\Delta{v_{\rm rms}}$ in Figure \ref{fig:elsa}b. Because ZEPHYR assumes equipartition, $\Delta{v_{\rm rms}}$ and $\Delta{b_{\rm rms}}$ are equivalent for that model.

\afterpage{%
\begin{figure}
\centerline{\includegraphics[height=0.8\textheight]{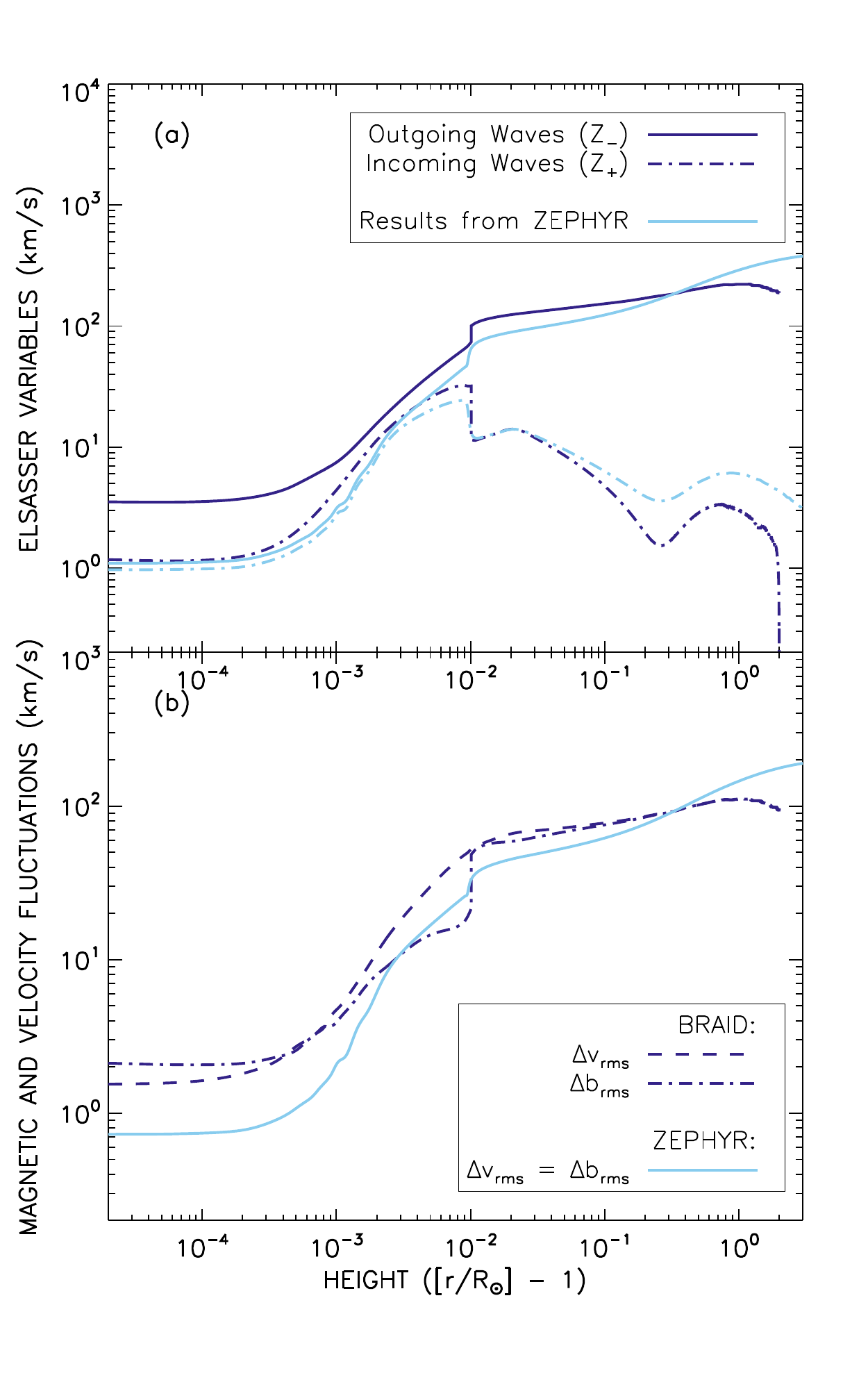}}
\caption[Els\"asser variables for PCH model]{For the PCH model, we show (a) the time-averaged amplitudes of incoming and outgoing Alfv\'en waves and (b) the rms amplitudes of magnetic and velocity fluctuations from BRAID and ZEPHYR.}
\label{fig:elsa}
\end{figure}
\clearpage
}

\afterpage{%
\begin{figure}
\centerline{\includegraphics[height=0.8\textheight]{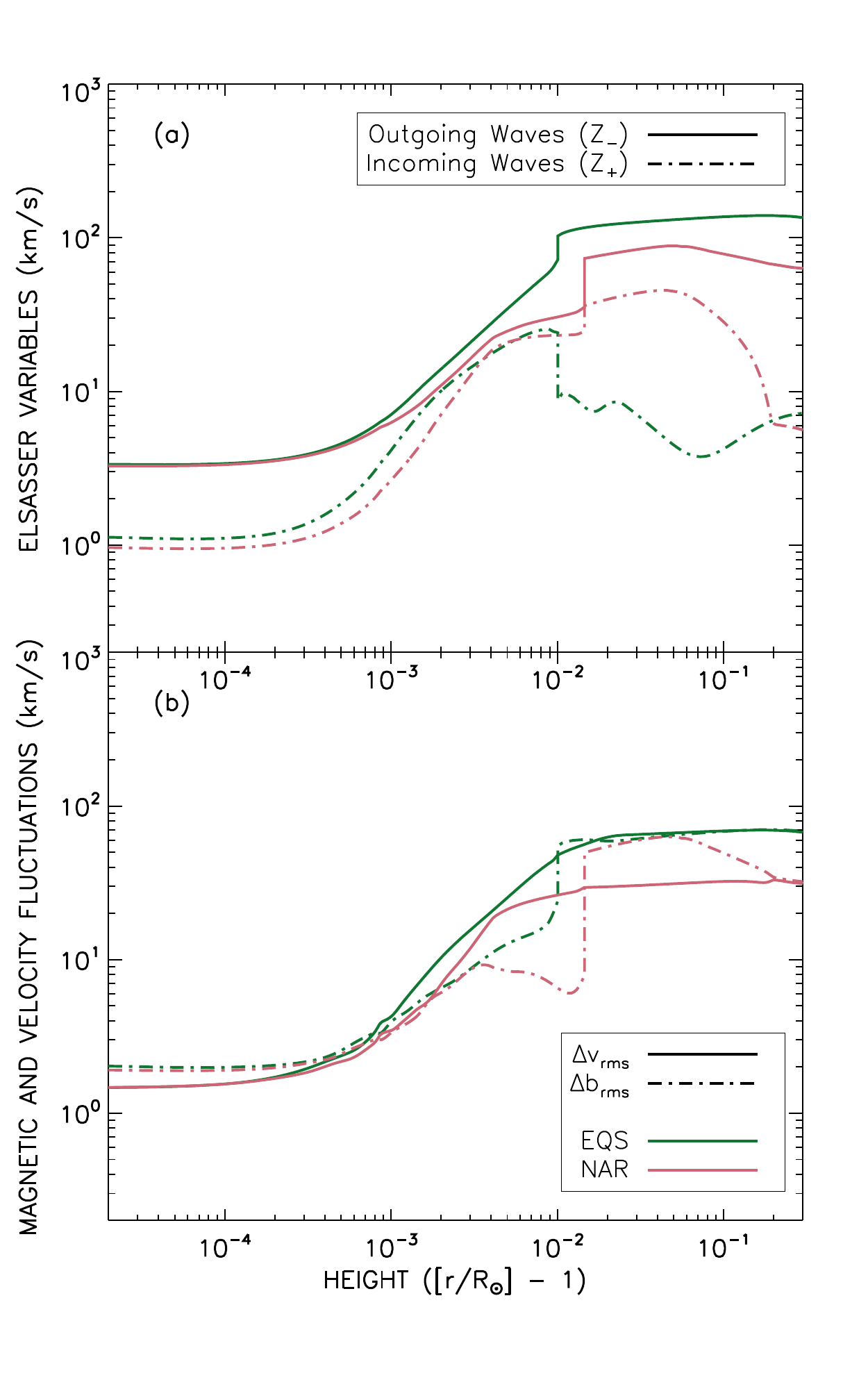}}
\caption[Els\"asser variables for EQS and NAR models]{Similar to Figure \ref{fig:elsa}; For the EQS and NAR models, we show the (a) outgoing and incoming Els\"asser variables and (b) the rms amplitudes of magnetic and velocity fluctuations.}
\label{fig:elsasser}
\end{figure}
\clearpage
}

Figure \ref{fig:elsasser} shows the Els\"asser variables and the amplitudes of magnetic and velocity fluctuations for both the equatorial streamer (EQS) and active region (NAR) models to a height of $z = 0.5 R_{\odot}$, where turbulence has had time to develop in the simulation. It is worthy of note that the NAR model shows an increase in $Z_{+}$ above the transition region where the PCH and EQS models show a decrease. For these models in Figure \ref{fig:elsasser}b and the results for the PCH (Figure \ref{fig:elsa}b), the shape of the $\Delta{b_{\rm rms}}$ is expected to have a sharp increase at the transition region, while the $\Delta{v_{rms}}$ doesn't show it \citep[see, e.g., Figure 9 of][]{2005ApJS..156..265C}. 

\subsection{Energy partitioning in the corona}

In our previous modeling using ZEPHYR (see Section 4.2.1), we assumed equipartition between kinetic and magnetic potential energy densities. With BRAID, we are able to investigate how far the PCH model differs from this simplifying assumption. Figure \ref{fig:part} shows the ratio of the time-averaged magnetic and kinetic energy densities. Above the transition region, equipartition is a valid assumption. However, at the base of the flux tube, magnetic potential energy dominates, and this transitions to a stronger dominance of the kinetic energy right up to the transition region.

Also plotted in Figure \ref{fig:part} are six curves showing predictions from non-WKB reflection with a range of frequencies between 0.7 and 4.0 mHz. While the general shape is consistent with the BRAID results, it is interesting to note that the linear method of computing non-WKB reflection does have its limits. These predictions were made using a previous coronal hole model as a basis \citep{2005ApJS..156..265C}. However, that model had a lower transition region ($z_{\rm TR} \approx 0.003 R_{\odot}$), so we have multiplied the height coordinate by a factor of 3.4 to match with the ZEPHYR/BRAID coronal hole model used in this project.

\afterpage{%
\begin{figure}
\centerline{\includegraphics[width=\columnwidth]{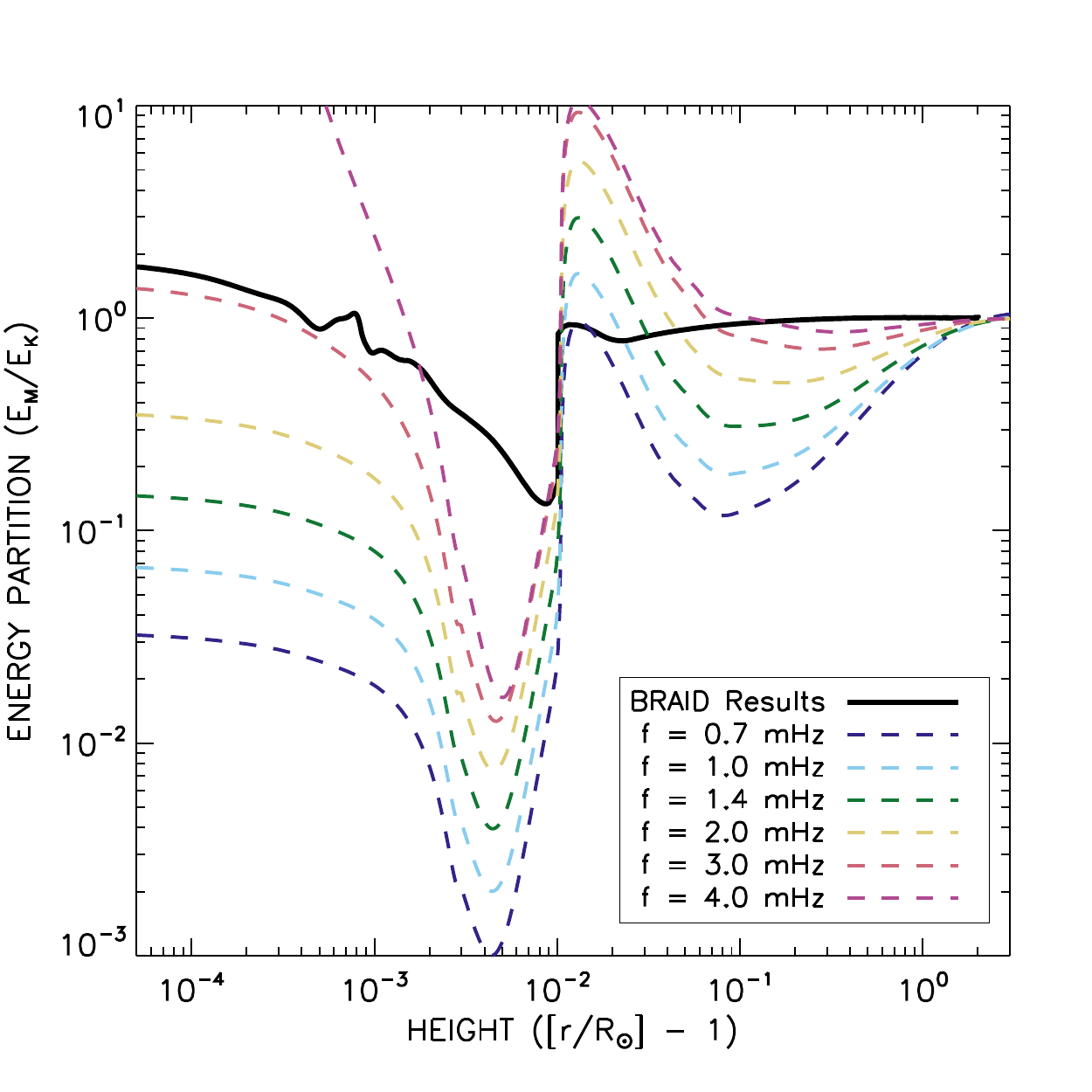}}
\caption[Investigating equipartition assumption of ZEPHYR using BRAID]{Ratio of magnetic to kinetic energy density differs from equipartition below the transition region.}
\label{fig:part}
\end{figure}
\clearpage
}

Non-WKB theory predicts the kinetic energy to dominate in the chromosphere (e.g., $E_{M}/E_{K} < 1$ at low heights). This was discussed as an after-effect of the transition from kink-mode MHD waves in the photosphere to volume-filling Alfv\'{e}n waves in the upper chromosphere \citep[Appendix A of][]{2005ApJS..156..265C}. At low frequencies, the kink-mode waves are partially evanescent, with two possible solutions for the height-increase of $\Delta v_{\rm rms}$. One of the solutions has $E_{M}/E_{K} < 1$ and the other has $E_{M}/E_{K} > 1$. However, only the solution with $E_{M}/E_{K} < 1$ has a physically realistic energy density profile (exponentially decaying with increasing height), and this solution also corresponds to a net upward phase speed \citep[see also][]{1995ApJ...444..879W}.

\subsection{Heating rates: comparison with time-steady modeling}
\label{sec:braiddetails}
We compare the time-averaged heating rates from BRAID with the results from the time-steady modeling using ZEPHYR. Figure \ref{fig:Qcompare}a shows the radial dependence of the heating rate $Q$, and Figure \ref{fig:Qcompare}b shows the ratio between the numerically computed heating rates $Q$ with the phenomenological heating rate $Q_{\rm phen}$, as well as comparisons between $Q$ and phenomenological heating rates with added correction factors ($Q_{\rm Z,07}$ and $Q_{\rm loop}$), described in the following paragraphs. The heating rate $Q_{\rm phen}$ is based on the result of a long series of turbulence simulations and models \citep{1980PhRvL..45..144D,1983A&A...126...51G,1995PhFl....7.2886H,1999ApJ...523L..93M,2001ApJ...548..482D,2002ApJ...575..571D,2009ApJ...701..652C}. The analytical expression for this base phenomenological rate is given by: 
\begin{equation}
Q_{\rm phen} = \rho_{0}\frac{Z_{-}^{2}Z_{+} + Z_{+}^{2}Z_{-}}{4L_{\perp}},
\label{eq:phen}
\end{equation}
where $\rho_{0}$ is the solar wind density, $Z_{-}$ and $Z_{+}$ are the Els\"asser variable amplitudes that represent incoming and outgoing Alfv\'en waves, and $L_{\perp}$ is the turbulent correlation length. We normalize $L_{\perp}$ to a value of 75~km at the photosphere \citep{2007ApJS..171..520C}. This allows us to write $L_{\perp} = 0.75 R$, where $R$ is the radius of the flux tube, which scales as $B^{-1/2}$. Below the transition region, $Q/Q_{\rm phen} \approx 0.2$, which is similar to what has been found for closed loops \citep{2011ApJ...736....3V}. Above the transition region, $Q/Q_{\rm phen} \gg 1$, which may be explainable if the actual correlation length $L_{\perp}$ expands {\it less rapidly} than we assumed from its proportionality with the flux tube radius $R$. Because the ratio $Q/Q_{\rm phen}$ strays as much as an order of magnitude away from unity, we further compare BRAID's computed heating rate with analytical expressions that contain correcting factors that take into account efficiency of turbulence as a function of height.

\afterpage{%
\begin{figure}
\centerline{\includegraphics[height=0.85\textheight]{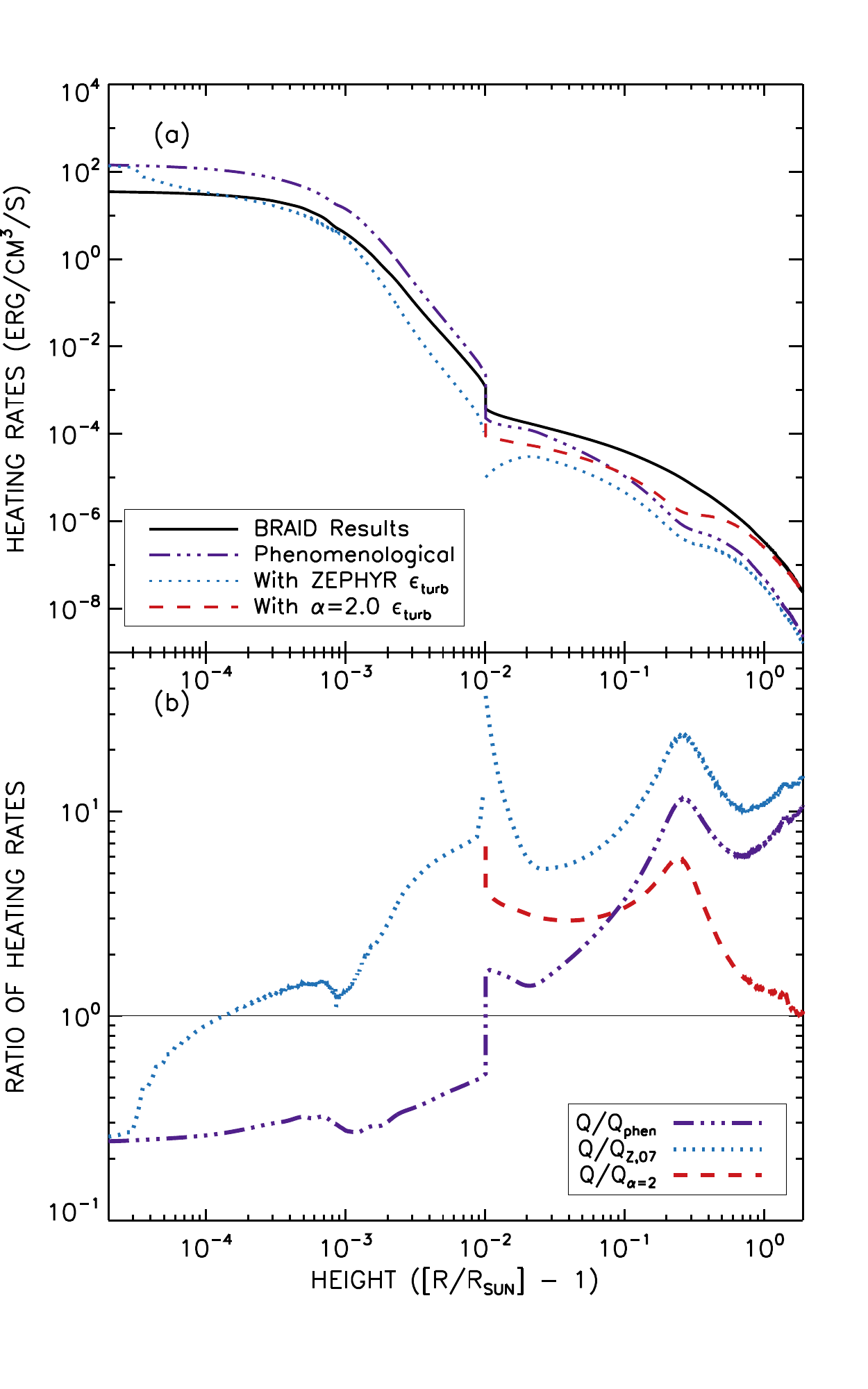}}
\caption[Comparing heating rates from BRAID]{We compare the time-averaged heating rate from BRAID for the PCH model with Equations \eqref{eq:phen}, \eqref{eq:qfull} and \eqref{eq:looprate} by (a) directly plotting them as a function of height and (b) plotting ratios of the rates, with a thin black line showing unity.}
\label{fig:Qcompare}
\end{figure}
\clearpage
}

The first of the two efficiency factors we use is based on prior studies \citep{2001ApJ...548..482D, 2003ApJ...597.1097D, 2007ApJS..171..520C}. The extended expression, $Q_{\rm Z,07}$ is given by: \begin{subequations}
\begin{align}
Q_{\rm Z,07} = \epsilon_{\rm turb} Q_{\rm phen},\\
\epsilon_{\rm turb} = \frac{1}{1+(t_{\rm eddy}/t_{\rm ref})^{n}},
\end{align}\label{eq:qfull}
\end{subequations} where $t_{\rm eddy}$ is the outer-scale eddy cascade time, $t_{\rm eddy} = L_{\perp}\sqrt{3\pi}/(1+M_{\rm A})/v_{\perp}$, and $t_{\rm ref}$ is the macroscopic Alfv\'en wave reflection timescale, $t_{\rm ref} = 1/|\nabla \cdot {\bf V_{\rm A}}|$ \citep{2007ApJS..171..520C}. In the expression for $t_{\rm eddy}$, the velocity $v_{\perp}$ is the amplitude of perpendicular fluctuations, defined previously in the BRAID results as $\Delta v_{\rm rms}$. When $t_{\rm eddy} \gg t_{\rm ref}$, turbulent heating is quenched. The turbulent efficiency factor $\epsilon_{\rm turb}$ accounts for regions where energy is carried away before a turbulent cascade can develop. The exponent $n$ is set to 1 based on analytical and numerical models by \cite{1980PhRvL..45..144D}, \cite{1989PhFlB...1.1929M}, and \cite{2006PhPl...13d2306O}. The efficiency factor works to make $Q_{\rm Z,07} < Q_{\rm phen}$, bringing the ratio $Q/Q_{\rm Z,07}$ up relative to $Q/Q_{\rm phen}$. At low heights, where the efficiency factor is low, the inclusion of the efficiency factor defined in Equation \eqref{eq:qfull}b does help to bring $Q$ and $Q_{\rm Z,07}$ into better agreement. At large heights, the efficiency factor is closer to 1, so $Q_{\rm phen}$ and $Q_{\rm Z,07}$ {\it both} underestimate the heating rate computed by BRAID.

An alternative efficiency factor has emerged from studies of closed coronal loops driven by slow transverse footpoint motions. In such models, magnetic energy is built up from the twisting and shearing motions of the field lines \citep{1972ApJ...174..499P}, and the energy dissipation appears to follow a cascade-like sequence of quasi-steady relaxation events. The time-averaged heating rate in these models can be parameterized as \citep{2009ApJ...706..824C}: \begin{equation}
  Q_{\rm loop} = \epsilon_{\rm loop} Q_{\rm phen}
  \,\,\, , \,\,\,\,\,\,\,\,\,
  \epsilon_{\rm loop} \, = \,
  \left( \frac{L_{\perp} V_{\rm A}}{v_{\perp} L_{\parallel}}
  \right)^{\alpha},
\label{eq:looprate}
\end{equation} 
where $L_{\parallel}$ is often defined as the loop length for closed magnetic structures, and $\alpha$ is an exponent that describes the sub-diffusive nature of the cascade in a line-tied loop. The quantity in parentheses is a ratio of timescales; this ratio is the nonlinear time over the wave travel time. To set the value of $L_{\parallel}$ in our open-field models, we follow prior work \citep{2004ApJ...615..512S}, which found that open and closed regions can be modeled using a unified empirical heating parametrization when the actual loop length $L$ is replaced by an effective length scale 
\begin{equation}  
L_{\parallel} \, \approx \, \frac{L}{1 + (L/L_{0})}, 
\end{equation} 
where $L_{0} = 50$~Mm. Thus, for open-field regions in which $L \gg L_0$, we use $L_{\parallel} = L_0$.

The exponent $\alpha$ describes how interactions between counter-propagating Alfv\'{e}n wave packets can become modified by MHD processes such as scale-dependent dynamic alignment \citep{2009ApJ...699L..39B}. The value $\alpha = 0$ corresponds to a classical hydrodynamic cascade. A previous model of MHD turbulence had $\alpha = 1.5$ \citep{2000SoPh..195..299G}, and a random-walk type model had $\alpha = 2$ \citep{1986ApJ...311.1001V, 2008ApJ...682..644V}. However, found from numerical simulations, $\alpha$ can occupy any value between 1.5 and 2, depending on the properties of the background corona and wave driving \citep{2008ApJ...677.1348R}. An analytic prescription for specifying $\alpha$ in a way that agrees with the \citep{2008ApJ...677.1348R} results was constructed \citep{2009ApJ...706..824C}. Subsequently, when modeling the coronal X-ray emission from low-mass stars, the longest loops---which seem to be most appropriate to compare with open-field regions---tend to approach the high end of the allowed range of exponents (i.e., $\alpha \approx 2$) \citep{2009ApJ...706..824C, 2013ApJ...772..149C}. Thus, we use $\alpha = 2$ here and note that the differences between $Q_{\rm loop}$ and $Q_{\rm phen}$ will be reduced for smaller values of the exponent. At the largest heights, $Q_{\rm loop}$ better matches the BRAID numerical results than either $Q_{\rm phen}$ or $Q_{\rm Z,07}$.

For additional comparison, in Figure \ref{fig:Qother} we plot the ratio of BRAID numerical results with the phenomenological heating rate for all three of our flux tube models. The behavior of the heating rate expressions with efficiency factors is similar, so we show only the ratio $Q/Q_{\rm phen}$ in Figure \ref{fig:Qother}. Note that the EQS model behaves similarly to the PCH model, but the NAR model exhibits a marked decrease in the ratio $Q/Q_{\rm phen}$ in the low corona before increasing again to approach the EQS model at the top of the grid. This behavior is reminiscent of closed-field models in which $Q/Q_{\rm phen}$ came back down to values of 0.2--0.3 in the coronal part of the modeled loops \citep{2011ApJ...736....3V}.

\afterpage{%
\begin{figure}
\centerline{\includegraphics[width=\columnwidth]{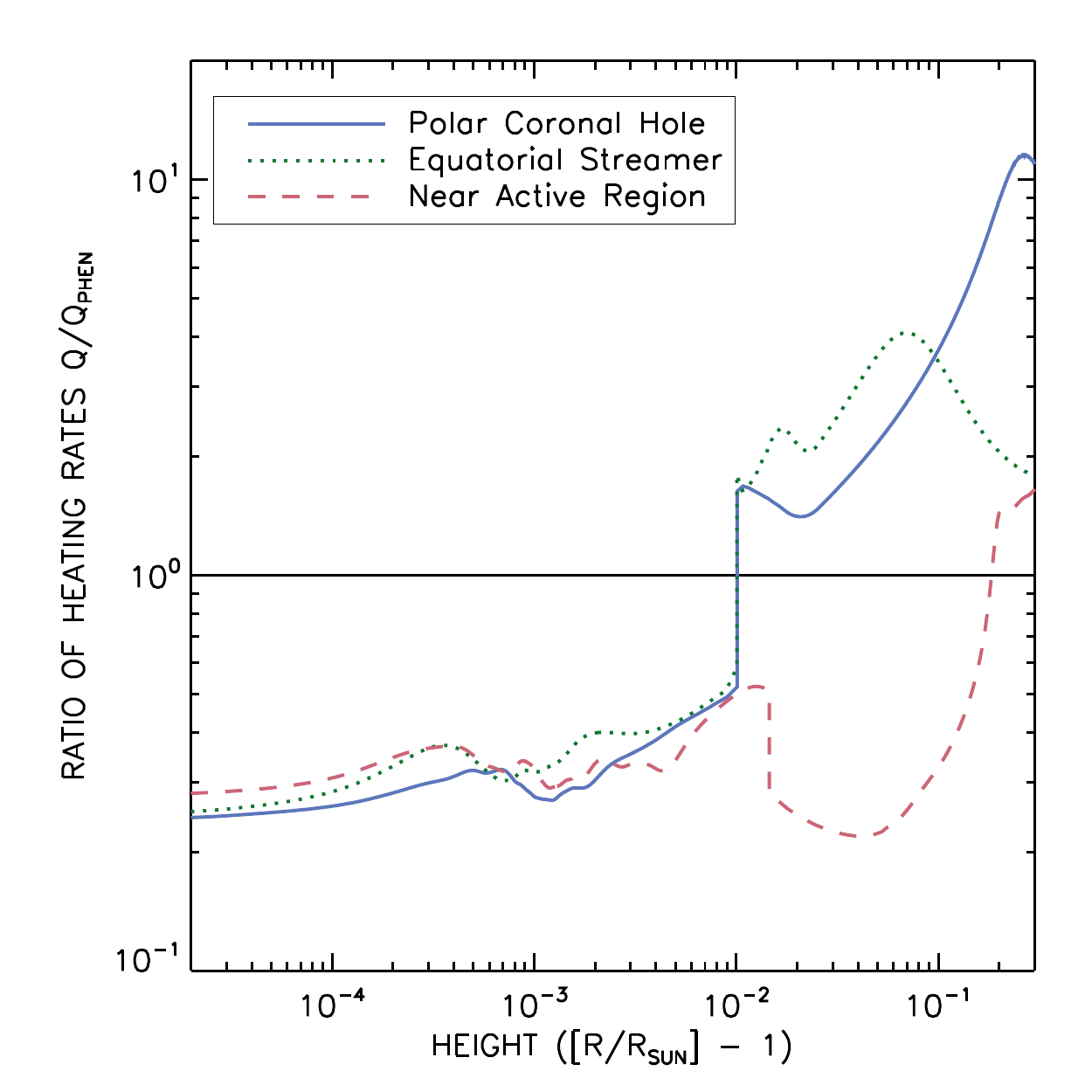}}
\caption[Ratio of BRAID results with phenomenological heating rates]{For each model, we show the ratio of the time-averaged heating rate from BRAID with the phenomenological expression given by Equations \eqref{eq:phen}.}
\label{fig:Qother}
\end{figure}
\clearpage
}

\section{Extended Analysis of Time Variability}

\subsection{Statistical variations as a source of multithermal plasma}

The turbulent heating simulated by BRAID was found to be quite intermittent and variable on small scales \citep{2011ApJ...736....3V}. Figure \ref{fig:bursty} illustrates some of this variability by showing the fluctuation energy density and heating rate volume-averaged over the low corona (i.e., between the transition region at $z = 0.01 \, R_{\odot}$ and an upper height of $0.5 \, R_{\odot}$). This is a similar plot as Figure 4 of \citep{2011ApJ...736....3V}.  Even with this substantial degree of spatial averaging, the nanoflare-like burstiness generated by the turbulence is evident in Figure \ref{fig:bursty}. There is a large body of prior work concerning such intermittent aspects of turbulent heating \citep[see, e.g.,][]{1996ApJ...457L.113E,1996ApJ...467..887H,1997ApJ...484L..83D,1998ApJ...505..974D,1999PhPl....6.4146E}.

\afterpage{%
\begin{figure}
\centerline{\includegraphics[height=0.8\textheight]{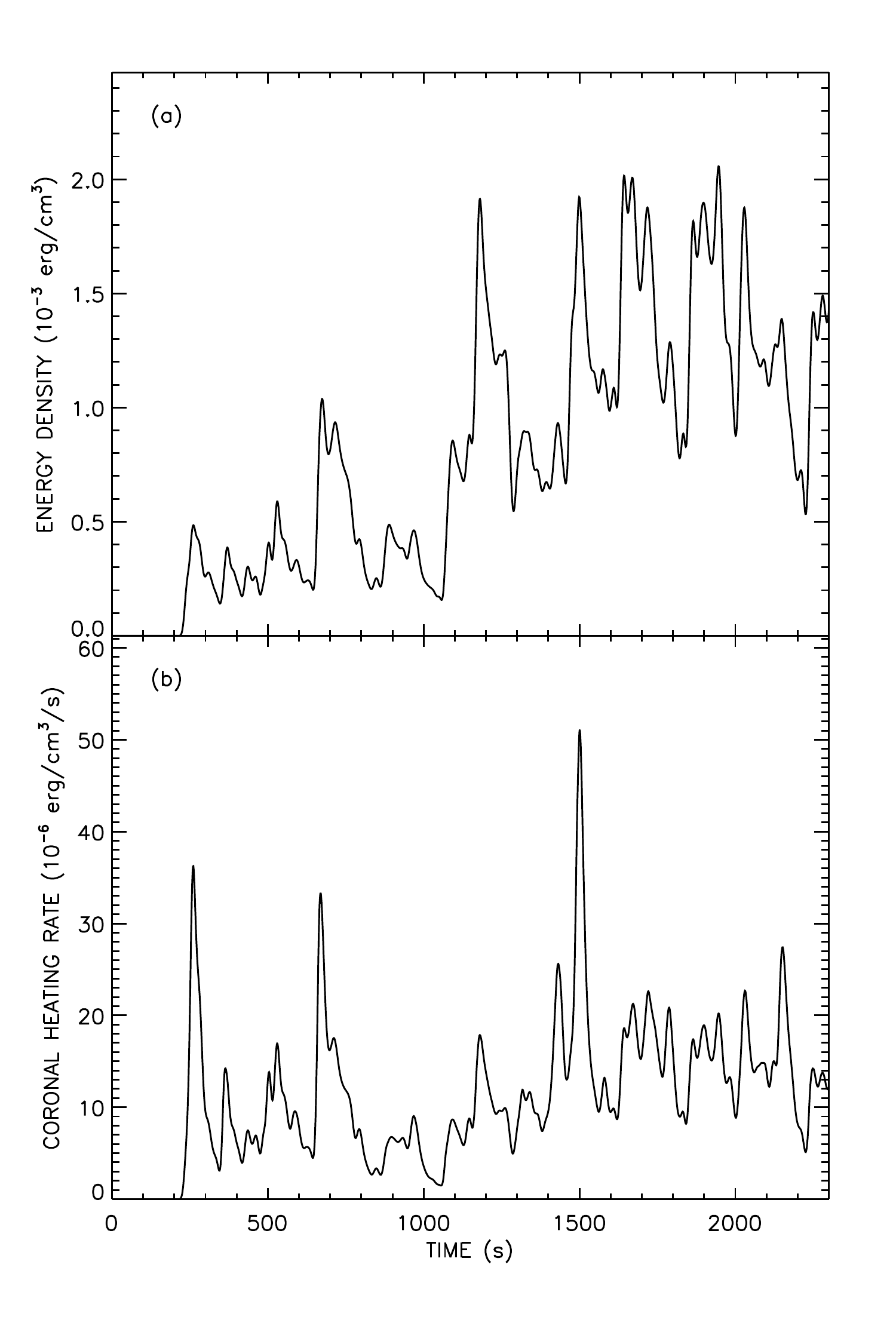}}
\caption[Volume-averaged coronal energy density and heating rate]{Due to the random jostling that generates the Alfv\'en waves, the (a) spatially-averaged energy density in the corona and (b) spatially-averaged heating rate per unit volume in the corona are bursty and strongly time-dependent in nature.}
\label{fig:bursty}
\end{figure}
\clearpage
}

The time-varying heating rate should also give rise to a similarly variable coronal temperature structure. We investigate the possibility that the resulting stochastic distribution of temperatures may be partially responsible for the observational signatures of {\em multithermal} plasmas---e.g., nonzero widths of the differential emission measure (DEM) distribution.  A study of the spatial and temporal response of a conduction-dominated corona to the simulated variations in $Q$ from BRAID found that conduction leads to a ``smeared out'' temperature structure that nevertheless retains much of the bursty variability seen in the heating rate \citep{2012ApJ...746...81A}. Here, we perform an even simpler estimate of the distribution of temperatures by taking the distribution of volume-averaged heating rates $\langle Q \rangle$ shown in Figure \ref{fig:bursty} and processing each value through a simple conductive scaling relation \citep{1978ApJ...220..643R}.  Thus, \begin{equation}
  \frac{\langle T \rangle}{\langle \bar{T} \rangle} \, = \, \left(
  \frac{\langle Q \rangle}{\langle \bar{Q} \rangle} \right)^{2/7}
\end{equation} where $\langle T \rangle$ is an estimated volume-averaged coronal temperature. The normalizing value of the heating rate $\langle \bar{Q} \rangle$ is assumed to be the mean value of $\langle Q \rangle$ seen in the BRAID simulation. For simplicity, we take the normalizing value of the temperature $\langle \bar{T} \rangle$ to be the maximum coronal temperature found in the corresponding ZEPHYR model \citep{2007ApJS..171..520C}. The PCH, EQS, and NAR models exhibited values of $\langle \bar{T} \rangle$ of 1.352, 1.224, and 1.675 MK, respectively.

Figure \ref{fig:dem}(a) shows the distribution of derived values of $\langle T \rangle$ for the PCH model. If the temperatures along this flux tube were measured by standard ultraviolet and X-ray diagnostics, with time integrations long in comparison to the scale of variability in the BRAID model, then this distribution would be equivalent to the DEM. For each simulated DEM, we measured its representative ``width'' in the same way as for observed loops \citep{2014ApJ...795..171S}; i.e., we used the points at which the DEM declined to 0.1 times its maximum value. For the PCH, EQS, and NAR models, we found widths of 0.2294, 0.2259, and 0.2839 in units of ``dex'' ($\log T$), respectively.

Figure \ref{fig:dem}(b) compares the properties of the three simulated DEMs with a selection of observationally derived coronal-loop DEMs \citep{2014ApJ...795..171S}. The BRAID models do appear to reproduce the observed multithermal nature of coronal plasmas, both in the absolute values of the widths (which fall comfortably within the range of the observed values) and in the overall trend for hotter models to have broader DEMs. Of course, the open-field models studied here only span a very limited range of central temperatures $\langle \bar{T} \rangle$ in comparison to the observed cases.

\afterpage{%
\begin{figure}
\centerline{\includegraphics[height=0.7\textheight]{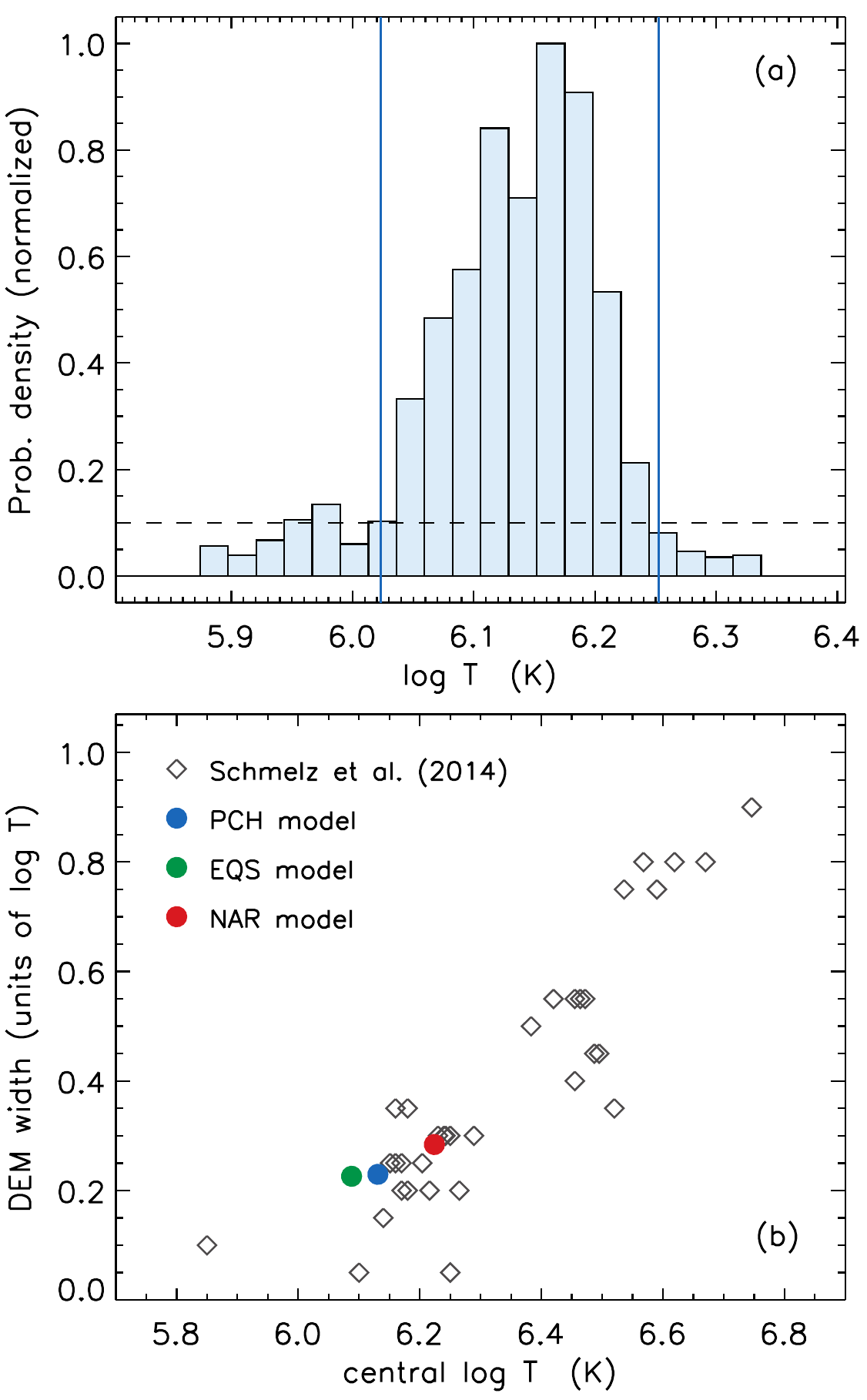}}
\caption[Simulated DEM and comparison with coronal loops]{(a) Probability distribution of coronal temperatures estimated from BRAID heating rates. Temperatures at which the distribution drops to 0.1 times the maximum value (dashed line) are indicated by solid red lines.
(b) Observed DEM widths from \citep{2014ApJ...795..171S}, plotted as the relative width in $\log T$ space and computed from the reported number of $\log T$ ``bins'' (gray diamonds), compared with simulated DEM widths from the PCH (red circle), EQS (green circle), and NAR (blue circle) models.}
\label{fig:dem}
\end{figure}
\clearpage
}

\subsection{Power spectrum of fluctuations}

For the PCH model, we investigated in detail the power spectrum of the velocity fluctuations caused by the Alfv\'en waves. To generate the Fourier transform, we assumed constant time spacing using the model results spanning from $t = \tau_{\rm A}$ to $t = 2 \tau_{\rm A}$. We subtracted the mean, doubled the length to make a periodic sequence, and then fed the cleaned quantity into a traditional FFT procedure. The power spectrum is the product of the result of that FFT procedure with its complex conjugate. This procedure gives us these spectra at each height $z$.

In Figure \ref{fig:contour}, we examine a contour plot of the power in these fluctuations. Certain frequencies in the $10^{-2}$ Hz to $10^{-1.5}$ Hz range have a relatively high amount of power at all heights, while at the higher frequencies, there is an increase in power as a function of height. This is more easily seen in Figure \ref{fig:spectra}. We show the basal input spectrum with a dashed line, and the power spectrum at the upper boundary of our model lies above the power spectrum at the photospheric base for the highest frequencies. Both boundaries show that there is a boost above the input spectrum for high frequencies.

\afterpage{%
\begin{figure}
\centerline{\includegraphics[width=\textwidth]{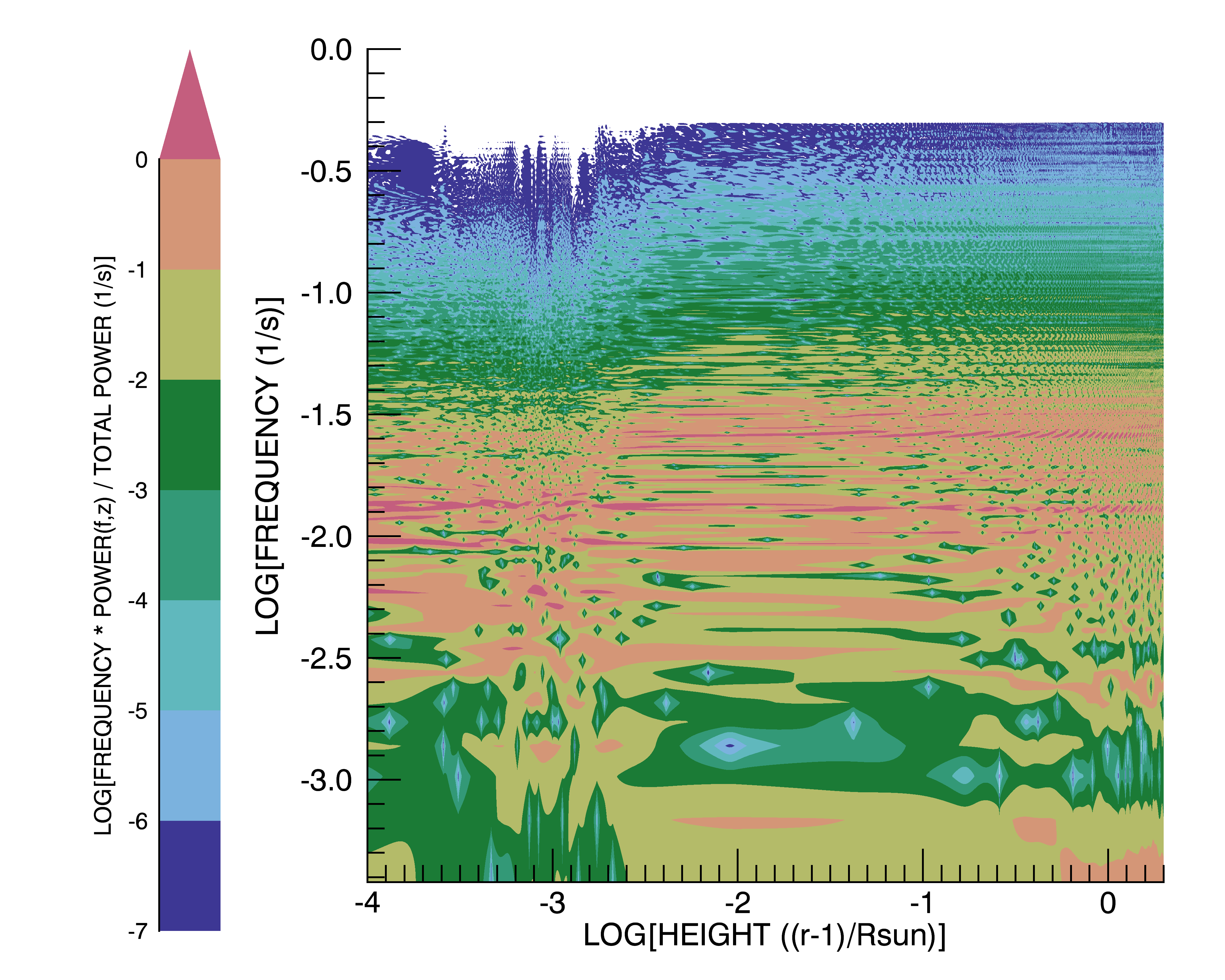}}
\caption[Contour plot of velocity fluctuation power spectrum]{Contour plot of the power spectrum in the PCH model showing power in color, as a function of height and of frequency.}
\label{fig:contour}
\end{figure}
\clearpage
}

Figure \ref{fig:spectra} shows that the high-frequency part of the BRAID turbulent power spectrum appears to be a power law ${\mathcal P} \propto f^{-n}$, where the value of $n$ varies a bit with height in the model. Looking only at the FFT data with $f \geq 0.03$~Hz, we found that $n \approx 4.5$ at the photospheric base, and then it steepens at larger heights to take on values of order 5.2--6.4 (i.e., a mean value of 5.8 with a standard deviation of $\pm$0.6) at chromospheric heights below the TR. In the corona, however, $n$ decreases a bit to a mean value of 4.9 and a smaller standard deviation of $\pm$0.3. Some of the quoted standard deviations are likely due to fitting uncertainties of the inherently noisy power spectra, but it is clear that the corona exhibits less variation in $n$ than the regions below the TR.

\afterpage{%
\begin{figure}
\centerline{\includegraphics[width=\columnwidth]{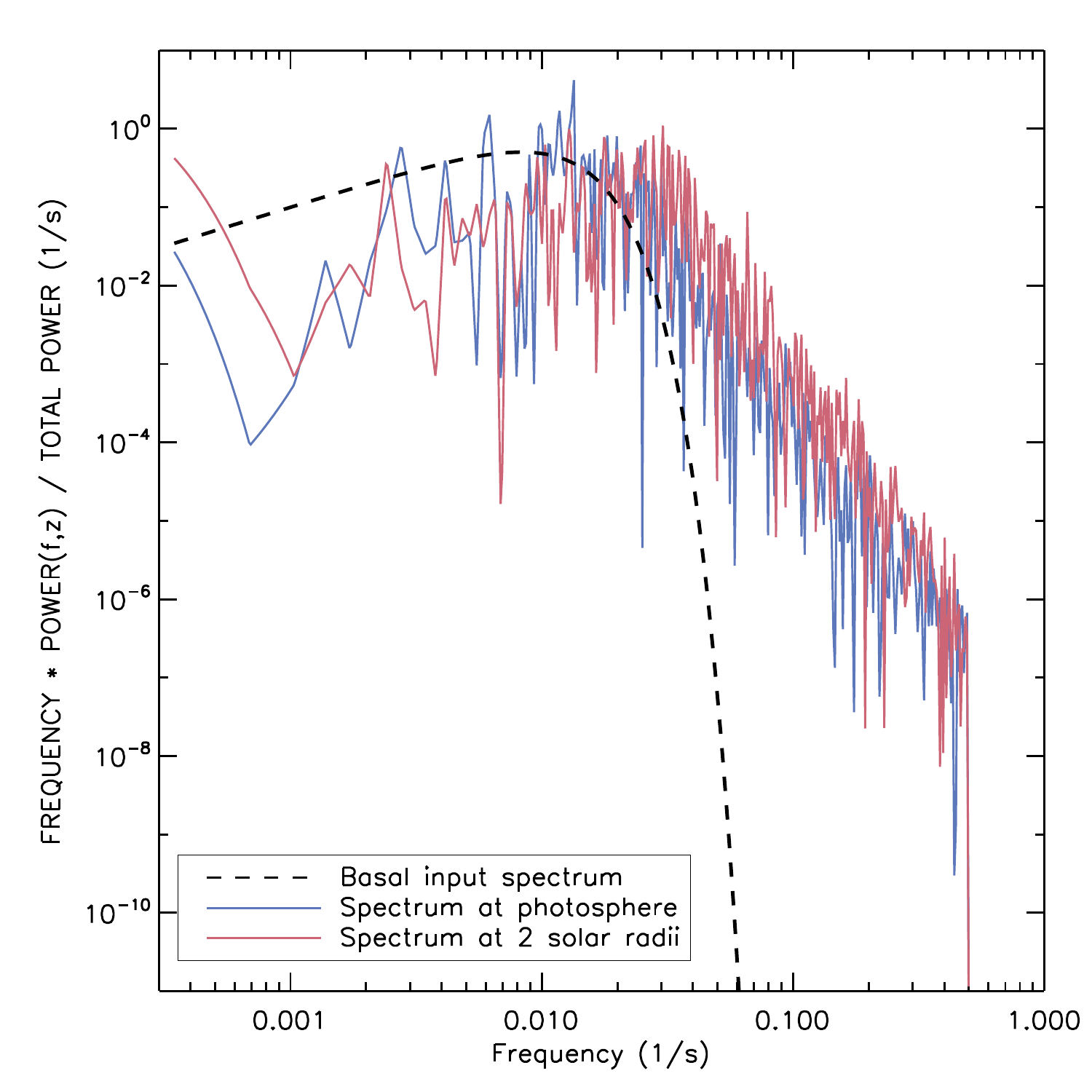}}
\caption[Power spectra of velocity fluctuations]{Power spectra for the upper and lower boundaries of the PCH model compared to the input spectrum show a boost in power at high frequencies.}
\label{fig:spectra}
\end{figure}
\clearpage
}

Despite several decades of spacecraft observations of power-law frequency spectra in the turbulent solar wind, it was not immediately obvious that the BRAID power spectrum should have exhibited such power-law behavior. Spacecraft-frame measurements are often interpreted as being a spatial sample through quasi-stationary wavenumber variations \citep[see, e.g.,][]{1938RSPSA.164..476T,2012SSRv..172..325H}. In strong-field MHD turbulence, the dominant wavenumber cascade is expected to be in the $k_{\perp}$ direction. However, the Alfv\'{e}nic fluctuations that make up an MHD cascade have a dispersion relation in which the frequency depends primarily on $k_{\parallel}$. Thus, MHD (especially RMHD) turbulence is typically described as ``low-frequency turbulence'' and the idea of an inherent power-law {\em frequency cascade} is met---usually, rightly so---with skepticism.

There has been one proposed model in which a power-law spectrum in frequency (i.e., in $k_{\parallel}$) occurs naturally and without the need for substantial parallel cascade: the so-called ``critical balance'' model \citep{1995ApJ...438..763G}. In this picture, strong mixing is proposed to occur between the turbulent eddies (primarily moving perpendicular to the background field) and Alfv\'{e}n wave packets (moving parallel to the field), such that the parameter space ``filled'' by a fully developed cascade
is determined by the critical balance parameter \begin{equation}\chi \, = \, \frac{k_{\parallel} V_{\rm A}}{k_{\perp} v_{\perp}}\end{equation} taking on values $\chi \lesssim 1$ \citep[see also][]{1984ApJ...285..109H}. The limit $\chi \ll 1$ corresponds generally to low-frequency fluctuations with $k_{\parallel} \ll k_{\perp}$ as is expected in anisotropic MHD turbulence. This parameter can also be interpreted as a ratio of timescales. 

Following the implications of critical balance led to a phenomenological expression of the time-steady inertial range \citep{1995ApJ...438..763G}, given here as \begin{equation} E (k_{\parallel},k_{\perp}) \, = \, \frac{V_{\rm A} \, v_{\perp}(k_{\perp})}{k_{\perp}^3} \, g (\chi) \label{eq:E3D}\end{equation} where this combines Equations (5) and (7) of \citep{1995ApJ...438..763G}. Above, $E$ is a three-dimensional power spectrum that gives the magnetic energy density variance (in velocity-squared units) when integrated over the full volume of wavenumber space, \begin{equation} \frac{\langle B_{\perp}^{2} \rangle}{8\pi\rho} \, = \, \int d^{3} {\bf k} \,\, E (k_{\parallel},k_{\perp}) \, = \, \int df \,\, {\mathcal P}(f) \,\, ,\label{eq:eq11}\end{equation} and we also define the frequency spectrum ${\mathcal P}(f)$ in a similar way. Note that $2\pi f = \omega = k_{\parallel} V_{\rm A}$ in the local rest frame of the plasma under the assumption that the fluctuations are Alfv\'en waves.

In Equation (\ref{eq:E3D}) above, $v_{\perp} (k_{\perp})$ is the reduced velocity spectrum, which specifies the magnitude of the velocity
perturbation at length scales $k_{\perp}^{-1}$ and $k_{\parallel}^{-1}$. This spectrum is often assumed to be a power-law with $v_{\perp} \propto k_{\perp}^{-m}$. The exponent $m$ has been proposed to range between values of 1/3 \citep[strong turbulence;][]{1995ApJ...438..763G} and 1/2 \citep[weak turbulence;][]{2000JPlPh..63..447G}. Lastly, the function $g(\chi)$ in Equation (\ref{eq:E3D}) is a ``parallel decay'' function that is expected to become negligibly small for $\chi \gg 1$. Because $g(\chi)$ is normalized to unity when integrated over all $\chi$, a simple approximation for it is a step function,
\begin{equation}
  g(\chi) \, \approx \, \left\{
  \begin{array}{ll}
  1 \,\, , & \chi \leq 1 \,\, , \\
  0 \,\, , & \chi > 1
  \end{array}
  \right. 
\end{equation} \citep[see also][]{2002ApJ...564..291C,2012ApJ...754...92C}. The above form for $g(\chi)$, combined with the assumption that the reduced spectrum $v_{\perp}(k_{\perp})$ extends out to $k_{\perp} \rightarrow \infty$, leads to the high-frequency end of the frequency spectrum obeying a power law, with \begin{equation} {\mathcal P}(f) \, \propto \, f^{(m+1)/(m-1)} \,\, .\end{equation} The strong turbulence case ($m=1/3$, ${\mathcal P} \propto f^{-2}$) has been studied extensively both observationally and theoretically \citep[see, e.g.,][]{2012SSRv..172..325H}.

Of course, the BRAID models highlighted in this chapter do {\em not} have reduced perpendicular velocity spectra that extend over large ranges of $k_{\perp}$ space. The models presented here (similar to those of \citep{2011ApJ...736....3V}) resolve only about one order of magnitude worth of an ``inertial range'' in $k_{\perp}$ space. For the step-function version of $g(\chi)$ given above, the imposition of a $v_{\perp}(k_{\perp})$ cutoff above an arbitrary $k_{\rm max}$ produces a frequency spectrum ${\mathcal P}(f)$ that is similarly cut off above a frequency $f_{\rm max}$ determined by critical balance ($\chi = 1$) at $k_{\perp} = k_{\rm max}$.

In an alternative to the step-function version of $g(\chi)$, \citep{2003ApJ...594..573C} found an analytic solution for $g(\chi)$ by applying an anisotropic cascade model that obeyed a specific kind of advection--diffusion equation in three-dimensional wavenumber space. The general form of $g(\chi)$ is reminiscent of a suprathermal kappa function \citep[see, e.g.,][]{2010SoPh..267..153P}, which is roughly Gaussian at low $\chi$ and a power-law at large $\chi$. For $\chi \gg 1$, $g(\chi) \propto \chi^{-(3s+4)/2}$, where $s$ is the ratio of the model's perpendicular advection coefficient to the perpendicular diffusion coefficient \citep{2003ApJ...594..573C}. It is still not known if MHD turbulence in the solar corona and solar wind exhibits a universal value of $s$, or even whether or not $s$ is even a physically meaningful parameter. Nevertheless, the wavenumber diffusion framework \citep{1990JGR....9514881Z, 2009PhRvE..79c5401M} has been shown to be consistent with a value of $s=2$ in this family of advection--diffusion equations. In a different model of coronal turbulence, a cascade of slow random-walk displacements of the field lines can be treated as the case $s=1$ \citep{1986ApJ...311.1001V}. On the basis of observations alone, if $s$ could be maintained at small values of order 0.1--0.3, there would be sufficient high-frequency wave energy to heat protons and minor ions via ion cyclotron resonance \citep{2003ApJ...594..573C, 2009ApJ...691..794L}.

No matter the value of $s$ or the reduced spectral index $m$, it can be shown that at large frequencies ($f > f_{\rm max}$), a power law of the form $g(\chi) \propto \chi^{-n}$ produces a power-law frequency spectrum ${\mathcal P} \propto f^{-n}$ with the same exponent (see the integration over $k_{\perp}$ in Equation \eqref{eq:eq11}). Thus, we postulate that measuring $n$ from the BRAID simulations {\em may} be a way to extract information about the exponent $s$, with \begin{equation} s \, = \, \frac{2n - 4}{3} \,\, .\end{equation} The values of $n$ reported above imply a photospheric value of $s \approx 1.7$, which increases to $s \approx 2.5$ in the chromosphere (with a relatively large spread) and then decreases to $s \approx 1.9$ in the corona. The similarities to the theoretical value of $s=2$ \citep[from, e.g.,][]{1990JGR....9514881Z, 2009PhRvE..79c5401M} are suggestive, but not conclusive.

\subsection{Nanoflare statistics of heating rate variability}
We investigated the variability of heating and energy as a function of height and time throughout the simulation. In a given finite ``zone,'' the energy lost via dissipative heating can be calculated using the heating rate as
\begin{equation}E(z,t) = Q(z,t)\Delta z (\pi R^{2})\Delta t,\end{equation}
where the zones are defined at a set of heights, $z$, with unequal spacing $\Delta z$, and at a set of times, $t$, with equal spacing $\Delta t = 0.25$ s. In Figure \ref{fig:cenergy}, we provide a contour plot of the energy, comparable with Figure 6b of \citep{2013ApJ...773..111A}. 

\afterpage{%
\begin{figure}
\centerline{\includegraphics[width=\columnwidth]{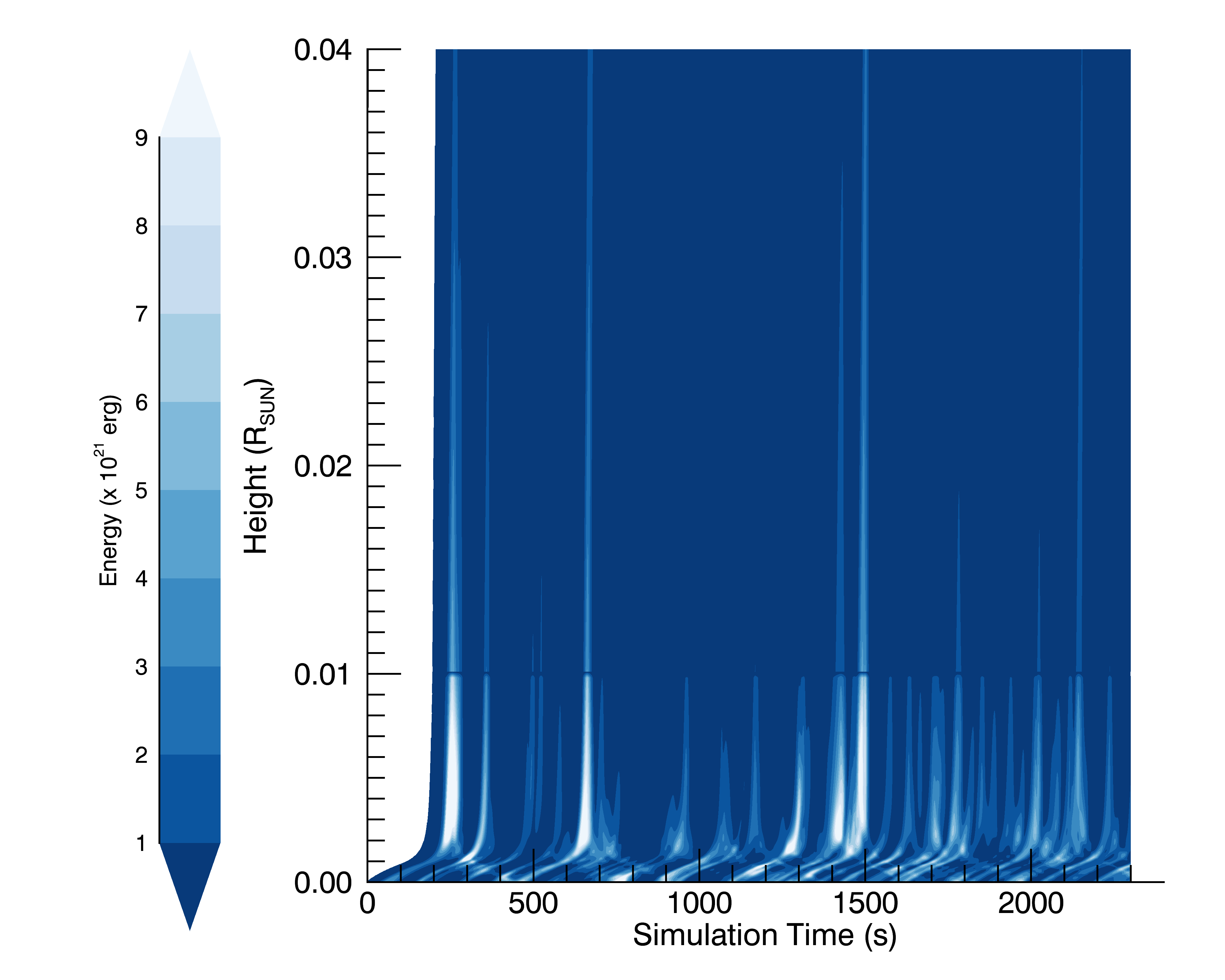}}
\caption[Contour plot of energy in corona]{Dependence of energy on height above photosphere (y-axis) and time (x-axis). The transition region is at 0.01 R$_{\odot}$.}
\label{fig:cenergy}
\end{figure}
\clearpage
}

Many of the impulsive heating events that result in spikes of energy over a short time frame stop at the transition region, which lies at $0.01 R_{\odot}$, but some extend to ten times that height. There is also a lack of energy at the beginning of the simulation ($t < \tau_{\rm A}$), where the information has not yet had time to propagate up through the model grid.

Following the method of Asgari-Targhi et al., (2013; \citep{2013ApJ...773..111A}), we use box capturing to get a statistical sense of the distribution of energy in these zones throughout the corona. In order to be directly comparable to their defined events, we also use boxes with a width of 19.4 seconds in simulation time and height of 19.4 seconds in Alfv\'en travel time (a proxy for height). This choice in box size results initially in 118 sections across the time dimension and 39 sections along the height dimension. However, the lowest 10 boxes are at heights below the transition region, and we plot only boxes in the corona in Figure \ref{fig:denergy}. Additionally, we take out the first 770 seconds corresponding to one Alfv\'en travel time in the PCH model (recall Table 1 and Equation \eqref{eq:traveltime}) to ensure that the waves have had time to propagate fully throughout the corona.

\afterpage{%
\begin{figure}
\centerline{\includegraphics[width=\textwidth]{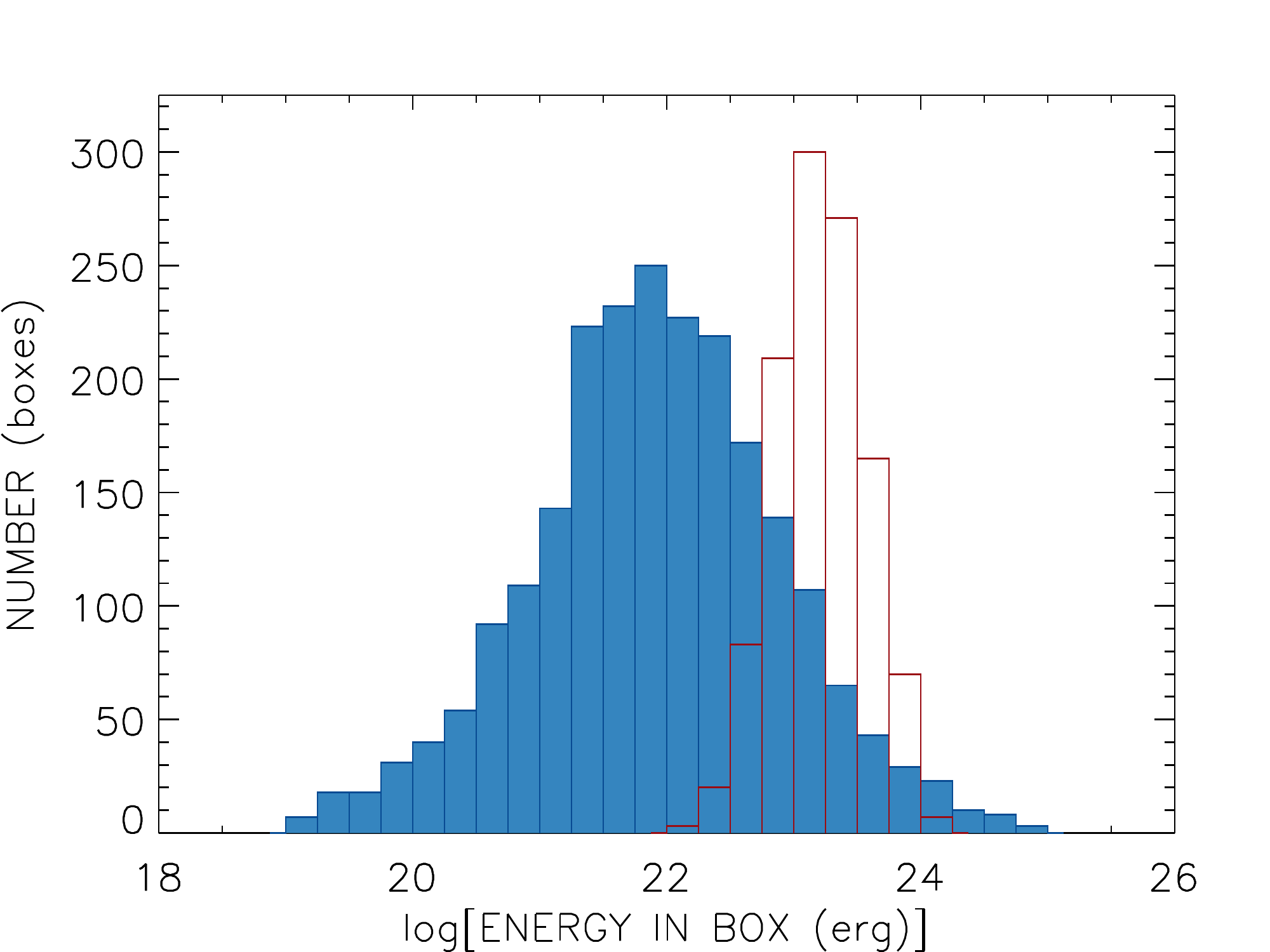}}
\caption[Histogram of energy dissipation events in coronal hole]{Filled histogram shows the statistical distribution of energy contained in boxes of space and time for the PCH model. Only boxes above the transition region are used. For comparison, red outlined histogram shows the results from closed magnetic loops \citep{2013ApJ...773..111A}.}
\label{fig:denergy}
\end{figure}
\clearpage
}

These cuts result in 78 time sections and 29 height sections, giving us a total of 2262 coronal boxes. The arithmetic mean in log-space of the energy contained within these boxes is 21.91 with a standard deviation of 0.97. The average energy contained in the boxes is lower than that found for coronal loops \citep{2013ApJ...773..111A}, since our model does not have two strong footpoints supplying separate sources of counter-propagating Alfv\'en waves, but it is still a significant amount of energy. Figure \ref{fig:denergy} also shows that we find several events that reach higher energies than the previous modeling, and that higher tail gets well into the classical nanoflare expected energies \citep{2011SSRv..159..263H}. The minimum-energy event contains 10$^{19.04}$ erg while the maximum-energy event contains 10$^{24.97}$ erg. It is worthy of note that the peak energies captured in these boxes fall well within the ``picoflare'' range \citep{1999SSRv...87...25A,2000ApJ...529..554P}. All of these events are a natural product of the total heating coming out of BRAID, suggesting that the physics contained within these models may lead to the formation of pico- and nanoflares.

\section{Discussion}

We have analyzed three typical open magnetic field structures using one-dimensional time-steady modeling and three-dimensional time-dependent RMHD modeling. These structures represent characteristic flux tubes anchored within a polar coronal hole, on the edge of an equatorial streamer, and neighboring a strong closed-field active region. We show that the time-averaged properties of the higher-dimensional BRAID models agree well with that of the less-computationally-expensive ZEPHYR models. 

We looked in detail at the energy partitioning, as BRAID imposes no assumptions or restrictions. At heights above the transition region, equipartition is shown to work well to describe the results from BRAID, and is assumed in the ZEPHYR algorithm. We also compared the energy partitioning with predictions from non-WKB reflection for a range of frequencies, showing the limitations of the linear method for such predictions. The time-averaged heating rates from BRAID are lower than the phenomenological expression in Equation \eqref{eq:qfull} for the heating rate below the transition region, rising sharply toward the upper boundary.

In BRAID, Alfv\'en waves are generated by random footpoint motions, whose properties and driving modes are described in Section 4.2.2. As the Alfv\'en waves propagate upward from the photosphere to the open upper boundary at a height of 2 solar radii, they partially reflect and cause turbulence to develop. With the time-dependence included in BRAID, we are able to show the bursty nature of turbulent heating by Alfv\'en waves. We show that this heating brings energy up into the corona and provide a statistical distribution of energy per event. The more energetic events (i.e., boxes with the most energy) fall within expected nanoflare values.

Overall, we show that time-steady modeling does a good job of predicting the time-averaged results from time-dependent modeling. There is, however, a bounty of information that can be found only by looking at changes in the heating rate over time. Moving from one dimension to three allows the model to contain more realistic physics. We have shown that these models of typical magnetic field structures provide additional compelling evidence to support the idea that Alfv\'en-wave-driven turbulence heats the corona and accelerates the solar wind.
%% forecasting_phd5.tex
\chapter{Magnetic Thresholds for Network Jets}
%%%%%%%%%%%%%%%%%%%%%%%%%%%%%%%%%%%%%%%%%%%%%%%%

This document has thus far focused on better understanding the particular physical mechanisms of coronal heating and solar wind acceleration by {\it modeling} those processes. The full cycle of science is incomplete if I do not also discuss the observational signatures predicted by modeling and compare various observations with model results. Chapter 1 introduced some of the overall correlations and relationships that exist between observables, and this chapter will focus on understanding those relationships by connecting solar observations with the modeling I have done during my graduate work. Specifically, I connect the bursty heating results from our time-dependent modeling with small-scale phenomena in the chromosphere, focusing on some of the most recent datasets available.

As I discussed in detail in Chapter 3, we used the reduced-MHD code called BRAID to model three-dimensional open flux tubes \citep{2015ApJ...811..136W}. The models contained discrete events from Alv\'en-wave-driven turbulent heating that are reminiscent of nanoflare-like heating. We also consider the ability of these discrete heating events to match solar spicule dynamics and the small-scale jets seen with {\it Interface Region Imaging Spectrograph} (IRIS, \citep{2014SoPh..289.2733D}) \citep{2015ApJ...812...71C}. To extend that line of thinking, we have turned to observations of these structures and the magnetic field properties in the regions. Before getting into a description of the resulting investigation, I will first take some time to discuss a short overview of the path to our current understanding of these observations.

\section{Generating Spicules and Network Jets} 

Before spacecraft were available to investigate the many layers of the Sun's atmosphere, ``small spikes'' were seen off the limb of the Sun during ground-based solar eclipse observations at least as far back as 1877 \citep{secchi1877}. These features in the lower chromosphere, later named spicules \citep{1945ApJ...101..136R}, are jetlike structures that are more pronounced near the poles of the Sun. The role that spicules play in heating is still a source of contention, though all early models proposed to describe the formation of these structures noted that the magnetic field must play an important role: ``Without this field, as enhanced by the supergranule and granule motions, there would probably be no spicules'' \citep{1972ARA&A..10...73B}. Whatever the cause, spicules represent chromospheric material that is lifted to coronal heights energetically. A more recent review by Sterling (2000, \citep{2000SoPh..196...79S}) categorized numerical models of spicule formation into four categories: 1) strong pulse in low chromosphere, 2) rebound shock, 3) pressure pulse in high chromosphere, and 4) Alfv\'en-wave models. Examples from all categories could match some aspects of spicule properties and dynamics, but none could match spicules entirely. Discrepancies were attributed to both simplified modeling and to poorly-constrained physical properties of spicules at that time.

With newer spacecraft and more sophisticated simulations, in the last decade it has become known that there are two distinct types of spicules. These types, named simply ``Type I'' and ``Type II'' \citep{2007PASJ...59S.655D}, have such distinct dynamic properties that it seems likely they are driven by different mechanisms\footnote{It should be mentioned that I use the definitions of the spicule types from the discovery paper by De Pontieu et al. (2007, \citep{2007PASJ...59S.655D}). Previous distinctions using the same terminology by, e.g., Beckers (1968, \citep{1968SoPh....3..367B}) do not reflect the understanding we have gained with more modern observations.}. Type I spicules have dynamics fully consistent with on-disk features like active-region fibrils and quiet-sun mottles \citep[and references therein]{2007PASJ...59S.655D}. Type II spicules, on the other hand, dominate coronal holes and have higher speeds and shorter lifetimes than Type I spicules. These properties made them unobservable before the spatial and temporal resolution of {\it Hinode} and more recent missions.

Once this new type of structure was discovered at the limb of the Sun, they were quickly connected to their own on-disk counterparts, rapid blueshifted events \citep{2008ApJ...679L.167L, 2009ApJ...705..272R}. However, there was early disagreement about the existence of a distinct new category of spicule. Zhang et al. (2012, \citep{2012ApJ...750...16Z}) argued that the classification should be used with caution, and that depending on the exact definition used, there seemed to be either only Type I or up to three or four different ``types'' of spicules in their analysis of observations. 

In response to the claims, Pereira et al. (2012, \citep{2012ApJ...759...18P}) added additional observations and quantified the types more clearly. While they found regional differences in Type II spicules between coronal holes (CH) and the quiet Sun (QS), those differences were small relative to comparisons with Type I spicules near active regions. It has now become widely accepted that the two types exist, though the role of Type II spicules in solving the coronal heating problem remains subject of some debate. Early observations of type II spicules show transverse motion, which suggests the presence of Alfv\'en waves with enough energy to accelerate the solar wind \citep{2007Sci...318.1574D}. However, alternate methods of analysis suggest that chromospheric nanoflares, and the resulting observational signatures as Type II spicules, may not be enough to be a primary heating source \citep[and references therein]{2015RSPTA.37340256K}. Part of the debate is whether Type II spicules are brightenings in loops or in open flux tubes \citep[see, e.g.,][]{2014A&A...567A..70P,2015ApJ...812...71C}.

Until the latest generation of spacecraft, spicules were observed with properties at spatial and temporal resolution limits. When IRIS was launched on 27 June 2013, it provided unprecedented views of spicules \citep{2014ApJ...792L..15P}. Shortly after first light, small jets that seemed to trace the supergranular network were discovered by Tian et al. (2014, \cite{2014Sci...346A.315T}). These chromosphere and transition region phenomena exhibit speeds between 80 and 250 km s$^{-1}$, lifetimes of 20 to 90 seconds, lengths of up to 10,000 km, and widths as narrow as 300 km. Due to their small size and transient nature, these jets had not been observed (nor were they observable) before the high spatial and temporal resolution of IRIS was used with wavelengths that probed below the corona into the transition region plasma. Evidence has started to accumulate to suggest that these network jets are the on-disk equivalent to Type II spicules \citep{2015ApJ...799L...3R}. The network jets are likely the visible structures that had been previously identified as rapid blue-shifted and red-shifted excursions identified in spectra \citep[see, e.g.,][]{2008ApJ...679L.167L,2013ApJ...764..164S}.

Explanations for the formation of spicules and network jets exhibit the same basic grouping into either reconnection-driven heating models or Alfv\'en wave models as discussed in Chapter 1 for solar wind acceleration. The argument for reconnection traces back to Parker's nanoflare model \citep{1988ApJ...330..474P} and the idea that the network jets should be considered as simply scaled-down versions of larger coronal jets, for which reconnection is indeed the likeliest explanation \citep[see, e.g.,][]{1995Natur.375...42Y,2009ApJ...691...61P}. On the other hand, there has been much work to model the role of waves in spicule formation, both with acoustic and Alfv\'en waves using linear and nonlinear mechanisms \citep[see, e.g.,][and references therein]{1982SoPh...75...35H, 1999A&A...347..696D, 1999ApJ...514..493K, 2009ApJ...703.1318K, 2010ApJ...710.1857M}. The persistent theme is the ability to lift the transition region, through the formation of an Alfv\'en wave resonant cavity or shocks. 

This is where the results of the BRAID modeling described in Chapter 3 come in. The time-dependent transverse Alfv\'enic fluctuations are, at times, at sufficient amplitudes for nonlinear interactions. The implications for this in the process of forming compressible fluctuations and shocks are described by Cranmer and Woolsey (2015, \citep{2015ApJ...812...71C}). At large amplitude, linearly polarized Alfv\'en waves can produce parallel fluctuations \citep{1971JGR....76.5155H}. The amplitude of these fluctuations, $\delta v_{\parallel}$, can be written as
\begin{equation}
\frac{\delta v_{\parallel}}{V_{A}} = N_{\beta}\left(\frac{\delta B_{\perp}}{B_{0}}\right)^{2},
\label{eq:convert}
\end{equation}
where $N_{\beta}$ encapsulates the dependence on the plasma beta parameter, for which we use the simplified definition $\beta = \left(c_{s}/V_{A}\right)^{2}$ \citep{2015ApJ...812...71C}, and an approximate function for $N_{\beta}$ based on the numerical results of Spangler (1989, \citep{1989PhFlB...1.1738S}) which falls between previous analytical derivations from MHD theory \citep{1971JGR....76.5155H,1996JGR...10113527V}:
\begin{equation}
N_{\beta} \approx \frac{0.25}{\sqrt{1+\beta^{2}}} + \frac{0.135 \beta^{2.4}}{0.305 + \beta^{4.6}}.
\end{equation}

The analysis of our BRAID results in the context of this mode conversion led to simulated levitation of the transition region with observable signatures directly in line with Type II spicules and the IRIS network jets. I therefore moved to my current investigation of archived observations of the network jets and their underlying magnetic field. This provides information that studies of spicules cannot, since we can measure magnetic field strength of the photosphere on the disk directly underneath the network jets.

\section{Observations}
IRIS has a slit-jaw imager (SJI) that can observe in four passbands: 1330 \AA\ (C II, transition region), 1400 \AA\ (Si IV, transition region), 2796 \AA\ (Mg II k, chromosphere), and 2830 \AA\ (Mg II wing, photosphere). Our work focuses on the 1330 \AA\ passband, where network jets appear most clearly \citep{2014Sci...346A.315T}. The imager has 0.166$''$ pixels, in contrast to both the Atmospheric Imaging Assembly (AIA, \citep{2012SoPh..275...17L}) and Helioseismic and Magnetic Imager (HMI, \citep{2012SoPh..275..207S}) on {\it Solar Dynamics Observatory} (SDO), which have a spatial resolution of 0.600$''$/pixel. For context, a network jet with a length of 10,000 km has an angular size of 13.8$''$, which translates to approximately 23 pixels in SDO/HMI and 83 pixels in IRIS/SJI.

\subsection{Data reduction}

After looking through the archival data using the Lockheed Martin Solar \& Astrophysics Laboratory web-based search at {\tt http://iris.lmsal.com/search/}, we focused on a large sit-and-stare observation with 9-second cadence from 2014 November 11 12:39--13:44 UT. The IRIS field of view is shown with an AIA 193\AA\ context image in Figure \ref{fig:iriscontext}. 

\afterpage{%
\begin{figure}[p!]
\includegraphics[width=\textwidth]{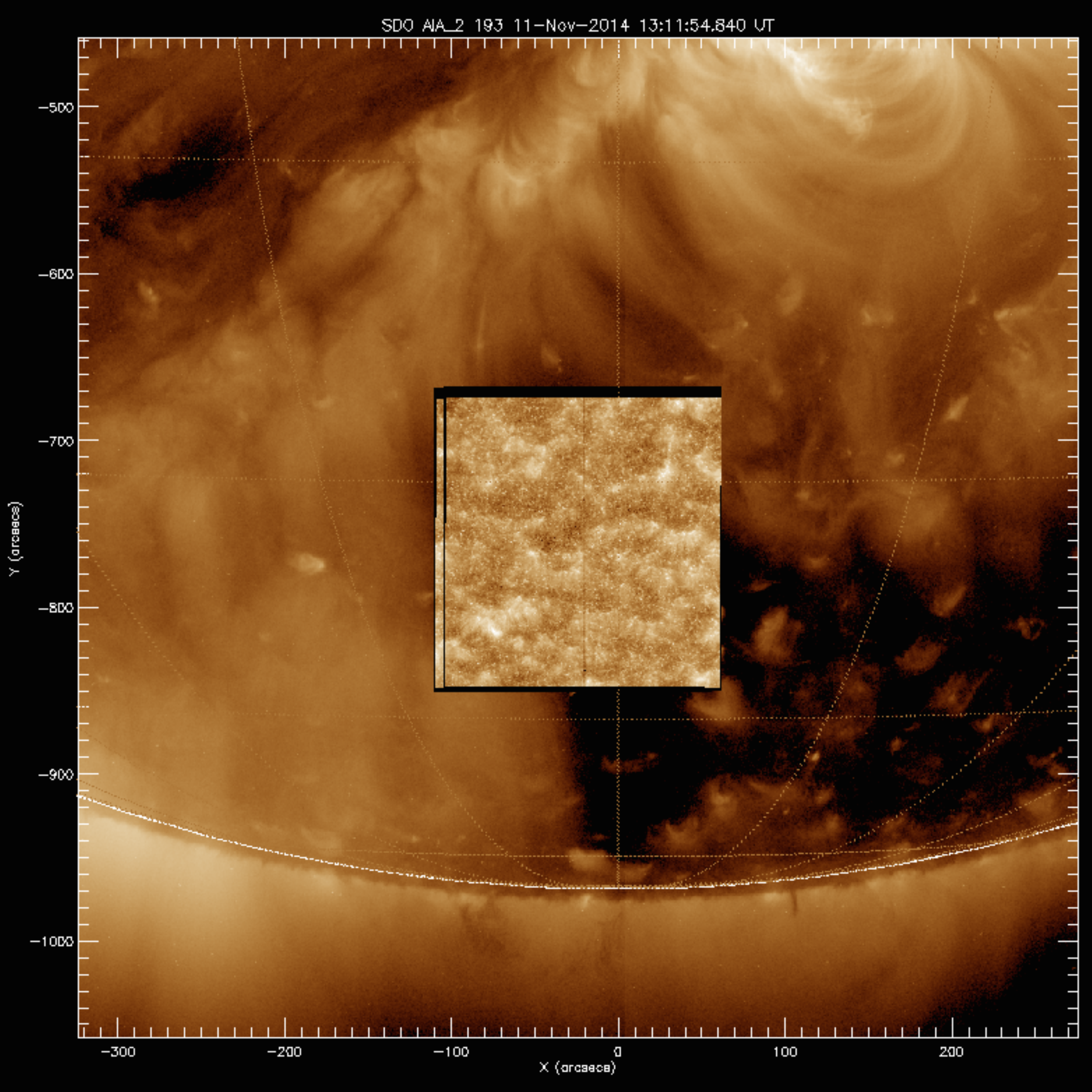}
\caption[Comparison of SDO and IRIS fields of view]{The IRIS field of view for our data set is shown in the larger context with AIA 193\AA\ data. The field is on the edge of a polar coronal hole, which appears dark in the AIA context image.}
\label{fig:iriscontext}
\end{figure}
\clearpage
}

The region of interest is on the edge of a polar coronal hole, in a ``sweet spot'' close enough to the limb to see the network jets in projection but far enough from the edge of the Sun to have reliable vector magnetogram data \citep{2015TESS....140204F}. We chose to do our initial study with HMI magnetic data, as SDO has had full-disk coverage during the entire lifetime of IRIS. See Appendix C for details of the search parameters and retrieval of SDO files and Level 2 IRIS data.

To align the HMI magnetograms and IRIS images, we used 1600\AA\ AIA images. I cut down the full-Sun images to a field of view in solar coordinates ranging from -150$''$ to 50$''$ in x and -900$''$ to -700$''$ in y. I used the sswidl procedure\footnote{Information on sswidl, a.k.a. SolarSoftWare for the IDL programming language, can be found at {\tt http://www.lmsal.com/solarsoft/}, formerly {\tt http://sohowww.nascom.nasa.gov/solarsoft/}.} aia\_prep to coalign the HMI magnetogram and 1600\AA\ data. Then, I used the pointing information for SDO and IRIS to estimate the alignment from AIA to IRIS 1330\AA. With manual inspection of the initial guessed alignment, I determined the additional relative shift required for the IRIS data, which was 18 pixels left and 15 pixels down. The shifted data is shown in Figure \ref{fig:irisshifts}, with contours of intensity to help guide the eye to the adjusted relative match of the data sets. The shifted image wraps around, so my analysis focuses on the lower central bright region of the IRIS image to avoid the edges of the shifted field of view.

\afterpage{%
\begin{figure}[p!]
\includegraphics[width=\textwidth]{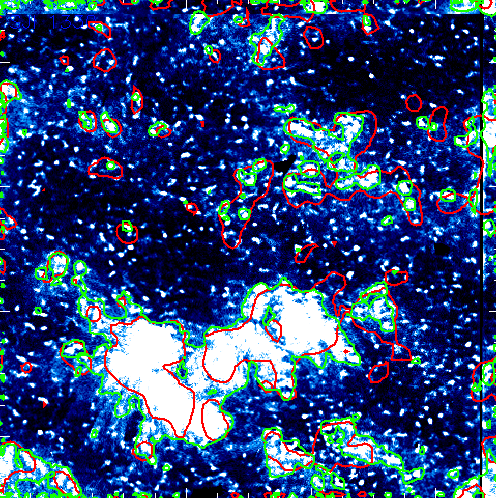}
\caption[Coalignment of SDO and IRIS]{Contours showing the shifted IRIS data, with green contours for IRIS and red contours for SDO. While there is not a direct match between the photospheric 1600\AA\ SDO image and the chromospheric 1330\AA\ IRIS image, there is strong correlation. Thanks to Chad Madsen for shifting algorithm.}
\label{fig:irisshifts}
\end{figure}
\clearpage
}

\subsection{Defining the search grid}

After reducing the data sets and coaligning them, I moved on to studying the network jets themselves. Along with Steve Cranmer, I attempted to figure out a method for automatically detecting jets in an image or running-difference movie. The primary difficulty with any algorithms we tried were with the bright grains that are ubiquitous in the chromosphere \citep{1991SoPh..134...15R}, visible throughout Figure \ref{fig:irisshifts}. These internetwork grains have lifetimes of 1.5-2.5 minutes, and undergo random-walk-like transverse motions in the wavelengths of interest to this project \citep[see, e.g.,][]{2010A&A...519A..58T, 2015ApJ...803...44M}. While these bright grains warrant study on their own, especially regarding their relation to the supergranular network and their apparant lack of a transition region counterpart, they prevented any attempt to search for linear features that exist on only 10s of second timescales, as the random alignment of these grains can occur with roughly the same regularity.

The timeline required to develop any reliable automatic detection of the short-lived jets was sufficiently long relative to our ability to show proof of concept using human detection. For my initial work, I therefore defined a grid of $4'' \times 4''$ boxes, shown in Figure \ref{fig:irisgrid}. The grid in solar coordinates is from -100$''$ to -44$''$ in x and -820$''$ to -788$''$ in y. For each grid cell, I focused my attention on the boxed area while running the movie made of the IRIS image frames. Based on the activity in the grid box over the hour-long observations, I categorized jet production using the following system:
\begin{itemize}
\item 0: No activity, and not near jet outflows or footpoints
\item 1: No activity, but within one grid cell of jet footpoints
\item 2: Likely contains footpoints, but no identifiable jets during observation
\item 3: Contains jet footpoints, and 1-2 identifiable jets during observation
\item 4: Contains jet footpoints, and 3 or more identifiable jets during observation
\end{itemize}

\afterpage{%
\begin{figure}[p!]
\includegraphics[width=\textwidth]{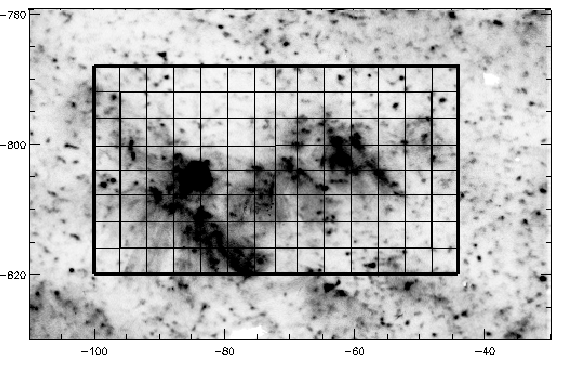}
\caption[Search grid in IRIS dataset]{For each grid cell, we looked at a movie of the 415 IRIS frames over the hour-long period to categorize the jet activity.}
\label{fig:irisgrid}
\end{figure}
\clearpage
}

It should also be noted that I applied additional shifts to account for a drift over the time of the IRIS observations that was not corrected by image cross-correlation. This drift downward of 6 IRIS pixels was smoothly corrected over the 415 IRIS frames. This is a rather subjective method, so to mitigate individual bias, two separate investigators went through this process (Sean C. McKillop and myself). In Figure \ref{fig:jetresults}, we show the individual results alongside the averaged categories. In many cases, we agree on the categorization, especially for the extreme cases of 0 or 4. 

The primary differences in our assigned categories is the specific definition of a jet event. I focused on transient, collimated brightenings; there may be cases where this could be visibly consistent with the heating of closed loops. McKillop focused on looking for bright blobs of material that could be tracked moving along a linear structure; he had used this method when contributing to the network jet discovery paper \citep{2014Sci...346A.315T}. Comparing our results directly, one notes the tendency for my definition to overestimate the jet productivity relative to McKillop's categories.

After defining the jet production category, we turned to the magnetic field data underlying the IRIS images. Our initial investigation uses the state of the magnetic field in a single 12-minute cadence magnetogram near the beginning of the hour of IRIS observations. This allowed us to maximize the signal to noise, since we are relatively close to the solar limb. 

The main properties we focus on are the net flux density and the absolute flux density. Magnetic flux along the line of sight, $\Phi_{\pm}$, is defined by
\begin{equation}
    \Phi_{\pm} = \int B_{\pm} \cdot dA,
\end{equation}
where the plus and minus refer to opposite polarity (toward or away from the observer). The flux imbalance, $\xi$, is then given by
\begin{equation}
    \xi = \frac{\rm net~flux}{\rm absolute~flux} = \frac{\left|\Phi_{+}\right| - \left|\Phi{-}\right|}{\left|\Phi_{+}\right| + \left|\Phi_{-}\right|}.
    \label{eq:xip}
\end{equation}
The quiet Sun exhibits typical values of $|\xi|$ between 0.1 and 0.5. Coronal holes, which are by definition large regions of open field, have $|\xi| \ge 0.7$. In the HMI data, each pixel contains the same $dA$ element. We can rewrite Equation \eqref{eq:xip} in terms of the line-of-sight magnetic field in our sample grid boxes:
\begin{equation}
    \xi_{\rm box} = \frac{B_{+} - |B_{-}|}{B_{+} + |B_{-}|} = \frac{\sum_{i=1}^{n} B_{i}}{\sum_{i=1}^{n} |B_{i}|},
    \label{eq:xib}
\end{equation}
where each grid box contains $n$ HMI pixels (as a reminder, we use an area of $4'' \times 4''$).

\afterpage{%
\begin{figure}[p!]
\includegraphics[width=\textwidth]{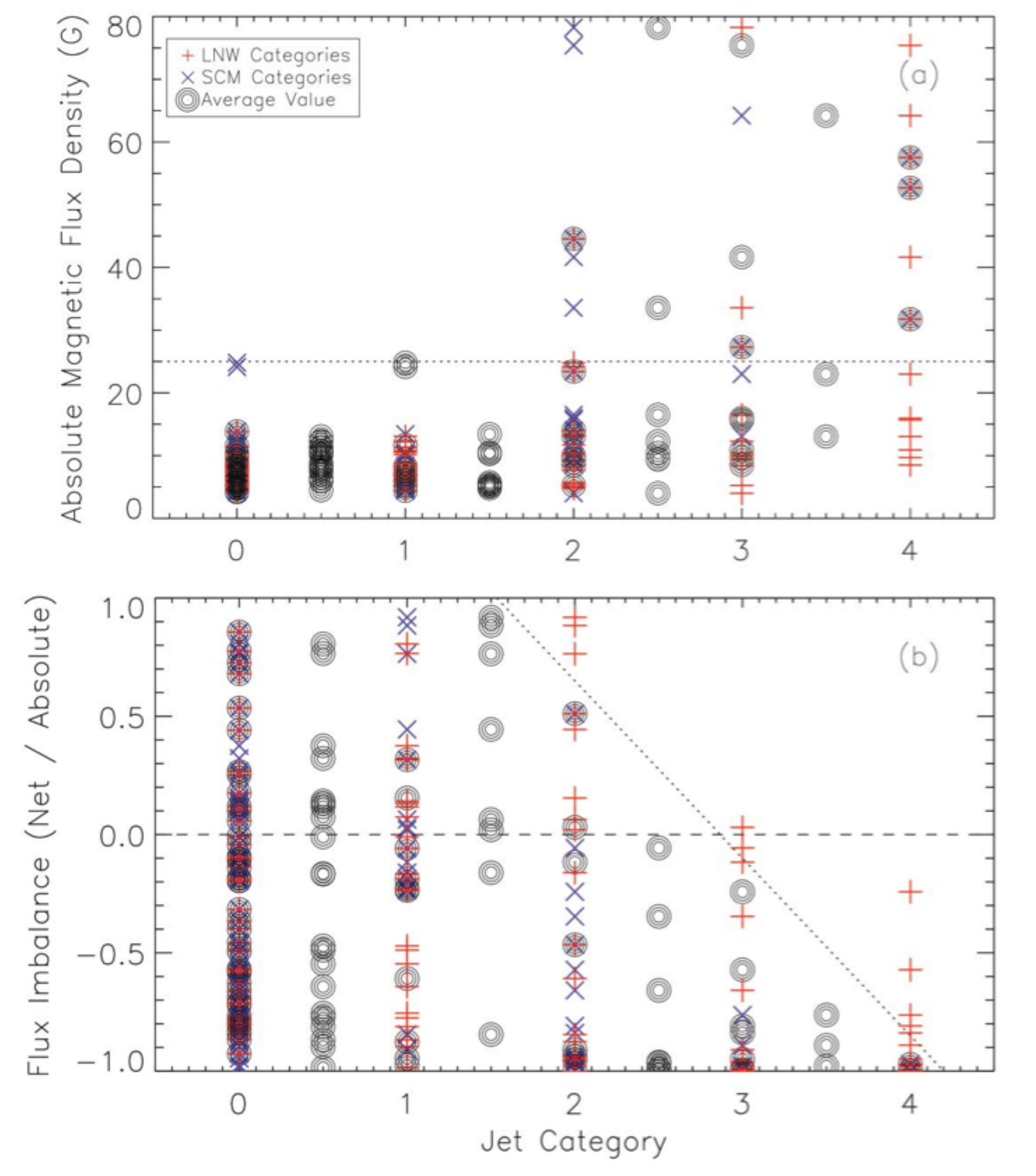}
\caption[Magnetic thresholds in jet production]{Two separate observers (initials shown in legend) gave categories to the grid cells. Here, results are plotted versus (a) absolute magnetic flux density and (b) flux imbalance; higher jet category refers to more active jet production. General trends discussed in text are shown with dotted lines.}
\label{fig:jetresults}
\end{figure}
\clearpage
}

It is important to note that both the 0 and 1 categories indicate no activity. Looking at the results in Figure \ref{fig:jetresults}a, there is a clear threshold in the absolute magnetic flux density at 25 G (25 Mx/cm$^{2}$), shown with a dotted line. If a grid box contains a higher flux density than that threshold, it exhibits some amount of activity, i.e. a category 2 or higher. There are also grid boxes with activity below the threshold, but the regions of strongest jet production (in the three boxes where both observers have marked Category 4) occur above this value.

In addition, there is a trend in the flux imbalance, defined from Equation \eqref{eq:xib}. In Figure \ref{fig:jetresults}b, zero net flux is shown with a dashed line. At higher categories, i.e. higher jet productivity, there is a distinct preference for strong, negative flux imbalance. The preferred negative polarity of the net flux matches the primary polarity of the neighboring coronal hole seen in Figure \ref{fig:iriscontext}. The three boxes with the strongest jet production are all at nearly complete flux imbalance (-1.0). We discuss the implications for these results in the following section.

\section{Discussion}

As this work currently stands, we have shown a proof of concept that there is a lot to learn and understand about the magnetic field environment where these network jets are produced. The main trends tell us that jets occur most frequently where there is high flux density ($|{\rm B}| > 25 ~{\rm Mx/cm}$) that is extemely imbalanced ($|\xi| > 0.9$ for highest category). This suggests that the locations where these network jets form do not contain much opposite polarity ``parasitic flux,'' and are therefore not conducive to magnetic reconnection. It lends weight to our suggested formation mechanism of heating from compressive waves generated by Alfv\'en waves \citep{2015ApJ...812...71C}. 

If the IRIS network jets are an on-disk counterpart to Type II spicules as we are assuming, we know that there are observations of transverse motion \citep{2007Sci...318.1574D} and torsional motion \citep{2012ApJ...752L..12D} which are indicative of the presence of Alfv\'en waves and the ability to form these structures through nonlinear coupling to other wave modes. Additionally, recent studies of Type II spicule temporal evolution \citep{2015ApJ...806..170S} provide constraints on future modeling.

New work by Narang et al. (2016, \citep{2016SoPh..tmp...56N}) studies IRIS network jets in CH and QS regions. The regions chosen are all close to the limb where magnetogram data becomes noisy and unreliable but jets are easier to see in projection. The jets seen in CH regions like ours are consistently observed at higher speeds, (185.6 $\pm$ 61.6) km/s versus (109.1 $\pm$ 39.2) km/s in QS, and reach longer maximum lengths, (4.88 $\pm$ 1.73) Mm in CH versus (3.53 $\pm$ 1.24) Mm in QS. The jets have similar lifetimes (roughly 30 seconds) and footpoint brightening (36\% increase) across all of the regions. These observations can provide additional constraints for models of physical mechanisms for the production of jets to go along with our magnetic constraints from the current work. 

As with any dissertation, the finished product documents what has been done through a graduate school career and marks a major milestone rather than the end of the road. I discuss the planned next steps for this project in the following chapter after a summary of what I have found over the course of this series of research projects.
%%conclusions_phd6.tex
\chapter{Discussion and Conclusions}
%%%%%%%%%%%%%%%%%%%%%%%%%%%%%%%%%%%%%%%%%%%%%%%%

\section{Summary}

The body of this dissertation has described three large investigations that can each stand alone, but also build successively upon one another to extend the community's overall understanding of how the Sun's magnetic field affects the solar wind. From a broad 1D parameter study \citep{2014ApJ...787..160W}, to focused 3D modeling \citep{2015ApJ...811..136W} to observations of small-scale phenomena \citep{2015ApJ...812...71C}, my dissertation has added new knowledge about solar wind acceleration to the field of solar physics.

The initial motivation for looking at a large grid of steady-state one-dimensional models in Chapter 2 was to see how much information is thrown away by current forecasting techniques that use one analytical expression to predict the wind speed from the magnetic field expansion factor \citep{1990ApJ...355..726W,2000JGR...10510465A}. I used potential field source surface extrapolations throughout the solar cycle to define a parameter space. I then sampled that parameter space to create 672 unique magnetic field profiles (see Figure \ref{fig:zt25}).

The results from ZEPHYR for the parameter study confirmed that any single analytical expression for predicting wind speed does a poor job of describing the true role that magnetic field plays. The study also showed that there exist other strong correlations, such as temperature at 1 AU (or maximum temperature) and magnetic field strength at the source surface, that could be used to understand the wind and predict its properties. Those correlations also allowed me to write TEMPEST, a python code that can be incorporated into space weather forecasting frameworks (see Section \ref{sec:tempesttool} for what is needed to make that happen) and can be used for undergraduate research projects (see Section \ref{sec:tempestclassroom} for details).

While I showed that the results from the parameter study agree in an overall statistical sense with observations, the project pushed me to consider more physically realistic modeling. In Chapter 3, I describe the targeted modeling that used the time-dependent, three-dimensional, reduced-magnetohydrodynamics code BRAID on three selected flux tubes. The first order of business was to compare the time-averaged results of the more complex code to the previous results from ZEPHYR. The results agreed well, which proves that less-computationally-intensive modeling like ZEPHYR continues to play an important role in testing new physics and gaining big-picture understanding.

However, the true meat of the BRAID project came from the new results when I looked at time-dependent quantities. Focusing on the coronal hole model, I discussed the power spectrum of fluctuations, multithermal plasma signatures, and the energy contained within discrete heating events. The biggest discovery of the project was that Alfv\'en waves can produce nanoflare-like bursty heating in an open flux tube. Nanoflares have traditionally been attributed to magnetic reconnection \citep{1988ApJ...330..474P}.

Following up on the results that Alfv\'en waves can create bursty turbulent heating, I searched for a way to predict the observational signatures of the heating. Initial work suggested Type II spicules and the recently-discovered chromospheric network jets as likely candidates \citep{2015ApJ...812...71C}. Chapter 4 describes my work to understand the magnetic fields of such structures.

My initial area of study used chromospheric imaging from the {\it Interface Region Imaging Spectrograph} and magnetograms from the Helioseismic and Magnetic Imager on the {\it Solar Dynamics Observatory} (see Appendix \ref{ap:data} for details of the specific datasets used). My proof-of-concept categorization method showed that network jets are produced preferentially in regions that contain a high amount of absolute flux density and that are at extreme flux imbalance. This provides more evidence in favor of Alfv\'en waves rather than magnetic reconnection, as there is a lack of the needed ``parasitic'' flux to cancel in reconnection events. As this is the project that is still ongoing, in Section \ref{sec:jetsnext} I describe the next steps planned for continuing the work that will be submitted for publication in the coming months.

The past hundred-odd pages has described the project on which I have spent my last five years working. Science does not have a stopping point, however, so now that I have summarized the work I have already done, I will describe in the following section what I have yet to do. Finally, I add a few final thoughts in Section \ref{sec:finalthoughts} about the big picture and what the solar physics field is likely to discover in the next few years on the subject.

\newpage
\section{Future Projects}

\subsection{Solar wind forecasting with TEMPEST}
\label{sec:tempesttool}

In Chapter 2, I developed TEMPEST to serve as a tool for the community. As I described in the introduction to this dissertation and in that chapter, the space forecasting community continues to rely heavily on very simple empirical relations for solar wind predictions, even when only to create boundary conditions for more complex models. TEMPEST can fill a hole in the current library of modeling options as it is more physically meaningful than an analytical, empirical expression and is less computationally expensive than the 3D full MHD models.

In order to become a more versatile tool for the community, TEMPEST needs to be improved and extended. I have not optimized TEMPEST in any way, and over the past few years I have learned much more about ways to speed up the code without sacrificing accuracy.

Alongside strictly computer-programming-based improvements, one of the most useful extensions I can make to TEMPEST will be to include stream-stream interactions. Co-rotating interaction regions (CIRs) are thought to be the structures that develop when high-speed solar wind streams catch up with slower streams, and have long been a subject of interest in heliophysics \citep[see, e.g.,][and references therein]{1976GeoRL...3..137S,1978JGR....83.1401G,1999SSRv...89..179C,2013LRSP...10....5O}. There are several ways to implement such stream mapping from near the Sun out to 1 AU, and Riley \& Lionello (2011, \citep{2011SoPh..270..575R}) compared the primary techniques and introduced a method for 1D upwinding. I intend to use this technique, because it sits between the simplest methods (e.g. ballistic approximation) and full 3D MHD, just as TEMPEST does overall.

Given a grid of open flux tubes and their wind properties, using magnetic profiles taken from extrapolations of a single photospheric magnetogram, I can define coordinates of the solar wind streams emanating from these flux tubes in relation to one another. The 1D Upwind Model neglects magnetic field, gravity, and pressure gradient effects, so one must be careful to use it only after TEMPEST solves the full momentum equation out to a reasonable height in the corona (e.g. 30 $R_{\odot}$) where these effects are small.

 The fluid momentum equation in a corotating frame of reference and in one dimension can be written as \citep{1978JGR....83.5563P,2011SoPh..270..575R}
\begin{equation}
-\Omega_{\rm rot}\frac{\partial v_{r}}{\partial \phi} + v_{r}\frac{\partial v_{r}}{\partial r} = 0,
\label{eq:upwind1}
\end{equation}
where $\Omega_{\rm rot}$ is the rotation period of the Sun at a given latitude and $v_{r}$ is the solar wind speed in the radial direction. Recasting this into a grid of $r$ and $\phi$ with indicies $i$, $j$ and taking the fact that $\Omega_{\rm rot} / v_{i,j}$ is always positive for the solar wind (i.e. no inflows), Equation \eqref{eq:upwind1} becomes
\begin{equation}
v_{i+1,j} = v_{i,j} + \frac{\Delta r \Omega_{\rm rot}}{v_{i,j}}\left(\frac{v_{i,j+1} - v_{i,j}}{\Delta \phi}\right)
\label{eq:upwind2}
\end{equation}
and provides an upwind estimate of the radial velocity based on the neighboring streams in longitude. Because of the simplifications used to get to Equation \eqref{eq:upwind2}, it may be necessary to include an estimate of residual acceleration \citep[and references therein]{2011SoPh..270..575R}:
\begin{equation}
v_{acc}(r) = \alpha v_{r0} (1-e^{-r/r_{h}}),
\end{equation}
where $v_{acc}$ is the adjustment to the speed, $\alpha$ is a parameter to affect how much acceleration is expected, set to 0.15 from previous simulations, $v_{r0}$ is the inner boundary speed, and $r_{h}$ is the scale length where the acceleration matters.

The comparison between observations at 1 AU from CR 2068 and the upwind mapping technique yielded a correlation coefficient of 0.98 \citep{2011SoPh..270..575R}. This is very promising for implementation with TEMPEST and is one of the first major improvements planned for the python code.

Once this improvement is added, it will be more physically meaningful to compare TEMPEST results directly to observations, rather than only taking the statistical understanding from our parameter study. I need to study how accurate TEMPEST is compared to exisiting models in the solar wind forecasting community. I also must determine the effect of using different methods for extrapolating magnetic field profiles. I have used potential field source surface models in this work, but predictions can vary greatly if other methods are used \citep{2012LRSP....9....6M,2015SoPh..290.2791E}. Exploring these aspects further will help confirm the validity of the model and make it a better tool for others to use and compare to existing options \citep{2008SpWea...6.8001O}. In the next section, I discuss uses for undergraduate or early graduate student projects.

\subsection{TEMPEST in classroom applications}
\label{sec:tempestclassroom}
As I will be starting a career teaching astronomy and physics at the college level in Fall 2016, I have often thought about many different ways to incorporate my research material in the classroom. This section reproduces my published essay on how TEMPEST could be used in undergraduate projects, either with my own students or for anyone who is interested in solar and space physics. Some of the suggested projects can be done with the current version of TEMPEST, while others may require improvements like those mentioned in the previous section.

The Sun is an astronomical object to which students in all types of physics courses can relate, since it is easily observable and has important connections to Earth that can provide motivation for its study. For example, high-speed solar wind streams produce a greatly increased electron flux in the Earth's magnetosphere and can disrupt satellite communications and power grids on the ground \citep{2011A&A...526A..20V}. Models like TEMPEST (described in Chapter 2) make progress towards the ability to predict the nature of plasma streams that will be rotating towards Earth long before they can damage space-based equipment or create geomagnetic storms. Whether as part of a course in physics, independent study, or directed research, students can use TEMPEST to further their understanding of the role of magnetic fields in this bigger picture.

The Sun goes through an 11-year cycle of high activity and low activity. During low activity periods, called solar minimum, the Sun's magnetic field looks considerably like a dipole field. The poles of the Sun during solar minimum are covered by large “coronal holes,” which are regions of open flux and lower plasma density in the corona. The equator is home to the streamer belt, where the opposite polarities of the two hemispheres join together. However, during solar maximum the magnetic field of the Sun is incredibly chaotic, with small or virtually non-existent polar coronal holes and many active regions where strong bundles of magnetic field have pushed up out of the solar interior. Figure \ref{fig:fourlines}a presents a sketch of some of the many types of structures that are seen in the corona.

\afterpage{%
\begin{figure}
\includegraphics[width=\textwidth]{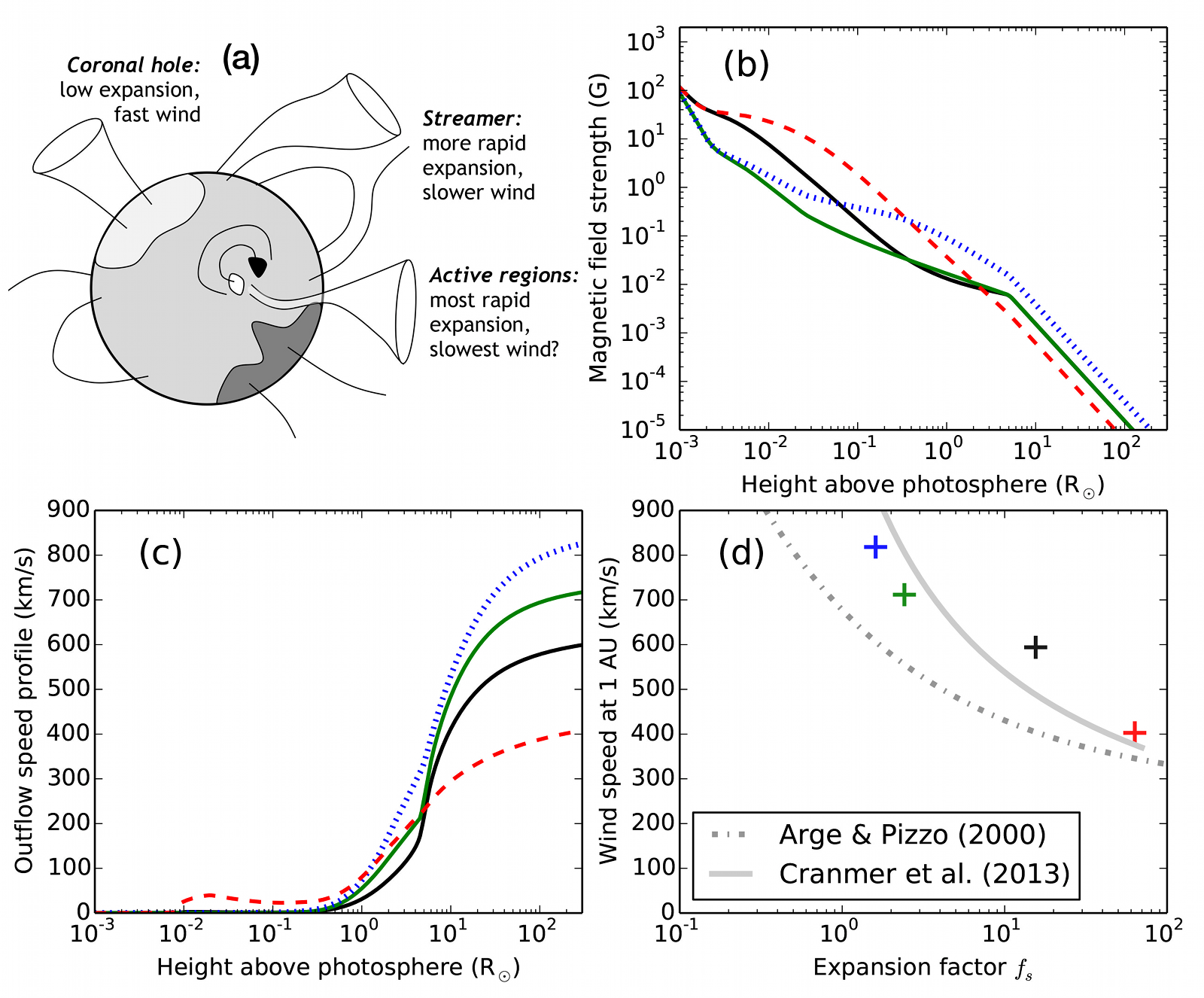}
\caption[Example set of flux tubes in TEMPEST]{(a) A cartoon of possible field structures in the corona, courtesy of S. R. Cranmer. The figure then shows TEMPEST results from four models: (b) magnetic field inputs, (c) wind speed outputs, (d) a comparison of TEMPEST with analytical relations with expansion factor.}
\label{fig:fourlines}
\end{figure}
\clearpage
}

The solar wind is present at all times, but as the Sun goes through different points of the cycle, the properties of the solar wind that reaches Earth change dramatically. Figure \ref{fig:fourlines} shows four models from a large published grid \citep{2014ApJ...787..160W}. Predictions of solar wind speeds used in space weather forecasting often rely on the Wang-Sheeley-Arge (WSA) model \citep{1990ApJ...355..726W,2000JGR...10510465A}, based on a defined quantity called the expansion factor. This factor is a measure of the amount of cross-sectional expansion from the photospheric base of a flux tube, to a “source surface” at a height of 1.5 solar radii above the Sun's surface, where field lines are set to be purely radial and defined as open to the heliosphere. An expansion factor of 1 refers to perfect radial expansion (i.e. an inverse-square relation of the magnetic field strength), while larger expansion factors mean more rapid “super-radial” expansion and vice versa. Several simple analytical relations between expansion factor and wind speed have been put forward. Two have been plotted in Figure \ref{fig:fourlines}d: one that maps the outflow speed at the source surface--the points all lie above it because the wind continues to accelerate above this height \citep{2000JGR...10510465A}--and another to fit results from the model on which TEMPEST is based \citep{2013ApJ...767..125C}.

Students can use TEMPEST to investigate magnetic field profiles throughout the solar cycle. Similar to the analysis on a large grid of synthetic models \citep{2014ApJ...787..160W}, students could investigate how different magnetic field profiles of their own creation can lead to varied solar wind solutions, and how the WSA model works for a variety of structures.

\begin{figure}[h!]
\includegraphics[width=\textwidth]{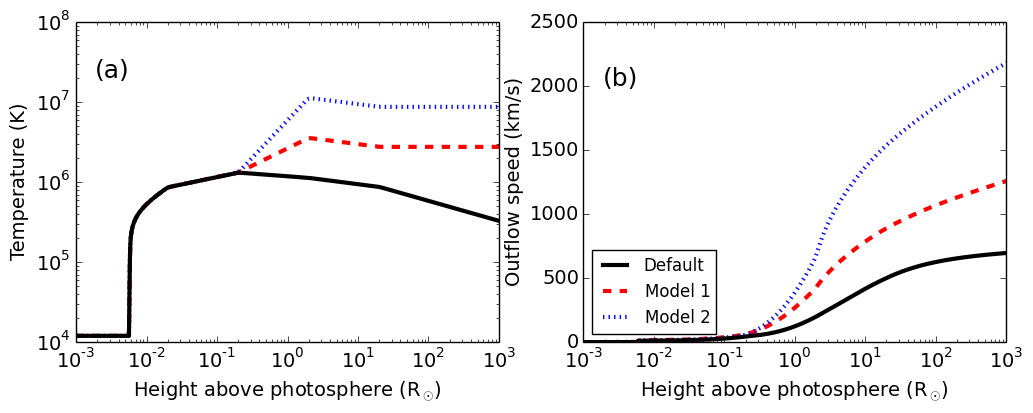}
\caption[Expected relations in TEMPEST from theory]{The relation between (a) temperature profile and (b) outflow speed follows original theory \citep{1958ApJ...128..664P}.}
\label{fig:temps4}
\end{figure}

TEMPEST does not solve the energy conservation equation, because it has set up temperature profiles based on the results from ZEPHYR. Students could investigate how a different temperature profile, due to possible other sources of heating, would affect the solar wind. Using the function in TEMPEST called Miranda, which does not include the wave pressure term, and a different temperature profile, students could accurately use the momentum conservation equation for different coronal heating sources, including mechanisms that do not use wave-driven turbulence. Figure \ref{fig:temps4} shows a quick study of changes to the default TEMPEST temperature profile and profiles that reach higher or lower temperatures at large heights. As Parker (1958; \cite{1958ApJ...128..664P}) originally demonstrated, a hotter corona generally produces a faster wind.

Students can investigate many aspects of stellar winds. How hot does the corona need to be in order to have a steady-state supersonic solar wind? This also allows TEMPEST to be used for other stars, if the gravitational term in the momentum equation was also modified based on the mass and radius of the star. While this goes beyond the scope of most physics courses, it could be a topic of independent study or undergraduate research.

A larger undertaking would be to use TEMPEST to compare real observations to the model output, exposing students to the process of science validation. The magnetic field profile would need to be determined from extrapolations of magnetograms, a map of magnetic field strength and polarity on the Sun's surface. This is most easily achieved using the solar models that can be obtained from the Community Coordinated Modeling Center. Once the magnetic field profiles of Earth-directed flux tubes are run through TEMPEST, predictions from the model of wind speed, density, and temperature at 1 AU can be compared to {\it in situ} spacecraft measurements.

TEMPEST models the steady-state solar wind from open flux tubes, using only the magnetic field profile of the flux tubes as input. While the topic of the model itself is specialized, the physics involved is covered in many basic undergraduate courses. I have discussed the fundamental physics used in TEMPEST (for further detail, see \cite{2014ApJ...787..160W}; for a list of general solar physics resources, see \cite{2010AmJPh..78..890P}), I presented the use in an example coronal hole, and I have provided a few possible student applications of full code and of the associated Python library.

While I have given several possible uses of TEMPEST for student projects, the beauty of science is that success can often be found at the end of a long and widely-branching path. There are likely countless other ways in which TEMPEST can be used for teaching and exploring interesting science; I hope by making it publicly available on GitHub\footnote{http://github.com/lnwoolsey/tempest} it can find such new uses.

\subsection{Next steps with IRIS network jets}
\label{sec:jetsnext}

The work I outlined in Chapter 4 represents ongoing research. Although this section appears after the descriptions of TEMPEST-related projects, the work I will describe here is what I will work on in the coming months with collaborators. Figure \ref{fig:jetresults} shows magnetic properties as a function of jet category, but one of the big things I have to do is quantify those categories as a rate or probability. To do this, we will use the categories to set upper limits somehow and revisit the grid cells that are considered category 4 to better count the number of jets per unit time. The grid cells that are category 3 or 4 can also help us determine what jets might look like in time-distance plots or other ways of looking at the data. This will allow us to revisit the possibility of creating a machine-learning algorithm to automatically detect jets in data sets.

While we chose the 11 November 2014 data set from the archive because it was the best targeted observation of network jets for our purposes, there may be other observations that are conducive to analyzing in parallel with magnetic field data. Such observations will come from the archive, because the throughput of the IRIS slit jaw imager has decreased over the lifetime of the spacecraft \citep{wulserposter}. The lower throughput means that exposures long enough to detect the small-scale jets will last too long to capture any dynamics of the jets from one frame to another, making them impossible to quantify. This is especially true while we still lack any automated detection algorithm.

To round out the project, we are also looking into a way to extend the work of Cranmer \& Woolsey (2015, \citep{2015ApJ...812...71C}) to take the SDO/HMI data as an input to a simple model that follows the logical steps laid out in that paper. Just as the cutoff of 2$''$ reflected naturally the extent of measurable Type II spicules and IRIS network jets, so too could the magnetic thresholds we are finding be a natural consequence of wave mode conversion. This may loop in the kind of modeling I did for the parameter study. For the project from Chapter 2, the boundary conditions on Alfv\'en waves and acoustic waves at the photosphere were held fixed. It may be worthwhile to take a single magnetic field profile, such as the coronal hole from Chapter 3, and run a series of ZEPHYR models with varied input wave properties, after learning more about how the power spectra change with time from the BRAID results.

\section{Final Thoughts}
\label{sec:finalthoughts}

In the current understanding of coronal heating and solar wind acceleration that I laid out in Chapter 1, I described the two populations of solar wind: The fast wind and the slow wind. The work I've done over the last five years has focused on understanding wind emanating from open flux tubes. Coronal holes, the sources of fast wind, are often defined by their predominantly {\it open} magnetic field geometries. I am confident in asserting that the physical processes I have focused on, wave and turbulence-driven models, are the drivers of the acceleration of the fast wind.

It is my opinion, though, that much of the dichotomy of the two types of wind stems from two separate physical mechanisms at work. The opening and closing of coronal flux as a way to release the slow wind is consistent with the FIP effect and charge state ratios. Ongoing work like that from Higginson et al. (2015, \citep{2015TESS....110804H}) successfully models the reconnection-driven slow wind in the larger context of the S-web \citep{2011ApJ...731..112A}. When we measure wind {\it in situ}, neighboring streams have already interacted, smearing out an even greater distinction of these two wind populations. Boundary winds like that described by Stakhiv et al. (2015, \citep{2015ApJ...801..100S}) may be a sharper transition in wind acceleration/release that has been smoothed out to fall between the typical two types.

Looking ahead, the upcoming generation of observatories will be game changing. The ground-based {\it Daniel K. Inouye Solar Telescope} (DKIST, \citep{2014SPIE.9147E..07E}) will drastically improve our view of photospheric magnetic fields. The field of view will be several arcminutes across, with spatial resolutions of 0.03$''$ to 0.08$''$. The research I am doing now with magnetic thresholds in network jet production can serve as a foundation for a larger-scale project when DKIST is fully online, and in fact the work reflects the goals of submited science use cases for the telescope.

Another highly anticipated mission in the near future is {\it Solar Probe Plus} (SPP, \citep{2015SSRv..tmp..105F}). This spacecraft will get within 9 $R_{\odot}$ of the Sun, far closer than we have ever gone before (the previous record is {\it Helios 2} with 62 $R_{\odot}$). The {\it in situ} measurements will be unprecedented and will supplement ground-based observations to give a full understanding of the 3D structure of magnetic fields and heating of plasma in the solar corona.

During my time in graduate school, I have seen the community come together on endeavors like the turbulence dissipation challenge \citep{2013arXiv1303.0204P}, and I have seen how new observations like those from IRIS and the Hi-C sounding rocket \citep{2013Natur.493..501C} have fundamentally change how we view the Sun. With more intricate modeling efforts using improved technologies and the data expected from DKIST and SPP, I look forward to what this field will look like another five years from now.

% ----- Bibliography and Appendix -----
%\nocite{}
\addcontentsline{toc}{chapter}{Bibliography}
{
\ssp
\bibliographystyle{plain}      %numbered by .bib file? (alphabetically)
\bibliography{woolsey_PhD_thesis}{}
}
\newpage

\appendix
\chapter{Weak Zeeman Effect in the Interstellar Medium}
\label{ap:ism}

This dissertation has focused on the role of magnetic fields in accelerating the solar wind and heating the solar corona. Magnetic fields, however exist in a wide range of astrophysical contexts at small scales and at large scales. Everywhere they appear, they can have far-reaching effects that are similarly unsolved. In this appendix, I describe a side project I worked on that takes a look at one of these contexts, the interstellar medium. Magnetic fields may strongly affect star formation, and one of the methods for making observations of the strength and direction of magnetic fields can be simulated with a simple widget for use in classroom demonstrations or public outreach.

When spectral lines come up in astronomy and physics courses, instructors are most commonly discussing the absorption and emission lines created when an electron changes energy levels in an atom \citep{1979rpa..book.....R}. There are, however, subtler processes that also create spectral lines. A well-known example in astronomy is created by neutral Hydrogen (HI; \citep{1990ARA&A..28..215D}). When the spin of the lone electron in an HI atom flips from being parallel with the proton in the nucleus to being anti-parallel, a photon with a wavelength at 21 cm is emitted. The spin-flip transition of HI and other atoms or molecules are often observed in the radio due to the small energy level differences, and I will discuss why long wavelengths are ideal for observing the Zeeman effect in the interstellar medium (ISM). ``When you can use it, it's gold, but there are only very limited places in the universe where Zeeman splitting actually works.''\footnote{Bryan Gaensler, colloquium at Harvard-Smithsonian Center for Astrophysics, 21 March 2013}

These subtle processes that can produce long-wavelength spectral lines are often polarized. Components that are elliptically polarized will be important for magnetic field measurements of the ISM. The amount of splitting between two angular momentum quantum numbers for a single line is given by \begin{equation}\Delta \nu_{Z} = \left(\frac{g \mu_{B} B}{h}\right),\label{eq:zeeman1}\end{equation} where $\mu_{B}$ is the Bohr magneton, $h$ is the Planck constant, and $g$ is the Lande g-factor \citep{1993prpl.conf..279H}.

The strong Zeeman effect, also known as Zeeman splitting, can be done in undergraduate labs \citep[see, e.g.,][]{2010AmJPh..78..503B}. When the magnetic field is not strong enough to split the spectral lines completely, the process is referred to as the weak Zeeman effect. There is an extensive list of candidates for observations of the weak Zeeman effect; I show the most commonly used candidates in Table \ref{tab:zeeman1} \citep{1993prpl.conf..279H}. Here, $b = (2 g \mu_{B} / h)$ Hz/${\mu}$G such that Equation \eqref{eq:zeeman1} can be rewritten as $\Delta \nu_{Z} = (b B / 2)$.

In Table \ref{tab:zeeman1}, the density $n_{\rm H}$ is representative of the density of the environment in which each species is observable, such as molecular clouds (C) or masers (M). Note that each of the lines listed, which are the best candidates for detecting Zeeman splitting, all have frequencies of 22 GHz or lower, which means very long wavelength observations (for comparison, visible light has frequencies on the order of 500 THz). 

\begin{table}[h]\begin{center}
\caption[Spectral lines used to probe interstellar magnetic fields]{Common candidate spectral lines for ISM Zeeman effect observations} \label{tab:zeeman1} 
\begin{tabular}{|l|l|l|l|l|}
\hline
\textbf{Species} & \textbf{Transition} & \textbf{Line $\nu$ (GHz)} & \textbf{b (Hz/$\mu$G)} & \textbf{$n_{\rm H}$ (cm$^{-3}$)} \\ \hline
HI       & $^{2}\Sigma_{1/2}$, F=1-0        & 1.42             & 2.8           & Low (100-300)          \\ \hline
OH       & $^{2}\Pi_{3/2}$, F=1-1        & 1.665            & 3.27          & C: 10$^{3}$, M: 10$^{7}$      \\ \hline
OH       & $^{2}\Pi_{3/2}$, F=2-2        & 1.667            & 1.96          & C: 10$^{3}$, M: 10$^{7}$      \\ \hline
H$_{2}$O & $6_{16} - 5_{23}$ & 22.235           &  0.0029             & M: up to 10$^{9}$ \\ \hline
\end{tabular}\end{center}
\end{table}

The environments that can produce measurable Zeeman effects are denser than much of the ISM. The outer parts of molecular clouds are traced by excess 21-cm emission for HI and have densities of a few hundred molecules per cubic centimeter. OH is found in molecular clouds and is observable for densities up to 2500 cm$^{-3}$. Zeeman splitting in OH emission has also been observed in masers with strong magnetic fields. Hyperfine transition lines of H2O are also seen in masers \citep{1992ASSL..170.....E}. In these extremely high density masers, the magnetic field can be strong enough to classically split the spectral line and the total magnetic field strength can be observed just as for solar magnetograms. 

However, the magnetic fields commonly observed in the interstellar medium in molecular clouds are low enough that the splitting of the spectral line is not great enough to produce a measurable effect. In this case, observers must turn to Stokes parameters. Right circular polarization and left circular polarization of spectral lines can be measured separately; the sum of these is called the Stokes I parameter and the difference is called the Stokes V parameter. With no magnetic field present, the Stokes V spectrum should be a flat line. If a magnetic field is present, the “V-spectrum” can be fit by the derivative of the “I-spectrum” scaled up by the strength of the magnetic field along the line-of-sight as follows:  \begin{equation}V = \Delta\nu_{Z} \frac{dI}{d\nu}\end{equation} \citep{1993prpl.conf..279H,2013ApJ...767...24S}. Thus, by comparing the observed Stokes I and V, a direct measurement of the magnetic field can be made using Equation \eqref{eq:zeeman1}.

To illustrate the way astronomers make these measurements, I have developed an interactive module to show the resulting Stokes parameters due to a set of input environmental conditions \citep{woolsey_module}. It has been written in Mathematica, converted to a Computational Document Format (CDF), and uses free software from Wolfram to run and use. The module has several options that the user can change. They are seen as a pull-down menu and five slider bars in Figure \ref{fig:module}:
\begin{enumerate}
\item Lines: the user can choose a spectral line from the list presented in Table 1.
\item Temperature: hotter environments produce thermally broadened spectral lines.
\item Turbulent Velocity: turbulence broadens spectral lines.
\item Line-of-Sight Magnetic Field: this is what produces the Zeeman splitting.
\item Y-axis Bounds for Stokes V: the magnitude of the Zeeman effect ranges depending on inputs, so the user can ``zoom in'' to where the effect is observable.
\item Signal-to-Noise Ratio: longer observations on a telescope produce a higher signal-to-noise, so the exposure time we need to see a clear signal for given inputs may differ, and this simulates the observational constraint.
\end{enumerate}

\afterpage{%
\begin{figure}
\includegraphics[width=\textwidth]{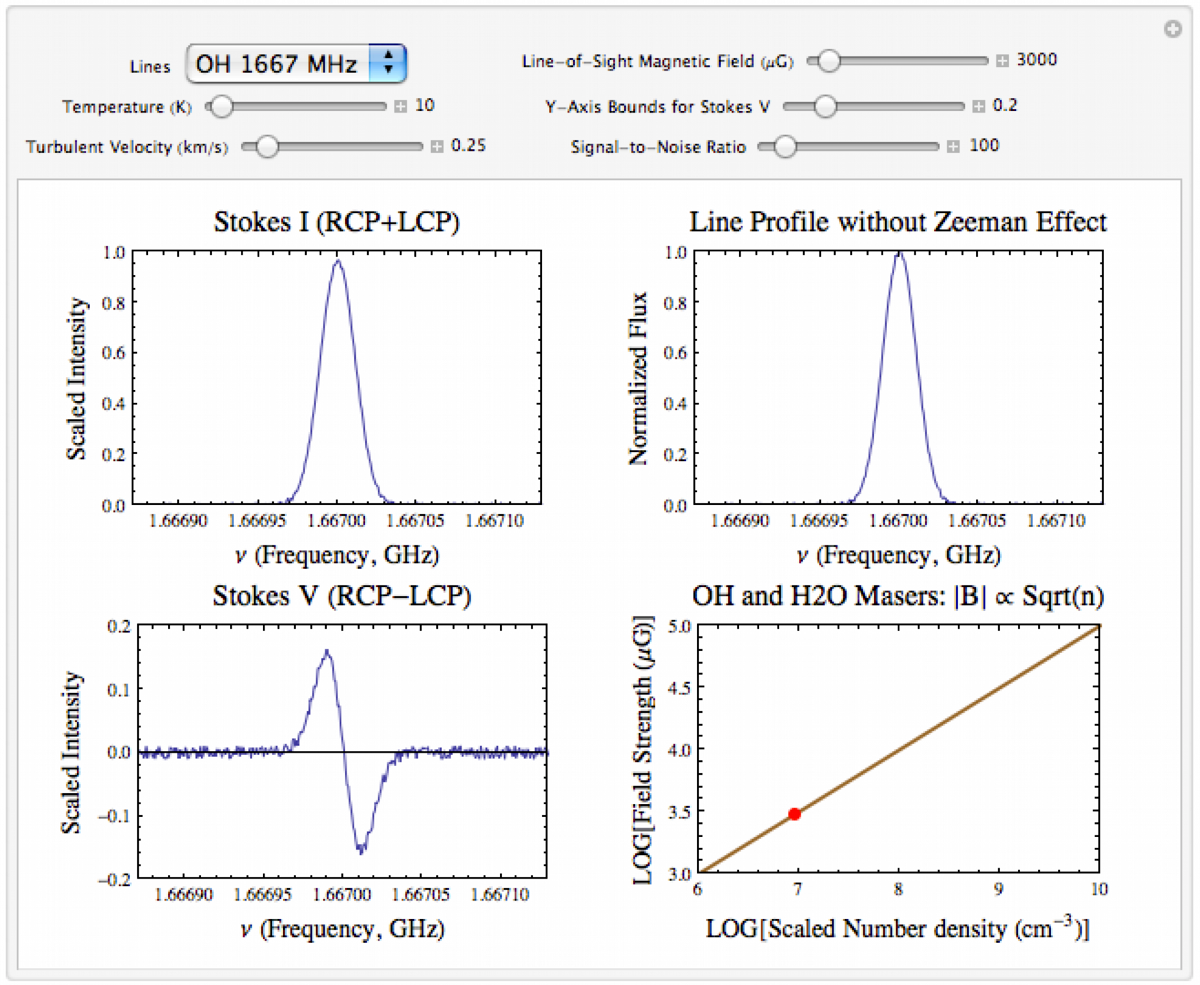}
\caption[Interactive module on the Zeeman effect in the ISM]{Screen capture of complete interactive module using OH at 1667 MHz.}
\label{fig:module}
\end{figure}
\clearpage
}

The module plots Stokes I and Stokes V, discussed previously. There is also a plot of the line profile without Zeeman effect, which is meant to provide a baseline for comparison with the Stokes I parameter. If there is no magnetic field, the top two plots (Stokes I and line profile without Zeeman effect) will be identical, and the Stokes V line will be flat. The flux is normalized to peak at a unitless value of 1 because we have not made any assumptions about the source brightness and distance or the telescope used. These details are beyond the scope of this module and are not necessary to understand the Zeeman effect in a general sense. The final plot represents a rough idea of how the magnetic field strength is related to the density of OH and H$_{2}$O masers. Observations indicate a weak proportionality of the magnitude of line-of-sight magnetic field strength with the square root of $n_{\rm H}$ \citep{2010ApJ...725..466C}. The relation arises from the effect of flux freeze-in, where magnetic field lines are dragged along by the dense medium above a threshold density.

This should serve as a good starting point to start using the module and discovering the difficulty of making different observations of the magnetic field in the interstellar medium. To provide additional guidance, Table \ref{tab:zeeman2} lists typical values of structures in the ISM. The turbulent velocity is approximated by $\sigma_{v} \approx 1.1$ (L [pc])$^{0.38}$ km/s, where L is the size of the structure \citep{1981MNRAS.194..809L}. The ISM has magnetic field strengths that typically range from 1 microgauss to tens of milligauss (0.1 nT to 100s nT).

\begin{table}[h]\begin{center}
\caption {Typical properties of common structures in the interstellar medium} \label{tab:zeeman2} 
\begin{tabular}{|l|l|l|l|l|}
\hline
\textbf{Region}       & \textbf{L (pc)} & \textbf{$n_{H}$ (cm$^{-3}$)} & \textbf{Temperature (K)} & \textbf{$\sigma_{v}$ (km/s)} \\ \hline
Giant Molecular Cloud & 100                   & 100                             & 50                       & 0.5                          \\ \hline
Dark Cloud            & 10                    & 1000                            & 20                       & 0.3                          \\ \hline
Core                  & 0.3                   & 10000                           & 10                       & 0.2                          \\ \hline
\end{tabular}\end{center}
\end{table}
Students can use the module to experiment with similar values in the interactive module to get a sense of how difficult observations of the Zeeman Effect in the interstellar medium can be. I have written a few example situations that students might investigate individually or during a short classroom investigation, but there are numerous uses of this module beyond these questions.
\begin{enumerate}
\item A typical Giant Molecular Cloud (GMC) might be 100 pc across, with an average temperature of 50 K (-370 degrees Fahrenheit!) and turbulent velocity of roughly 0.4 km/s (900 mph!). If line-of-sight magnetic fields are measured to be on the order of 200 microgauss by observing neutral hydrogen's 21-cm line, what is the ratio between peak intensity of Stokes V to Stokes I?
\item Astronomers set out to observe the OH 1665 MHz line in a maser. If the temperature is 300 K (typical room temperature on Earth, relatively hot for the interstellar medium) and there is no turbulence, at what magnetic field strength can we distinguish two completely separate spectral lines in the Stokes I spectrum? What if the turbulent velocity were 0.5 km/s? What if the turbulent velocity were as high as 1 km/s?
\item Let's see how difficult using the Water line at 22 GHz is for measuring the Zeeman effect. Play around with the five sliding controls until you have the clearest Stokes V signal. What were your values? Explain the effect of changing each slider on the signal you observe.
\end{enumerate}

Students should take note of the relative intensity of the Stokes V to the Stokes I and the amount of signal-to-noise required for a clear signal in the Stokes V. Regions of the ISM where such observations are possible make up only a small fraction of the total volume. This module can therefore give students a sense of the difficulties faced in observations compared to the clear and straightforward theory they learn in classes. For advanced students, the module can also be used to reproduce scientific results by determining the magnetic field strength of a specific region based on the shape of a measured signal and the associated environmental properties.

In conclusion, the module is an effective way for students to understand both the ways that magnetic field can be measured and the methods that scientists use to learn about the interstellar medium. It can easily augment classroom lectures and undergraduate labs in spectroscopy. The module was created using Mathematica and the powerful Computing Document Format (CDF) from Wolfram, which can be read with free software (Wolfram CDF Player). While I hope this specific module can find ongoing uses in astronomy courses, I also want it to serve as an example of the capabilities of CDF. For students comfortable with Mathematica, the Wolfram Demonstrations Project is an incredible resource for learning modules like this \citep{woolsey_module}.
\chapter{Temperature and Reflection Coefficient Profiles in TEMPEST}
\label{ap:profiles}
One of the primary differences between the ZEPHYR and TEMPEST codes is the way in which internal energy conservation is handled. ZEPHYR finds a self-consistent solution for the equations of mass, momentum, and energy conservation, including the physical processes of Alfv\'en wave-driven turbulent heating. TEMPEST is meant to be a stand-alone code that runs faster by making reasonable assumptions about these processes. To do so, we use correlations with the magnetic field to set up the temperature and Alfv\'en wave reflection coefficient profiles. We describe this process here.\\ \indent The temperature profile can be described by a relatively constant-temperature chromosphere that extends to the transition region height, $z_{\rm TR}$, a sharp rise to the location of the temperature peak, $z_{max}$, followed by a continued gradual decrease. We found that $z_{\rm TR}$ was best correlated with the strength of the magnetic field at $z_{B} = 2.0$ R$_{\odot}$, and this fit (with Pearson correlation coefficient $\mathcal{R} = -0.29$) is given by \begin{equation}z_{\rm TR} = 0.0057 + \left(\frac{7 \times 10^{-6}}{B(2.0 ~R_{\odot})^{1.3}}\right) R_{\odot}\end{equation}.
\begin{subequations}
We then chose evenly-spaced heights $z_{T}$ in log-space to set the temperature profile according to the following set of linear fits in log-log space with the location along the magnetic field profile that best correlated (Pearson coefficients $\mathcal{R}$ given for each):
\begin{align}
\log_{10}(T(0.02 {\rm ~R}_{\odot})) = 5.554 + 0.1646\log_{10}(B(0.00314 {\rm ~R}_{\odot})) ~~~~~\mathcal{R} = 0.51 \\
\log_{10}(T(0.2 {\rm ~R}_{\odot})) = 5.967 + 0.2054\log_{10}(B(0.4206 {\rm ~R}_{\odot})) ~~~~~\mathcal{R} = 0.83 \\
\log_{10}(T(2.0 {\rm ~R}_{\odot})) = 6.228 + 0.2660\log_{10}(B(2.0 {\rm ~R}_{\odot})) ~~~~~\mathcal{R} = 0.91 \\
\log_{10}(T(20 {\rm ~R}_{\odot})) = 5.967 + 0.2054\log_{10}(B(3.0 {\rm ~R}_{\odot})) ~~~~~\mathcal{R} = 0.87 \\
\log_{10}(T(200 {\rm ~R}_{\odot})) = 5.967 + 0.2054\log_{10}(B(3.0 {\rm ~R}_{\odot})) ~~~~~\mathcal{R} = 0.84
\end{align}
\end{subequations}
\begin{subequations}
We looked for correlations in the residuals of each of these fits, and found well-correlated ($\mathcal{R} > 0.45$) terms for the first two heights, at $z_{T} = 0.02$ R$_{\odot}$ and $0.2$ R$_{\odot}$. We added the following terms to the $\log(T)$ estimates given above in order to improve the overall correlation,
\begin{align}
\log_{10}(T_{resid}(0.02 {\rm ~R}_{\odot})) = 0.0559 + 0.13985\log_{10}(B(0.662 {\rm ~R}_{\odot}))\\
\log_{10}(T_{resid}(0.2 {\rm ~R}_{\odot})) = -0.0424 + 0.09285\log_{10}(B(0.0144 {\rm ~R}_{\odot})).
\end{align}
\end{subequations}
After adding these terms, the correlation coefficients improved to $\mathcal{R} = 0.78$ at $z_{T} = 0.02{\rm ~R}_{\odot}$ and $\mathcal{R} = 0.94$ at $z_{T} = 0.2{\rm ~R}_{\odot}$. With all of these fitted values, we then constructed the temperature profile with the following continuous piecewise function, where the chromosphere is at a constant temperature $T_{\rm TR} = 1.2 \times 10^{4}$ K and $aT \equiv log_{10}(T)$:
\footnotesize
\[T(z) = \left\{
  \begin{array}{lr}
	T_{\rm TR} & : z \le z_{\rm TR}\\
	\left[T_{\rm TR}^{3.5} + \left(\frac{T(0.02 {\rm ~R}_{\odot})^{3.5} - T_{\rm TR}^{3.5}}{(0.02 {\rm ~R}_{\odot})^{2} - z_{\rm TR}^{2}}\right)(z^{2} - z_{\rm TR}^{2})\right]^{2/7} & : z_{\rm TR} < z \le 0.02 {\rm ~R}_{\odot}\\
	10^{x}, x = \left(aT(0.02 {\rm ~R}_{\odot}) + \frac{aT(0.2 {\rm ~R}_{\odot}) - aT(0.02 {\rm ~R}_{\odot})}{\log_{10}(0.2) - \log_{10}(0.02) }(\log_{10}(z) + 1.7)\right) & : 0.02 {\rm ~R}_{\odot} < z \le 0.2 {\rm ~R}_{\odot}\\
	10^{x}, x = \left(aT(0.2 {\rm ~R}_{\odot}) + \frac{aT(2.0 {\rm ~R}_{\odot}) - aT(0.2 {\rm ~R}_{\odot})}{\log_{10}(2.0) - \log_{10}(0.2) }(\log_{10}(z) + 0.7)\right) & : 0.2 {\rm ~R}_{\odot} < z \le 2.0 {\rm ~R}_{\odot}\\
	10^{x}, x = \left(aT(2.0 {\rm ~R}_{\odot}) + \frac{aT(20 {\rm ~R}_{\odot}) - aT(2.0 {\rm ~R}_{\odot})}{\log_{10}(20) - \log_{10}(2.0) }(\log_{10}(z) - 0.3)\right) & : 2.0 {\rm ~R}_{\odot} < z \le 20 {\rm ~R}_{\odot}\\
	10^{x}, x = \left(aT(20 {\rm ~R}_{\odot}) + \frac{aT(200 {\rm ~R}_{\odot}) - aT(20 {\rm ~R}_{\odot})}{\log_{10}(200) - \log_{10}(20) }(\log_{10}(z) - 1.3)\right) & : z > 20 {\rm ~R}_{\odot}
  \end{array}
\right.
\]\\
\normalsize
\indent We proceeded with a similar method to create the Alfv\'en wave reflection coefficient $R$ used in TEMPEST. For the same evenly spaced in log-space heights ($0.02,~0.2,~2.0,~20,~200$ R$_{\odot}$), we found linear fits between $\log_{10}(B)$ and $\log_{10}(R)$. They are (with Pearson coefficients $\mathcal{R}$):
\begin{subequations}
\begin{align}
\log_{10}(R(0.02 {\rm ~R}_{\odot})) = -1.081 + 0.3108\log_{10}(B(0.011 {\rm ~R}_{\odot})) ~~~~~\mathcal{R} =  0.64\\
\log_{10}(R(0.2 {\rm ~R}_{\odot})) = -1.293 + 0.6476\log_{10}(B(0.573 {\rm ~R}_{\odot})) ~~~~~\mathcal{R} =  -0.20\\
\log_{10}(R(2.0 {\rm ~R}_{\odot})) = -2.238 + 0.0601\log_{10}(B(0.0.315 {\rm ~R}_{\odot})) ~~~~~\mathcal{R} = 0.70 \\
\log_{10}(R(20 {\rm ~R}_{\odot})) = -2.940 - 0.2576\log_{10}(B(3.0 {\rm ~R}_{\odot})) ~~~~~\mathcal{R} = -0.27 \\
\log_{10}(R(200 {\rm ~R}_{\odot})) = -3.404 - 0.4961\log_{10}(B(3.0 {\rm ~R}_{\odot})) ~~~~~\mathcal{R} = -0.38
\end{align}
\end{subequations}
It is important to note that the correlations are not as strong for this set of fits as they were for the temperature profile. However, we found no additional strong correlations in the residuals, and the effect due to the difference in the reflection coefficient between ZEPHYR and TEMPEST is small. Finally, with these fits we defined the following continuous piecewise function, $aR \equiv \log_{10}(R)$:
\footnotesize
\[R(z) = \left\{
  \begin{array}{lr}
	\frac{B(0.00975 {\rm ~R}_{\odot})}{0.7 + B(0.00975 {\rm ~R}_{\odot})} & : z \le z_{\rm TR}\\
	10^{x}, x=\left(   aR(z_{\rm TR}) + \frac{aR(0.02 {\rm ~R}_{\odot}) - aT(z_{\rm TR})}{\log_{10}(0.02) - \log_{10}(z_{\rm TR}) }(\log_{10}(z) - \log_{10}(z_{\rm TR}))    \right) & : z_{\rm TR} < z \le 0.02 {\rm ~R}_{\odot}\\
	10^{x}, x=\left(aR(0.02 {\rm ~R}_{\odot}) + \frac{aR(0.2 {\rm ~R}_{\odot}) - aT(0.02 {\rm ~R}_{\odot})}{\log_{10}(0.2) - \log_{10}(0.02) }(\log_{10}(z) + 1.7)\right) & : 0.02 {\rm ~R}_{\odot} < z \le 0.2 {\rm ~R}_{\odot}\\
	10^{x}, x=\left(aR(0.2 {\rm ~R}_{\odot}) + \frac{aR(2.0 {\rm ~R}_{\odot}) - aT(0.2 {\rm ~R}_{\odot})}{\log_{10}(2.0) - \log_{10}(0.2) }(\log_{10}(z) + 0.7)\right) & : 0.2 {\rm ~R}_{\odot}< z \le 2.0 {\rm ~R}_{\odot}\\
	10^{x}, x=\left(aR(2.0 {\rm ~R}_{\odot}) + \frac{aR(20 {\rm ~R}_{\odot}) - aT(2.0 {\rm ~R}_{\odot})}{\log_{10}(20) - \log_{10}(2.0) }(\log_{10}(z) - 0.3)\right) & : 2.0 {\rm ~R}_{\odot}< z \le 20 {\rm ~R}_{\odot} \\
	10^{x}, x=\left(aR(20 {\rm ~R}_{\odot}) + \frac{aR(200 {\rm ~R}_{\odot}) - aT(20 {\rm ~R}_{\odot})}{\log_{10}(200) - \log_{10}(20) }(\log_{10}(z) - 1.3)\right) & : 20 {\rm ~R}_{\odot}< z \le 200 {\rm ~R}_{\odot}\\
R(200 {\rm ~R}_{\odot}) & : z > 200 {\rm ~R}_{\odot}\\
  \end{array}
\right.
\]\\
\normalsize
The final step we followed to set up both the temperature and reflection coefficient profiles was to smooth each of the piecewise functions with a Bartlett window $w(x)$ of width $N = 15$, where \begin{equation}w(x) = \frac{2}{N-1}\left(\frac{N-1}{2} - \left|x - \frac{N-1}{2}\right|\right).\end{equation} These final profiles are presented in Figures \ref{fig:tempest_temps}b and \ref{fig:refl}b.
\chapter{IRIS and SDO Data Files for 2014.11.11}
\label{ap:data}
It is important to catalog the specific data files used in any observational project. This is especially true when the databases are publicly accessible and the work was done with archival data, making it straightforward for others to access the same data if desired. I include here as much information about accessing this data as possible, in case anyone would like to replicate the results. The data retrieval was all done prior to October 2015. 

IRIS is a NASA small explorer mission developed and operated by LMSAL with mission operations executed at NASA Ames Research center and major contributions to downlink communications funded by ESA and the Norwegian Space Centre. Figure \ref{fig:iriscontext} in this dissertation is courtesy of NASA/SDO and the AIA, EVE, and HMI science teams.

\subsection*{Interface Region Imaging Spectrograph}
Using the IRIS data search (http://iris.lmsal.com/search/) with ``network jets'' in the description box for the time period between launch and October 2015, there are 21 observations where network jets were a stated goal. We then limited our results to A) regions far enough from the limb for vector magnetograms to be useable, i.e. within $60^{\circ}$ viewing angle of disk center, or within $830''$ in solar-xy coordinates, and B) close enough to the limb for network jets to be seen in projection.

There were several good candidate observations in November 2014. These all fell under a coordinated observing plan with Hinode, called HOP270. While we intend to extend our work to the other portions of this coordination effort, we focused on an observation in the middle of a set of large sit-and-stares. With a goal of ``HOP 270 Coronal Hole Jets,'' the observation we use in Chapter 5 has an OBSID of 3844259554, runs from 12:39 to 13:44 on 11 November 2014 with a cadence of 9 seconds, is centered at (x,y) = $-27'',-759''$ in solar coordinates, and uses the 1330\AA\ wavelength in the slit-jaw imager. The level 2 FITS (iris\_l2\_20141111\_123921\_3844259554\_SJI\_1330\_t000.fits) and the others in HOP 270 were downloaded in May 2015.

\subsection*{Solar Dynamics Observatory: AIA and HMI}
I initially used Virtual Solar Observatory\footnote{Available at {\tt http://sdac.virtualsolar.org/cgi/search}.} to search for observations from Solar Dynamics Observatory for this time frame. I looked for both vector magnetic field data from Helioseismic and Magnetic Imager (HMI) and the Atmospheric Imaging Assembly (AIA). AIA was used to get the coronal context (193\AA) and to align the IRIS and SDO fields of view (1600\AA). However, to get the data, I turned to Joint Science Operations Center (JSOC)\footnote{Available at {\tt http://jsoc.stanford.edu}.} to get the HMI Milne-Eddington vector magnetograms (hmi.ME) and the AIA level 1 data sets. The hmi.B data series is full-disk disambiguated vector magnetic field data, available for dates after December 2013. Below are the JSOC export request IDs as searched:

\begin{itemize}
\item {\bf JSOC\_20150716\_283} \\
op=exp\_request\&ds=hmi.B\_720s[2014.11.11\_12:35\_TAI-2014.11.11\_13:50\_TAI]\\ 
\&sizeratio=1\&process=n=0|no\_op\&requestor=Lauren Woolsey\\
\&notify=lwoolsey@cfa.harvard.edu\&method=url\&filenamefmt=hmi.B\_720s.\\
\{T\_REC:A\}.\{segment\}\&format=json\&protocol=FITS,compress Rice

\item {\bf JSOC\_20150701\_275} \\
op=exp\_request\&ds=hmi.ME\_720s\_fd10[2014.11.11\_12:35\_TAI-\\
2014.11.11\_13:50\_TAI]\&sizeratio=1\&process=n=0|no\_op\\
\&requestor=Lauren Woolsey\&notify=lwoolsey@cfa.harvard.edu\\
\&method=url\&filenamefmt=hmi.ME\_720s\_fd10.\{T\_REC:A\}.\{segment\}\\
\&format=json\&protocol=FITS,compress Rice\&dbhost=hmidb2

\item {\bf JSOC\_20150627\_1703} \\
op=exp\_request\&ds=aia.lev1\_uv\_24s[2014-11-11T12:39:15/30s]\\
\&sizeratio=1\&process=n=0|no\_op\&requestor=Lauren Woolsey\\
\&notify=lwoolsey@cfa.harvard.edu\&method=url\\
\&filenamefmt=aia.lev1\_uv\_24s.\{T\_REC:A\}.\{WAVELNTH\}.\{segment\}\\
\&format=json\&protocol=FITS,compress Rice\&dbhost=hmidb2
\end{itemize}

\newpage
\thispagestyle{empty}
\vspace*{\fill}
\begin{center}
{\it END.}
\end{center}
\vspace{\fill}
%\mbox{}
\end{document}